\documentclass[
 reprint,
 superscriptaddress,
 amsmath,amssymb,
 aps,
]{revtex4-2}

\usepackage{graphicx}
\usepackage{dcolumn}
\usepackage{bm}
\usepackage{hyperref}

\usepackage{tcolorbox}

\newcounter{boxcounter}

\newtcolorbox[auto counter]{mybox}[2][]{%
    title=Box~\thetcbcounter: #2, #1}

\usepackage{color}
\definecolor{blue}{rgb}{0.00,0.00,0.95}

\begin{document}

\preprint{APS/123-QED}

\title{bioSBM: a random graph model to integrate epigenomic data in chromatin structure prediction}

\author{Alex Chen Yi Zhang}
 \email{alexchenyi.zhang@ist.ac.at}
\affiliation{%
  Institute of Science and Technology Austria (ISTA), Klosterneuburg AT-3400, Austria
}%
\affiliation{%
  Scuola Internazionale Superiore di Studi Avanzati (SISSA), Via Bonomea 265, 34136 Trieste, Italy
}%
\author{Angelo Rosa}
 \email{anrosa@sissa.it}
\author{Guido Sanguinetti}
 \email{gsanguin@sissa.it}
\affiliation{%
  Scuola Internazionale Superiore di Studi Avanzati (SISSA), Via Bonomea 265, 34136 Trieste, Italy
}%
\date{\today}

\begin{abstract}
The spatial organization of chromatin within the nucleus plays a crucial role in gene expression and genome function.
However, the quantitative relationship between this organization and nuclear biochemical processes remains under debate.
In this study, we present a graph-based generative model, bioSBM, designed to capture long-range chromatin interaction patterns from Hi-C data and, importantly, simultaneously link these patterns to biochemical features.
Applying bioSBM to Hi-C maps of the GM12878 lymphoblastoid cell line, we identified a latent structure of chromatin interactions, revealing $7$ distinct communities that strongly align with known biological annotations.
Additionally, we infer a linear transformation that maps biochemical observables, such as histone marks, to the parameters of the generative graph model, enabling accurate genome-wide predictions of chromatin contact maps on out-of-sample data, both within the same cell line, and on the completely unseen HCT116 cell line under RAD21 depletion. 
These findings highlight bioSBM's potential as a powerful tool for elucidating the relationship between biochemistry and chromatin architecture and predicting long-range genome organization from independent biochemical data. 
\end{abstract}

\maketitle

\section{Introduction}\label{sec:level1}
A characteristic feature of a eukaryotic cell, as opposed to archaea and eubacteria, is the sequestration of the cellular genome in a tight cellular space called the nucleus. In humans, the approximately two-meter-long chromatin filament is tightly packed in a nucleus of 5-10 $\mu m$ in diameter.
This packing is highly organized, as demonstrated by immunofluorescence microscopy experiments that show the heterogeneous subnuclear localization patterns of various proteins and histone marks thus hinting at the existence of functionally distinct compartments within the nucleus~\cite{matera_nuclear_2009,banani_biomolecular_2017,wang_histone_2019,ladouceur_clusters_2020}.

The advent of Chromatin Conformation Capture (3C)~\cite{dekker_capturing_2002}, particularly its now popular derivative Hi-C~\cite{lieberman-aiden_comprehensive_2009}, allowed the mapping of chromatin contacts at a genome-wide level.
Thanks to the fixation of nuclei with formaldehyde, which preserves information about the spatial proximity of linearly distal DNA loci, Hi-C experiments generate contact frequency maps that display non-trivial interaction motifs.
One notable feature of the organization of chromatin unveiled by Hi-C is the segregation of the genome into two classes of domains dubbed A and B compartments, which are characterized by distinct interaction patterns.
On top of these connectivity differences, A and B compartments have been shown to correlate with epigenetic marks associated with active (A) and silenced (B) transcriptional states~\cite{lieberman-aiden_comprehensive_2009}.
Despite these correlations, more detailed microscopy studies and information from epigenetics ({\it e.g.}, histone modifications and preferential binding of transcription factors~\cite{nichols_principles_2021,rowley_organizational_2018,rao_3d_2014,hedstrom_enhancer-insulator_2024, ernstChromatinstateDiscoveryGenome2017, boixRegulatoryGenomicCircuitry2021}) suggest that the transcriptional state of the genome is more nuanced and the binary classification into A/B compartments may be excessively oversimplified.
In particular, while A/B compartments and the sub-compartments defined by Rao {\it et al.}~\cite{rao_3d_2014}(further studied in numerous other works~\cite{wangSPINRevealsGenomewide2021, xiongRevealingHiCSubcompartments2019, spracklinDiverseSilentChromatin2023})seem to point out the existence of a direct statistical correlation between epigenetic marks and 3D chromatin interaction patterns, there is comparably much less quantitative understanding of the ``microscopic'' processes at the origin of these correlations. 
In general, several attempts to explain the emergence and spatial organization of 3D compartments have invoked polymer-based models~\cite{bianco_polymer_2018,di_pierro_novo_2017,jost_modeling_2014,zhang_bottom-up_2024}, which aim at rationalizing the observed 
structure as the consequence of direct, sequence-specific interactions between distal chromatin loci. 
While such models 
connecting the biochemistry of epigenetics 
to 
genome folding provide valuable insights, their reliance on polymer simulations results in high computational cost, which becomes especially problematic when one tries to scale simulations up to the size of a typical chromosome~\cite{zhang_bottom-up_2024}.  
In contrast, more recent deep learning models scale efficiently and have achieved impressive success in predicting Hi-C contact maps.
Early models make predictions using DNA sequence alone, but these predictions do not account for cell-type-specific variability~\cite{fudenberg_predicting_2020,schwessinger_deepc_2020,zhou_sequence-based_2022}.
More recently, deep learning approaches have started to incorporate 1D epigenetic signals~\cite{yang_epiphany_2023}, improving predictive accuracy across different cellular conditions.
However, these models remain largely uninterpretable, making it difficult to connect their predictive power with underlying biological mechanisms.

In recent years, ideas from the field of network or graph theory~\cite{newman_networks_2016} have emerged as a promising paradigm to study chromatin organisation at the mesoscopic level.~\cite{ashoorGraphEmbeddingUnsupervised2020} 
These methods avoid the computational overheads of microscopic polymer-based models by abstracting chromatin structure as a network of interactions, where DNA loci are treated as nodes and their contacts as edges.
Such graph-based approaches not only were successfully used to reveal structural patterns~\cite{cabreros_detecting_2015,norton_detecting_2018, hedstrom_identifying_2024} but also provided interpretable insights into the relationship between chromatin architecture and biological function~\cite{pancaldi_integrating_2016,pancaldi_chromatin_2021,pancaldi_network_2023}.


In this paper, we propose bioSBM, an interpretable network model that directly links chromatin structure with biochemical features.
bioSBM is based on the stochastic block model (SBM)~\cite{wang_stochastic_1987,snijders_estimation_1997}, a class of generative network models that partition the network into communities based on interaction patterns, making them highly suitable for uncovering latent structures in chromatin interaction maps. The first SBM attempt to model long-range chromatin contacts was proposed in 2015 by Cabreros {\it et al.}~\cite{cabreros_detecting_2015}.
bioSBM builds on this previous study by modulating this community structure by considering biochemical covariates such as histone modifications and transcription factor binding, therefore constructing a quantitative framework to understand the relationship between 3D chromatin organization and biochemical processes, a problem already studied in previous works such as~\cite{lanIntegrationHiCChIPseq2012}.
Unlike traditional SBM's, which assign each node to a single community, bioSBM allows for mixed memberships, enabling genomic regions to participate in multiple communities simultaneously, thus capturing the context-dependent nature of chromatin interactions.

We apply our model to Hi-C data from the GM12878 lymphoblastoid cell line, where we identify interpretable community structures that include, but go beyond, the conventional A/B compartmentalization and subcompartments~\cite{rao_3d_2014}.
In addition to community detection, we can infer the map from biochemical features to the community composition of entire chromosomes and we learn the interaction patterns that link the various communities.
Finally, the results of our inference allow us to show that bioSBM can serve as a generative model capable of predicting Hi-C maps for unseen chromosomes and cellular conditions, further demonstrating its robustness and utility.

The paper is organized as the following:
In Section~\ref{sec:model_and_methods} we provide a detailed overview of the \textit{vanilla} stochastic block model and discuss how it can be adapted to better describe different types of analyzed data.
In particular, we provide details of the data we utilize (distance-corrected, or {\it observed-over-expected}, Hi-C maps) in Sec.~\ref{sec:SBMs}, and our customized version of the SBM in Sec.~\ref{sec:bioSBM_model}.
Then, in Sec.~\ref{sec:inference_algo} we introduce the main features of the inference algorithm specifically developed to compute posterior probabilities for our model, while leaving the mathematical details of its derivation in the Supplemental Material (SM~\cite{SMnote}).
We present our main results in Sec.~\ref{sec:results}, demonstrating the biological relevance and predictive power of our bioSBM model.
Finally, in Sec.~\ref{sec:DiscConcls} we discuss our results in the context of chromosome organization and conclude by highlighting, in particular, possible future applications.

\section{Model and Methods}\label{sec:model_and_methods}

\subsection{SBM and its generalization}\label{sec:SBMs}
SBM's are a particular class of random graphs. 
A graph $\mathcal{G}=(\mathcal{V},\mathcal{E})$ consists of a set of vertices $\mathcal{V}$, representing entities $1,\ldots,N$ and a set of edges $(i,j)\in\mathcal{E}\subset \mathcal{V}\times\mathcal{V}$, denoting pairwise interactions between these entities.
Edges can be binary, indicating the presence or the absence of a link, or they can be weighted, reflecting the {\it strength} of interactions.
Random graphs~\cite{frieze_introduction_2015,bollobas_random_2001} denote graphs whose edges (or edge values) are generated according to a probability distribution, making the graph structure itself random.

SBM's have their roots in the world of social sciences~\cite{lee_review_2019,daudin_mixture_2008}, where they were used to model populations divided into sub-populations or {\it communities}.
The central idea is that interactions between individuals are influenced by their community, creating a non-trivial structure in the interaction graph.
This simple yet powerful idea made block models into popular models to study general types of relational data. 

Numerous algorithms have been developed to detect community structures in complex networks and to make sense of them~\cite{abbe_community_2018,zhou_sequence-based_2022,yang_epiphany_2023,cabreros_detecting_2015}. 
Box~\ref{box:stochastic_block_models} provides a schematic overview of the main ideas behind SBM's and of some of their most common extensions.

In our case, the relational data is derived from Hi-C contact frequency maps.
Early Hi-C experiments demonstrated that contact frequency between genomic loci is strongly influenced by {\it linear proximity} or genomic distance, with the contact probability exhibiting a power-law decay as a function of this distance~\cite{lieberman-aiden_comprehensive_2009}.
This scaling behavior can be explained based on fundamental polymer physics mechanisms that shape the three-dimensional folding of chromosomes~\cite{Grosberg1993,LangowskiPRE1998,RosaEveraersPlosCB2008,rosa_looping_2010,grosberg_how_2012,halverson_melt_2014}.
Instead of using raw Hi-C data, it is often more insightful to study the so-called {\it observed-over-expected} (OE) Hi-C maps.
These OE maps are derived as the logarithmic ratio between the actual contact frequencies recorded in Hi-C matrices and the expected contact frequencies based on genomic distances.
These maps effectively highlight interaction patterns of chromatin by accounting for and removing the global polymeric effects that contribute to the power-law scaling.
Significant interactions can be observed across large genomic scales, sometimes spanning entire chromosomes.
To uncover a latent structure in these interaction patterns, we employ a weighted SBM, and to capture the complexity of community structures we use a so-called {\it mixed membership} version of the SBM (MMSBM, see Sec.~\ref{sec:bioSBM_model}) that allows the same genomic regions to belong to multiple communities.

\onecolumngrid
\begin{mybox}[label=box:stochastic_block_models, parbox=false,width=\textwidth]{Flavours of Stochastic Block Models (SBM's)}
SBM's are a particular type of generative models used in network theory to describe the structure of networks by dividing nodes into communities or blocks.
Each block represents a group of nodes that have a similar pattern of connections.
The SBM assumes that the probability of a connection between any two nodes depends only on the blocks to which the nodes belong.
This model helps understand the network's underlying structure and is commonly used for community detection~\cite{abbe_community_2018,lee_review_2019,daudin_mixture_2008}.
The generative process defined by an SBM is as follows:
\begin{itemize}
    \item
    Determine the communities and their total number, $K$;
    \item
    Assign each node to one of the $K$ communities;
    \item
    For each pair of nodes, generate an edge with a probability that depends on the communities of the nodes.
    Specifically, generate a Bernoulli random variable with the parameter of the distribution depending on the colors of the two nodes.
\end{itemize}
The left panel of the figure shows a stochastic block model with $K=3$ communities.
In this example, the graph is an {\it assortative} SBM, meaning that intra-community edge probabilities are higher than inter-community ones.   

\begin{minipage}{\linewidth}
        \centering
        \includegraphics[width=.9\linewidth]{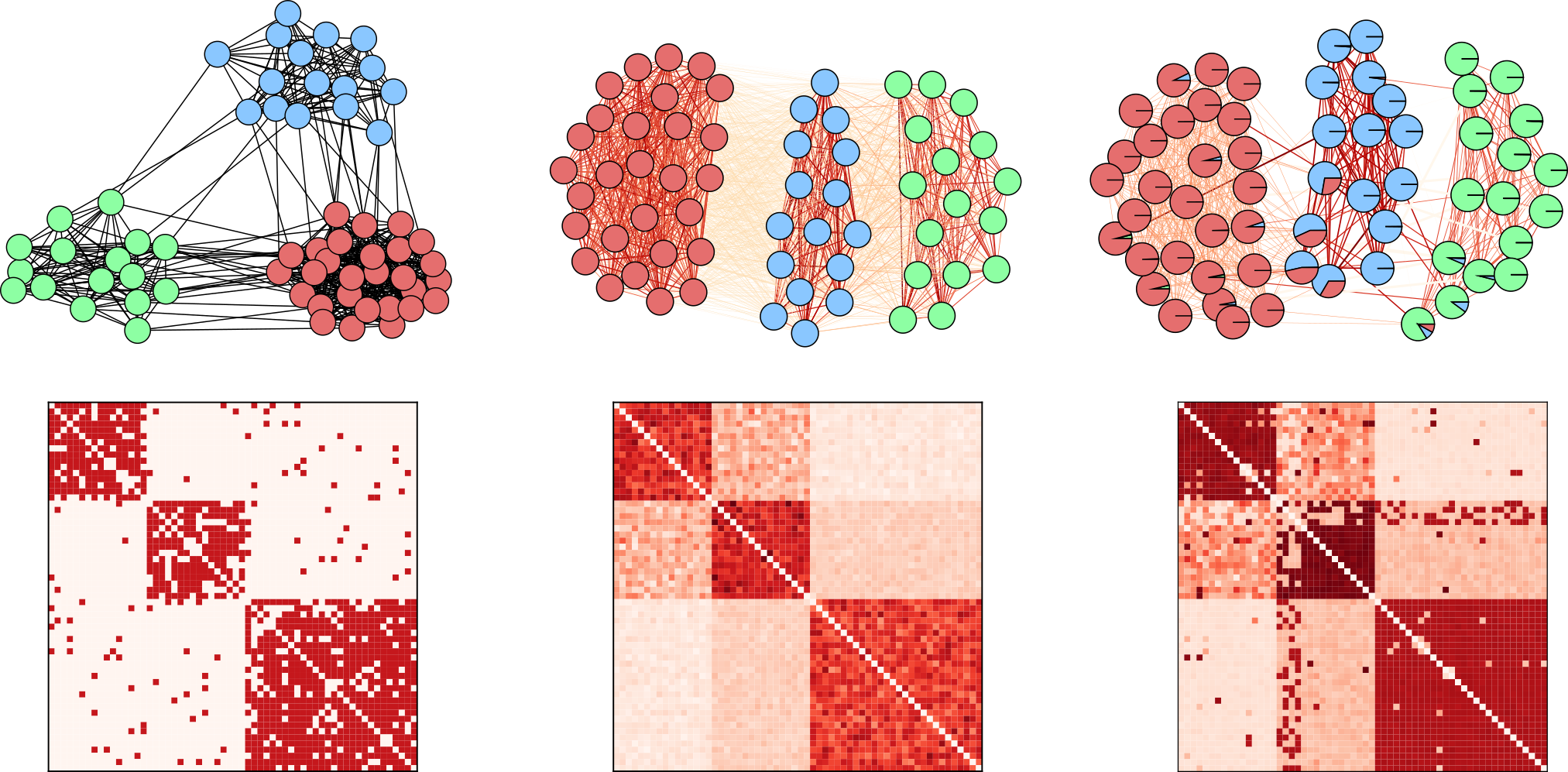}
        \label{fig:enter-label}
    \end{minipage}
In its basic version, the SBM is binary, {\it i.e.} the edges between nodes are either present or absent.
However, many real-world networks involve weighted edges, where the connections between nodes have different strengths or capacities.
To adapt the binary SBM for valued (weighted) graphs, we can modify the probability distribution of the edges given the colors or communities to which the two involved nodes belong. 
Instead of a Bernoulli random variable, we might use Poisson random variables for integer-value edges, or Gaussian random variables for real-value edges by specifying the means and variances of the distributions for each pair of distinct communities~\cite{mariadassou_uncovering_2010}.
The central panel shows an instance of weighted SBM with Gaussian edges.

Another aspect we can tweak is the fact that in the traditional SBM, each node belongs to a single community or block.
In many networks, nodes may exhibit characteristics of multiple communities.
The {\it mixed membership} SBM (MMSBM)~\cite{airoldi_mixed_2008} addresses this by allowing nodes to belong to multiple communities by specifying a probability distribution over communities, or membership proportions.
In this paper, we will work with an SBM that has real-value edges and nodes with mixed membership proportions.
The right panel shows an example of weighted MMSBM.
\end{mybox}

\twocolumngrid

\subsection{bioSBM: a covariate dependent MMSBM for long-range chromatin contacts}\label{sec:bioSBM_model}

\begin{figure*}[htbp]
    \centering
    \includegraphics[width=0.85\linewidth]{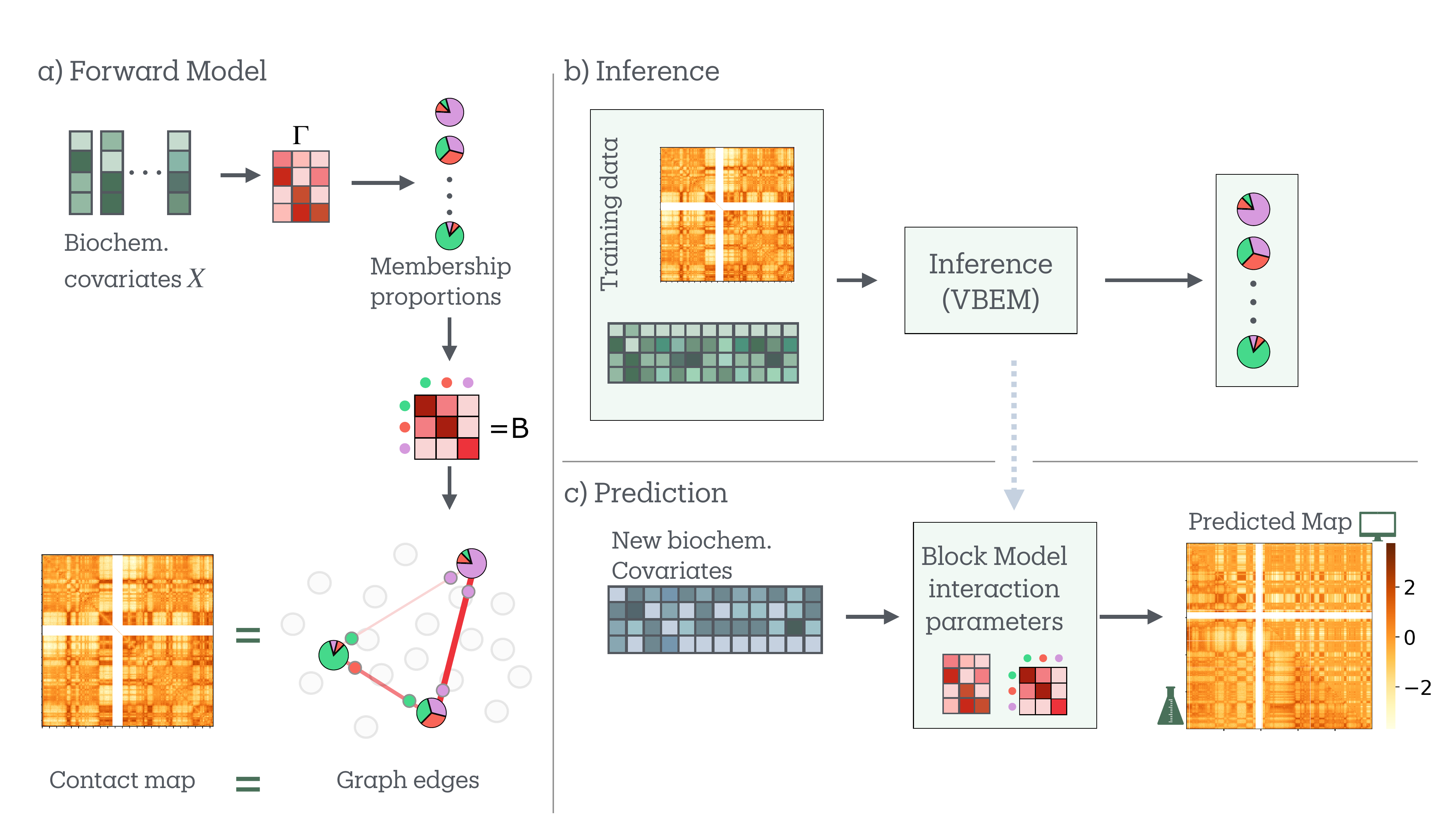}
    \caption[bioSBM model and inference]{(a) Forward model of the bioSBM model. Biochemical covariates associated with each node or genomic region are mapped into membership proportions. The membership proportions, together with the interaction strengths between pairs of communities (encoded in the block matrix $\mathbf{B}$), determine the connectivity patterns of the graph. (b) Schematic representation of the inverse inference problem. Given a set of contact maps and covariates matrices, our inference procedure generates estimates of the latent membership proportion for the studied genomic loci, and infers the interaction parameters that characterize the model. (c) the model parameters inferred on training data, is then used to make de novo predictions of contact maps, from biochemical features, for unseen data.}
    \label{fig:bioSBM_genprocess}
\end{figure*}

A key difference between standard relational data and the biological setting we consider is the availability of a wealth of additional data in biology.
While Hi-C measures contact patterns between chromosomal regions, a variety of biochemical assays, such as ChIP-seq~\cite{johnson_genome-wide_2007}, provide 1D genomic maps of specific epigenomic marks at such regions.
In our model, we integrate the information from ChIP-seq data as a vector of biochemical covariates associated with each node, and which modulates the probability of each node belonging to the different communities.

Formally, bioSBM is a hierarchical Bayesian model~\cite{allenby_hierarchical_2005,gelman_bayesian_2014} over weighted graphs.
Graph nodes $i\in\{1,\ldots,N\}$ represent a set of contiguous genomic regions of fixed length.
The observed weighted network has adjacency matrix $Y$, with $Y_{ij}$ representing the logarithmic OE Hi-C contact frequency (Sec.~\ref{sec:SBMs}) between region $i$ and $j$.
The observed weighted network is assumed to be generated according to the latent distributions of group memberships for each node/genomic region, as well as the matrices that specify group-group interaction strengths.
Each node $i$ has an associated membership proportions' vector $\theta_i$, where $\theta_{ig}$ denotes the probability of node $i$ belonging to group $g$, allowing nodes to belong to multiple communities and display interactions that are context-dependent. 

The group-group interaction strengths are defined by matrices $B$ and $\sigma^2$, where $B_{kg}$ and $\sigma^2_{kg}$ represent, respectively, the mean and variance of the strength of interaction between community $k$ and community $g$.
For each pair of nodes $(i,j)$, discrete variables $z_{ij}$ and $z_{ji}$ denote the group membership of $i$ when interacting with $j$, and viceversa.

Then, the edge weight is sampled from a Gaussian distribution parameterized by $B$ and $\sigma^2$ matrices.
Putting everything together, the generative process for bioSBM procedes as follows:
\begin{itemize}
\item
Each node has an associated vector of $P$ features $x_i$, with $X \in \mathbb{R}^{P \times N}$ denoting the covariate matrix.
The covariates correspond to biochemical data that can be associated with the different genomic regions, such as data from ChIP-seq assays.
\item
For every node $i$, we sample the distribution $\theta_i$ from the logistic normal distribution~\cite{aitchison_statistical_1982,blei_correlated_2007} with mean $\mu_i = \Gamma(x_i)$ and global covariance $\Sigma$, {\it i.e.}
\begin{eqnarray}
\eta_i & \sim & {\mathcal N}(\Gamma (x_i), \Sigma) \, , \\
\theta_{ik} & = & \frac{\exp(\eta_{i k})}{\sum_{k'=1}^K \exp(\eta_{i k'})} \, .
\end{eqnarray}
$\Gamma : \mathbb{R}^P \rightarrow \mathbb{R}^K$ is the parametric function that maps biochemical features to probabilities over group memberships.
In our specific implementation, $\Gamma$ is simply a linear transformation encoded in a $K \times P$ matrix.
Notice that $\Gamma$ and $\Sigma$ are global parameters shared among all nodes.
\item
For every pair of nodes $(i,j)$ with $i=1, \ldots, N$ and $j=1, \ldots, i-1$, we define:
\begin{equation}
z_{i j}  \sim  \text{Mult}(\theta_i) \, , \qquad z_{j i}  \sim  \text{Mult}(\theta_j) \, ,
\end{equation}
with $z_{i j}$ being the membership of node $i$ interacting with node $j$ and viceversa, sampled from the multinomial distribution with probabilities $\theta_i$ and $\theta_j$ respectively.
\item
Once the memberships of $i$ and $j$ are sampled, the weight $Y_{ij}$ of the corresponding edge is sampled from a Gaussian whose parameters are encoded in the global $B$ and $\sigma^2$ matrices:
\begin{equation}
P(Y_{ij}|z_{i j}=k, z_{j i}=g, B, \sigma^2) = \mathcal{N}(Y_{i j} | B_{k g}, \sigma^2_{k g}) \, ,
\end{equation}
\end{itemize}
where the notation (used throughout the whole text) ``$P( \cdot | \cdot)$'' stands for the conditional probability of the variable(s) on the left given the variable(s) on the right.
Schematically, the bioSBM model is illustrated as a graphical model in Fig.~\ref{fig:bioSBM_genprocess}.

\subsection{Posterior inference}\label{sec:inference_algo}
To uncover the latent structure of chromatin interactions, we developed a posterior inference algorithm tailored to bioSBM.
This algorithm estimates the latent parameters that best explain the observed Hi-C interaction data, integrating both chromatin interaction frequencies and biochemical covariates.

Our approach is based on variational inference, a method well-suited for complex probabilistic models like the bioSBM, where exact inference is intractable.
We optimize a variational lower bound on the model evidence, commonly referred to as the Evidence Lower Bound (ELBO), which enables us to approximate the posterior distribution of the latent variables.

The variational inference procedure optimizes the ELBO, defined as:
\begin{equation}
    \mathcal{L}(q,\Psi) = \mathbb{E}_q\left[\log P(Y,\eta_{1:N}, Z|\Psi, X) - \log q(\eta_{1:N}, Z)\right] \, ,
\end{equation}
where $Y$ and $X$ are for the OE Hi-C interaction data and the biochemical covariates respectively, $\eta_{1:N}$ are the latent membership vectors ($\theta_{1:N}$ are the normalized versions), $Z$ represents the community assignments for edges, $\Psi$ includes the global model parameters, and the symbol $\mathbb{E}_q(\cdot)$ denotes the expectation value of the bracketed quantity with respect to the variational distribution $q$.
The variational distribution $q(\eta_{1:N},Z)$ approximates the true posterior distribution $P(\eta_{1:N}, Z | Y,X,\Psi)$.

Then, the algorithm proceeds in two main steps:
\begin{enumerate}
\item \textbf{Variational E-step}:
We update the variational distributions of the latent variables, $\eta_i$ and $z_{ij}$, by maximizing the ELBO with respect to the variational parameters. The factorized variational distributions take the form:
\begin{align}
q(\eta_i) 
& \propto \exp \left\{ \log P(\eta_i|\mu_i, \Sigma ) + \mathbb{E}_{q(Z)}[\log P(Z|\eta_i )] \right\} \label{eq:eta_update_main} \\
q(z_{ij})
& \propto \exp \biggl\{ \mathbb{E}_{q(z_{ji})} \left[ \log P(Y_{ij} |z_{ij} , z_{ji}, B) \right] \notag \\
& \quad + \mathbb{E}_{q(\eta_i )}\left[\log P( z_{ij}|\eta_i )\right] \biggr\} \label{eq:zeta_update_main}
\end{align}
Here, $\eta_i$ are the continuous latent membership vectors, and $z_{ij}$ are the discrete community assignments for edges.
\item \textbf{Variational M-step}:
This step involves optimizing the model parameters $\Psi \equiv (\Sigma ,\Gamma ,B, \sigma^2)$ with the current estimates of the variational distributions. 
The matrix $\Gamma$ maps the biochemical covariates to the latent space, while $\Sigma$ is the covariance matrix capturing the variability in the latent memberships.
The matrices $B$ and $\sigma^2$ describe the mean interaction strengths and variances between communities.
\end{enumerate}
The iterative process of alternating between the E-step and the M-step continues until convergence, at which point the model parameters and variational distributions jointly provide an interpretation of the chromatin interaction patterns.
For the detailed mathematical derivations and specific parameter update rules, refer to Sec.~S1 in SM~\cite{SMnote}.

\section{Results}\label{sec:results}
To train the model, we have performed posterior inference using pairs of biochemical covariates and Hi-C matrices $(X^{\mu}, Y^{\mu})_{\mu=1}^M$ for $M$ chromosomes ($M=11$ was chosen as the number of chromosomes in each training set in the experiments).
More specifically, we employ the two sets of odd-numbered human chromosomes from 1 to 21 and the even-numbered human chromosomes from 2 to 22, and we use the model trained on one set to make predictions on the other and vice versa computing approximate posterior distributions over per-node latent membership vectors $\theta_i$ and the model parameters.
We have applied the inference algorithm to data from the GM12878 lymphoblastoid cell line at a resolution of 100 kilo-basepairs; then, through Bayesian model selection, using evidence lower bound (ELBO) as a criterion, we determined that the optimal number of communities for our model is $K=7$ (see Fig.~S1 in SM~\cite{SMnote}).
These results align with recent efforts to describe chromatin organization extending beyond the conventional A/B compartmentalization~\cite {lieberman-aiden_comprehensive_2009} and encompassing more nuanced frameworks, including subcompartments~\cite{rao_3d_2014} and Interaction Profile Groups (IPGs) from Spracklin {\it et al.}~\cite{spracklinDiverseSilentChromatin2023}.
Notably, a recent orthogonal approach based on polymer modeling by Esposito {\it et al.}~\cite{esposito_polymer_2022} showed that their model could recapitulate Hi-C contact patterns with a set of ``binding domains'', which could be clustered into 9 statistically significant, epigenetically distinct groups. The similarity in the number of inferred domain types, despite the methodological differences, highlights the consistency and biological relevance of both approaches.

The {\it maximum-a-posteriori} (MAP) estimates of the vectors $\theta_i$ provide the most plausible community membership proportions for each node, based on the observation of experimental Hi-C maps and associated biochemical covariates.
Along with the $\theta_i$ values, the inference process also estimates the global parameters that characterize the generative model.
A key parameter is the matrix $\Gamma$ (Sec.~\ref{sec:inference_algo}), which represents the linear transformation mapping biochemical features to the probabilities of belonging to each community, offering insights into how biochemical factors shape chromatin structure.
Additionally, the matrix $B$ encodes the interaction strengths between all pairs of communities.

\begin{figure*}
    \centering
    \includegraphics[width=1\linewidth]{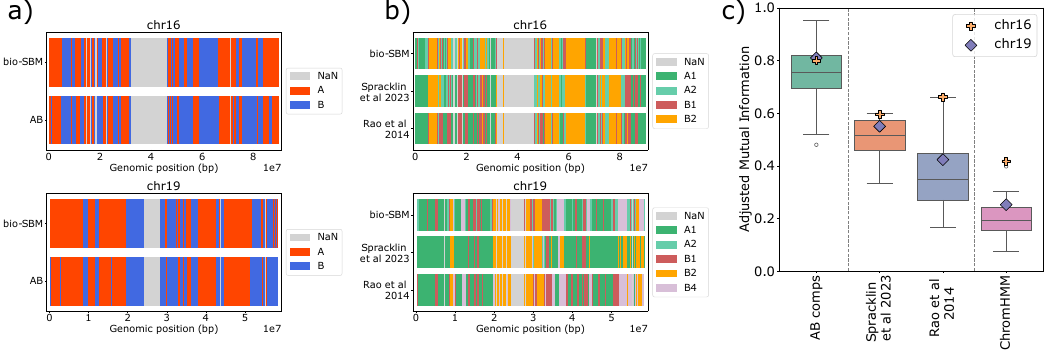}
    \caption{
    The latent representation found through inference of the bioSBM model is biologically meaningful.
    The top rows in (a) and (b) represent the clusters obtained by applying $k$-means on the MAP membership vectors inferred through our algorithm.
    The bottom rows are biological annotations.
    (c) Adjusted Mutual Information score between the clustering obtained by bioSBM's membership vectors and annotations previously reported in the literature, for the example chromosomes 16 and 19 and for all other chromosomes from 1 to 22.
    }
    \label{fig:clustering_of_membership_vectors}
\end{figure*}
%
\begin{figure*}[htbp]
    \centering
    \includegraphics[width=0.85\linewidth]{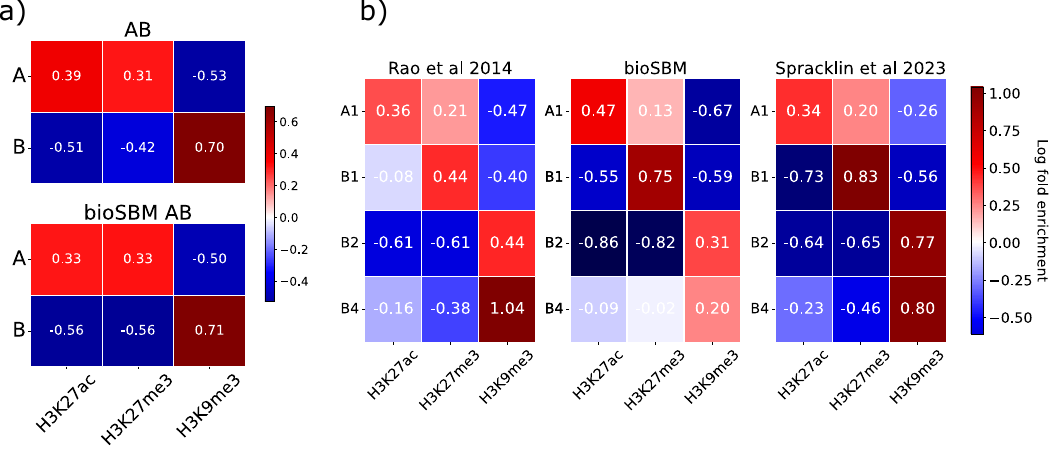}
    \caption{
    Log-fold enrichment in biochemical features (such as the presence of histone marks) for chromosome 19.
    The patterns of enrichment for our clusters are in good agreement with the enrichment found for A/B compartments, subcompartments, and IPGs from~\cite{spracklinDiverseSilentChromatin2023}. The naming for the cluster labels for the clustering obtained applying $k$-means to bioSBM membership vectors, and for IPGs, have been mapped to match the nomenclature of subcompartments by~\cite{rao_3d_2014}.
    }
    \label{fig:enrichment_with_covariates}
\end{figure*}
\subsection{bioSBM explains the hierarchical organization of the chromatin in terms of epigenomic marks}
The first output of the inference procedure is a set of $K$-dimensional ($K=7$ for GM12878) {\it mixed-membership} probability vectors $\theta_i$ for all the genomic bins in the training data. These vectors represent the probabilistic community membership of each genomic region.

To facilitate biological interpretation, we perform a separate clustering step on these vectors using $k$-means clustering~\cite{lloydLeastSquaresQuantization1982}, to obtain discrete labels.
Clustering into two groups allowed us to compare these clusters to known A/B compartments, while setting the number of clusters to 6 enabled comparisons both with the subcompartments defined by Rao {\it et al.}~\cite{rao_3d_2014} and the Interaction Profile Groups (IPGs) computed with the algorithm by Spracklin {\it et al.}~\cite{spracklinDiverseSilentChromatin2023} (see Sec.~S2.E in SM~\cite{SMnote}).

The results of the clustering showed a significant overlap with the established biological annotations.
Fig.~\ref{fig:clustering_of_membership_vectors} illustrates this comparison for chromosomes 16 and 19.
The binary subdivision in A/B compartments as well as the more granular classification in subcompartments or IPGs can be captured from the full mixed membership vectors inferred by our model (see Fig.~S9 in SM~\cite{SMnote} a comparison of centrome distances of the various clusters.). 
Further validation of the inferred communities was performed by assessing the enrichment (see Sec.~S2.C in SM~\cite{SMnote} for details) of each $k$-mean-derived cluster in the biochemical features $x_i$.
These enrichments were compared (see Fig.~\ref{fig:enrichment_with_covariates}) to those observed in A/B compartments, subcompartments, and annotations from~\cite{spracklinDiverseSilentChromatin2023}, where we can see a near-perfect agreement of the discretized bioSBM results with the enrichment of the binary A/B classification and a very good agreement with the enrichment of the subcompartments defined by~\cite{rao_3d_2014}and the IPGs..

We extended this analysis to other chromosomes (see Sec.~S2.A, in particular Table~S1, in SM~\cite{SMnote} for details on datasets used and Figures~S3-S4 in SM~\cite{SMnote} for enrichment results for all chromosomes) and computed the similarity between the subdivision found by clustering the membership vectors and those based on the biological annotations using the adjusted mutual information (AMI) score (see Sec.~S2.D in SM~\cite{SMnote}).
We obtained a median AMI score (which ranges from 0 to 1) of AMI$^{\rm A/B} \simeq 0.76$ for the binary clustering, AMI$^{\rm IPG} \simeq 0.52$ for the comparison with Spracklin {\it et al.}~(2023)~\cite{spracklinDiverseSilentChromatin2023} and AMI$^{\rm subcomp} \simeq 0.35$ for the comparison with Rao {\it et al.}~(2014)~\cite{rao_3d_2014}. These results suggest that the community structure inferred by bioSBM more closely resembles IPGs~\cite{spracklinDiverseSilentChromatin2023} than subcompartments, though it does not correspond one-to-one with either.
Additionally, we performed a comparison with the 15-state annotations by ChromHMM~\cite{ernstChromatinstateDiscoveryGenome2017} (panel (c) in Fig.~\ref{fig:clustering_of_membership_vectors}).
While there is a degree of overlap between the community structure detected by our model and the ChromHMM segmentation (median AMI$^{\rm chromHMM} \simeq 0.18$), the level of accordance is notably lower than the comparison we made between bioSBM and other labelings based on Hi-C data.
This difference is not entirely surprising, as bioSBM does incorporate biochemical covariates, but one of its core components is community detection based on chromatin contacts.
In contrast, ChromHMM performs a genome segmentation based solely on epigenomic data.

\begin{figure*}
\centering
\includegraphics[width=0.90\linewidth]{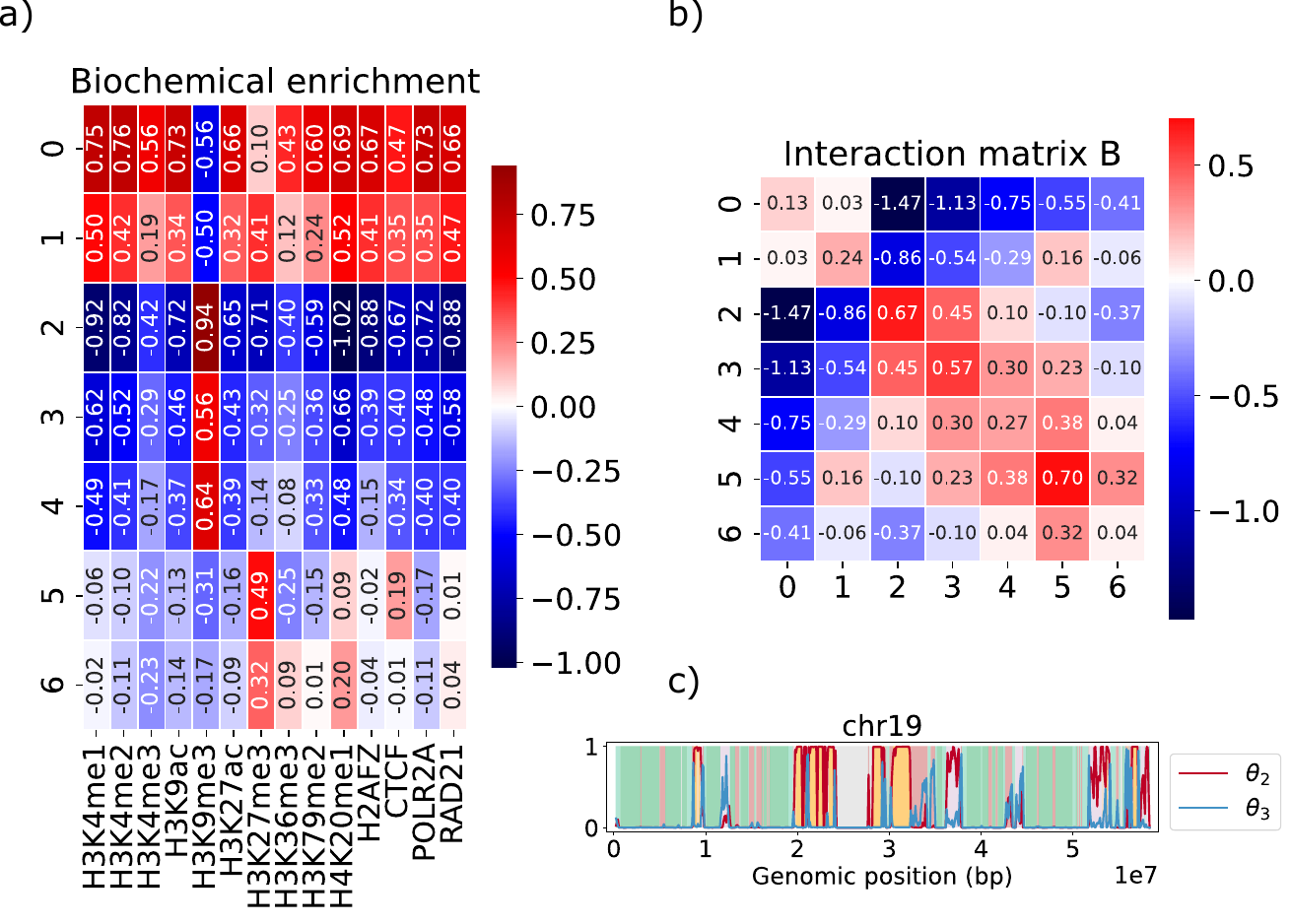}
\caption{
(a) Enrichment of mixed membership communities in the biochemical features used in the model. Each community correlates with distinct epigenetic features.
(b) Interaction intensities between all pairs of communities.
(c) Example of membership proportions of communities 4 and 5, for whole chromosome 19.
}
\label{fig:topics_and_B}
\end{figure*}

Importantly, the model goes beyond merely segmenting chromatin regions based on their epigenetic features; it also illustrates how these different communities interact, as represented by the matrix $B$ where each entry $B_{kg}$ encodes the interaction strength between community $k$ and $g$.
Panel (a) in Fig.~\ref{fig:topics_and_B} shows that each of the $7$ inferred communities is associated both with distinct epigenomic patterns and with interaction patterns between communities.
Interestingly, despite the apparent redundancy for some of the communities at the level of epigenomic profiles, there are clear distinctions between the interaction patterns of different communities.
For instance (see panel (b) in Fig.~\ref{fig:topics_and_B}), communities 0 and 1 have similar epigenetic profiles but regions predominantly associated with community 0 interact mainly with community 0 and community 1, while nodes predominantly associated to community 1 interact also with community 5.
Another notable observation is that community 2 interacts positively only with the epigenetically similar communities 3 and 4, but community 3 interacts positively also with the very different community 5.
The inset of panel (c) in Fig.~\ref{fig:topics_and_B} shows that regions with non-zero probabilities of belonging to communities 2 and 3 overlap with clusters corresponding to the B2 and B4 compartments.
Here, however, nodes do not need to be categorized into one community or another, as they can share properties of multiple communities in different proportions (see also Fig.~S2 in SM\cite{SMnote} for some results characterizing the importance and degree of the nodes' {\it mixed-membershipness}). 

Altogether, these observations show that the bioSBM representation of chromatin interaction patterns provides a more nuanced description than the one provided by simple segmentation of the genome in different clusters.

\subsection{bioSBM's predictive power}
The previous section focussed on an analysis of the interpretability of bioSBM, showing that the inferred model parameters recapitulate and extend previous observations on the epigenetic state of the chromatin and its compartments.
In this Section, we leverage the generative structure of bioSBM to test whether the simple representation of the genome contacts in terms of interactions between a limited number of communities and the fact that these communities can be determined starting from independent measurements of biochemical covariates is enough to reproduce long-range genome-wide chromatin contact patterns.

Specifically, given matrices $B$ and $\Gamma$, inferred from some training chromosomes in some conditions, and biochemical measurements for test chromosomes (either different chromosomes from the same cell line, or chromosomes from a different cell line), we can compute their community structure as $\theta_i = \Gamma x_i$.
With the predicted community structure, we use the matrix $B$ to sample contact maps and compute their expectations, which can then be compared to experimentally obtained Hi-C maps (refer to Sec.~S2.B in SM~\cite{SMnote} for technical details and panels (b) and (c) in Fig.~\ref{fig:bioSBM_genprocess} for a schematic representation of the pipeline).

\subsubsection{bioSBM assessment strategy}
Testing machine learning models in biology is non-trivial, because the meaning of generalisation is often unclear (see {\it e.g.} Schreider {\it et al.}~\cite{schreiber_pitfall_2020} for an in-depth discussion): one may seek to predict for unseen regions in the same chromosome (hence, with similar local environment), or for different chromosomes in the same biological condition (cell line, tissue, etc.), or seek to extrapolate to completely new unseen conditions.
In our experiments, we stress-test bioSBM under the two most stringent conditions: first, we learn model parameters using ChIP-seq and Hi-C data from a subset of chromosomes in the lymphoblastoid cell line GM12878.
Then, we test model performance in predicting Hi-C data on the remaining chromosomes, using as input the relative ChIP-seq data (panels (b) and (c) in Fig.~\ref{fig:bioSBM_genprocess}).
We then seek to assess the performance of bioSBM on the task of predicting Hi-C data in two completely unseen cell lines: the K562 leukemia cell line, using ENCODE ChIP-seq data, and the colorectal carcinoma line HCT116, with and without a major external perturbation (the rapid degradation of the RAD21 loop extrusion factor~\cite{rao_cohesin_2017}).
On the HCT116 RAD21-cell line, we initially perform the prediction task using the parameters learned from the GM12878 data set and ENCODE ChIP-seq data from HCT116, including tracks for RAD21.
We then seek to simulate the ability of bioSBM to predict perturbation by repeating our prediction experiment (still with the same parameters), but simply setting to zero the ChIP-seq tracks for RAD21 (see panels (b) and (c) in Fig.~\ref{fig:hic_correlations}).
This enables us to assess bioSBM predictions both under covariate shift (moving from GM12878 to HCT116 RAD21-cells), and its ability to model perturbations by changing the inputs in a deterministic way (setting to zero the RAD21 tracks). \\
\subsubsection{bioSBM predicts the structure of the chromatin in unseen cell lines}

\begin{figure*}[htbp]
\centering
\includegraphics[width=0.95\linewidth]{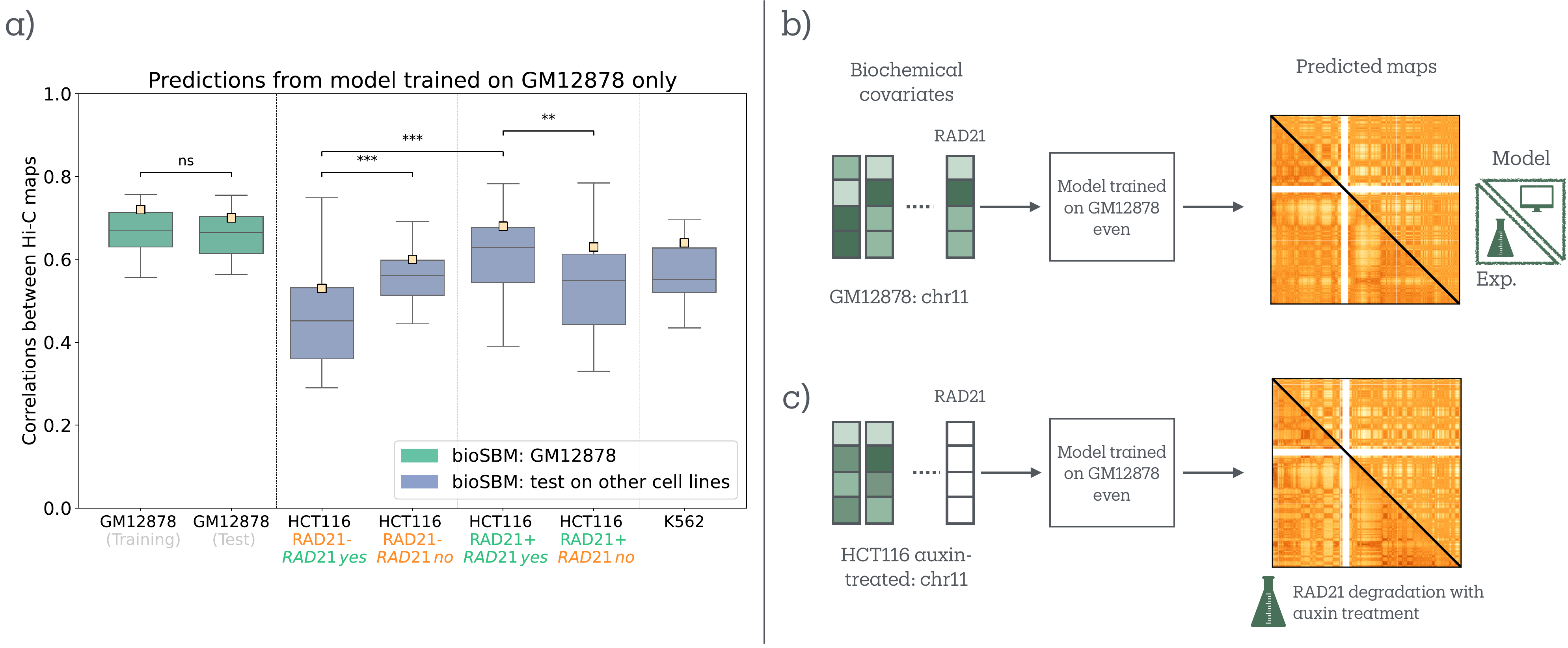}
\caption{
(a)
Box plots of correlations between experimental maps and model generated maps, where the individual data points are single chromosomes. Ivory squares are the correlation values for an example chromosome 11.
To test whether the training performance was significantly higher or not than the test on different chromosomes on the same cell line we performed a one-tailed Mann-Whitney U test, that yielded a negative answer ($p$-value = 0.82).
To test whether the improvement of predictions granted by removal of the RAD21 feature, for predictions on the AID-treated HCT116 cell line, was significant, we used the same test, which yielded a $p\text{-value}=1.20\cdot 10^{-4}$ (significant difference). Also the predictive accuracy on HCT116 without AID treatment is significantly higher ($p\text{-value}=2.33\cdot 10^{-5}$).
The orange box plots refer to {distance-corrected correlations} reported from Esposito {\it et al.}~\cite{esposito_polymer_2022} for a polymer-based model on the same dataset, see also Fig.~S6 in SM~\cite{SMnote} for a comparison with the results obtained with a neural network parametrization of $\Gamma$.
(b) and (c) Schematization of the test performed on the AID-treated HCT116 cell line.
}
\label{fig:hic_correlations}
\end{figure*}

The results of the tests described above are reported in panel (a) in Fig.~\ref{fig:hic_correlations}. The first 2 columns show Pearson correlation between experimental log O/E Hi-C maps, and model-generated maps, for the training chromosomes, and for test chromosomes on the same cell line. Interestingly, the difference in accuracy between the training and test sets was marginal, with a one-tailed Mann-Whitney U test yielding a $p$-value of 0.82 (meaning the difference in accuracy is not statistically significant), see Fig.~S5 in SM~\cite{SMnote} for a comparison between predicted and real maps for all chromosomes.
Therefore, bioSBM effectively generalizes across different chromosomes within the same cell line, suggesting that the inferred associations $B_{kg}$ between communities and the biochemistry-to-structure map $\Gamma$ reflect genuine chromatin interactions. 
Despite some differences in methodology and evaluation metrics, the correlation values we observe align well with the ``distance-corrected'' Pearson correlation values reported by Esposito et al.~\cite{esposito_polymer_2022}. Their polymer-based approach, grounded in physical modeling of chromatin structure, offers valuable and direct mechanistic insight. While their model performs well on training data, it exhibits a noticeable decrease in predictive accuracy on unseen chromosomes. In contrast, bioSBM maintains robust predictive performance across chromosomes, highlighting its strong generalization capability. These complementary strengths illustrate the potential for integrating diverse modeling strategies to better understand chromatin organization.


%
Additionally, we apply the model trained on half the chromosomes of GM12878, to make predictions on contacts on the other half of the chromosomes of the cancer cell lines HCT116 and K562 (last and third to last columns in panel (a) of Fig.~\ref{fig:hic_correlations}), from covariate tracks downloaded from the ENCODE~\cite{noauthor_encode_2024} database. The results show that the model trained on GM12878 chromosomes is able to explain a large fraction of the variance in the data from HCT116 and K562.
Finally, we tested the model on a dataset for the HCT116 cell line, where auxin-inducible degron (AID) technology was used to degrade RAD21~\cite{rao_cohesin_2017}. 

The tests on HCT116 are especially informative, as Rao et al.~\cite{rao_cohesin_2017} showed that  RAD21 depletion in HCT116 cells preserves the global compartmentalization pattern--quantified in terms of the first eigenvector of the Hi-C correlation maps--but introduces {\it quantitative} changes, such as a marked increase in compartment strength.
To quantify this effect, we computed chromosome-wise correlations between treated and untreated Hi-C maps for the HCT116 cell line (data from Rao et al. 2017)~\cite{rao_cohesin_2017}, yielding an average Pearson correlation of $\sim 0.78$. This confirms that while the plaid pattern remains globally consistent, substantial quantitative differences exist that a predictive model like bioSBM should aim to capture.

We make predictions of contact frequencies with the model trained on GM12878 and independent covariates from the ENCODE Project~\cite{noauthor_encode_2024} obtained from experiments on the HCT116 cell line (without AID-induced RAD21 degradation). We tested two scenarios: one where all covariates were used (denoted {\it ``RAD21 yes''}, and one where the RAD21 track was zeroed out to simulate degradation ({\it ``RAD21 no''}). We then compare these predicted maps to experimental maps of HCT116 cells without AID treatment (and therefore still have RAD21) or cells after six hours of treatment, leading to RAD21 depletion. We denote the first experimental dataset as RAD21+, and the second one as RAD21- (see Table~S1 in SM~\cite{SMnote}, for data accession codes).

In the {\it RAD21 yes} {\it vs} RAD21+ setup, we observe a decrease in predictive accuracy, compared to predictions made on the same cell line as the training data, likely due to the mismatch between the learned map $\Gamma$, which includes RAD21, and a biological sample where the latter has been degraded.
Additionally, RAD21 deletion could have induced slight changes also in the other covariates, further affecting predictive performance.
Notably, when the RAD21 covariate was excluded from the input ({\it ``RAD21 no''}), performance on the RAD21- experimental data improved significantly ($p$-value = $1.20 \cdot 10^{-4}$) (see Fig.~S8 in SM~\cite{SMnote} for the analysis stratified by genomic distance). Conversely, the {\it in silico} RAD21 exclusion, reduced the predictive accuracy for the RAD21+ data (see Fig.~\ref{fig:hic_correlations}). 

These results underscore the flexibility and robustness of bioSBM in capturing chromatin interactions under different cellular conditions (see also Fig.~S7 in SM~\cite{SMnote} for an example of predictions with a model retrained on a subset of covariates).

\section{Discussion and conclusions}\label{sec:DiscConcls}
The bioSBM model introduced in this study offers a novel approach for modeling long-range chromatin interactions by integrating Hi-C data with biochemical covariates such as histone modifications and binding of transcription factors.
By employing a mixed membership stochastic block model, we capture a more refined and nuanced view of chromatin structure, extending beyond the traditional binary A/B and sub-compartments framework.
Our results show that the 7 communities identified by bioSBM correlate with known epigenetic features, reinforcing the idea that chromatin interactions are closely tied to the biochemical landscape of the genome.
The partial agreement with subcompartments suggests that our model may capture additional layers of chromatin interaction complexity that may be missed by conventional classification methods. 

Although more abstract than polymer models, which explicitly take into account (to varying degrees of detail) the physical nature of the linear chromosomes, the latent representation learned by bioSBM remains biologically interpretable.
The inferred associations between biochemistry and structure encoded in the linear map $\Gamma$ and the communities interactions contained in $B$ provide a starting point for systematically exploring more mechanistic descriptions of how chromatin folding is affected by nuclear biochemical processes.
Importantly, bioSBM does not incorporate the underlying polymeric nature of chromatin, which is known to influence 3D genome architecture.
Factors such as the dynamics of loop extrusion, maintained by SMC complexes, and the local mechanical state of the chromatin fiber can lead to distinct contact patterns that are not always explained by epigenomic marks alone.
This has been demonstrated both experimentally (e.g.,~\cite{Davidson2021}) and via polymer simulations (e.g.,~\cite{leidescherSpatialOrganizationTranscribed2022}).
These properties -- absent from our current modeling framework -- may account for some of the structural distinctions inferred by bioSBM, even among communities with seemingly similar epigenomic profiles.

The predictive power of the model is another important contribution.
By leveraging biochemical covariates, bioSBM can accurately predict chromatin contacts across different chromosomes and cell lines, comparing favorably with state-of-the-art polymer approaches~\cite{esposito_polymer_2022}.
Notably, the model's robustness is highlighted by its performance on the HCT116 RAD21-cell line, where the removal of the RAD21 input covariate improved predictive accuracy, indicating that bioSBM can adapt to different chromatin environments and capture interactions under varying conditions, such as RAD21 depletion.
Interestingly, bioSBM also compares favorably with recent deep-learning models, though their objectives and designs differ. While some deep learning approaches rely solely on DNA sequence input~\cite{fudenberg_predicting_2020,schwessinger_deepc_2020}, limiting  their ability to model cell-type specific variations, bioSBM explicitly integrates epigenetic features for this purpose.
A recent model, Epiphany~\cite{yang_epiphany_2023}, also incorporated epigenomic data and reports predictive performance on held-out chromosomes in GM12878 that is in a similar range to what we observe for bioSBM. However, we note that a direct comparison is not straighforward, as Epiphany operates at a higher resolution and focuses on fine-scale perturbations. Additionally, its evaluation in other cell types is based on structural features such as insulation scores rather than whole-map correlation.

While our model cannot operate at this fine-grained scale, it remains effective in predicting the effects of larger-scale manipulations such as the AID-induced RAD21 reported above. 
Despite these differences, both approaches highlight the value of integrating epigenomic information into 3D genome prediction models.
Crucially, bioSBM stands out for its interpretability: every parameter has a clear probabilistic semantic, allowing their analysis to uncover a clear biological meaning. This makes bioSBM a valuable, flexible tool for studying chromatin organization in diverse cellular contexts.

While this study focuses on Hi-C data, the so-called enrichment methods~\cite{jerkovic_understanding_2021}, such as Promoter Capture Hi-C (PCHi-C)~\cite{schoenfelder_promoter_2018} or ChIA-PET~\cite{li_chia-pet_2010,li_chromatin_2014}, may offer more functionally relevant perspectives on chromatin structure. PCHi-C, for example, highlights promoter region interactions, while ChIA-PET focuses on interactions involving specific proteins.
These methods present challenges for polymer models because they capture interactions between non-contiguous regions. However, bioSBM, being graph-based, could accommodate these non-contiguous interactions with some adaptation.
Though some graph-based studies exist that link nuclear biochemistry to chromatin structure as assayed by these enrichment-based data~\cite{pancaldi_integrating_2016,pancaldi_chromatin_2021}, a fully generative predictive model mapping biochemistry to the chromatin interaction patterns, such as that provided by bioSBM for Hi-C, has yet to be developed.
With appropriate adjustments, bioSBM could readily be extended to accommodate these data types.

In conclusion, the bioSBM model successfully balances interpretability and scalability, offering a valuable tool for understanding the relationship between chromatin structure and its biochemical underpinnings.
By incorporating biochemical features and allowing for mixed memberships, bioSBM provides a more flexible and biologically meaningful representation of chromatin interactions.
The model's predictive power and adaptability across different cellular contexts underscore its potential for further applications, such as exploring chromatin dynamics in different developmental stages or disease states.

\section{Code availability}
An implementation of the code for the inference algorithm described in this paper is available at \url{https://github.com/alex-chenyi-zhang/bioSBM-code}.

\begin{acknowledgments}
GS acknowledges co-funding from Next Generation EU, in the context of the National Recovery and Resilience Plan, Investment PE1 - Project FAIR “Future Artificial Intelligence Research”. This resource was co-financed by the Next Generation EU [DM 1555 del 11.10.22]. AR acknowledges financial support from PNRR Grant CN 00000013 CN-HPC, M4C2I1.4, spoke 7, funded by Next Generation EU. AG acknowledges financial support by MIUR PRIN-COFIN2022 grant 2022JWAF7Y.
\end{acknowledgments}

\bibliography{references}

\begin{thebibliography}{72}%
\makeatletter
\providecommand \@ifxundefined [1]{%
 \@ifx{#1\undefined}
}%
\providecommand \@ifnum [1]{%
 \ifnum #1\expandafter \@firstoftwo
 \else \expandafter \@secondoftwo
 \fi
}%
\providecommand \@ifx [1]{%
 \ifx #1\expandafter \@firstoftwo
 \else \expandafter \@secondoftwo
 \fi
}%
\providecommand \natexlab [1]{#1}%
\providecommand \enquote  [1]{``#1''}%
\providecommand \bibnamefont  [1]{#1}%
\providecommand \bibfnamefont [1]{#1}%
\providecommand \citenamefont [1]{#1}%
\providecommand \href@noop [0]{\@secondoftwo}%
\providecommand \href [0]{\begingroup \@sanitize@url \@href}%
\providecommand \@href[1]{\@@startlink{#1}\@@href}%
\providecommand \@@href[1]{\endgroup#1\@@endlink}%
\providecommand \@sanitize@url [0]{\catcode `\\12\catcode `\$12\catcode `\&12\catcode `\#12\catcode `\^12\catcode `\_12\catcode `\%12\relax}%
\providecommand \@@startlink[1]{}%
\providecommand \@@endlink[0]{}%
\providecommand \url  [0]{\begingroup\@sanitize@url \@url }%
\providecommand \@url [1]{\endgroup\@href {#1}{\urlprefix }}%
\providecommand \urlprefix  [0]{URL }%
\providecommand \Eprint [0]{\href }%
\providecommand \doibase [0]{https://doi.org/}%
\providecommand \selectlanguage [0]{\@gobble}%
\providecommand \bibinfo  [0]{\@secondoftwo}%
\providecommand \bibfield  [0]{\@secondoftwo}%
\providecommand \translation [1]{[#1]}%
\providecommand \BibitemOpen [0]{}%
\providecommand \bibitemStop [0]{}%
\providecommand \bibitemNoStop [0]{.\EOS\space}%
\providecommand \EOS [0]{\spacefactor3000\relax}%
\providecommand \BibitemShut  [1]{\csname bibitem#1\endcsname}%
\let\auto@bib@innerbib\@empty
\bibitem [{\citenamefont {Matera}\ \emph {et~al.}(2009)\citenamefont {Matera}, \citenamefont {Izaguire-Sierra}, \citenamefont {Praveen},\ and\ \citenamefont {Rajendra}}]{matera_nuclear_2009}%
  \BibitemOpen
  \bibfield  {author} {\bibinfo {author} {\bibfnamefont {A.~G.}\ \bibnamefont {Matera}}, \bibinfo {author} {\bibfnamefont {M.}~\bibnamefont {Izaguire-Sierra}}, \bibinfo {author} {\bibfnamefont {K.}~\bibnamefont {Praveen}},\ and\ \bibinfo {author} {\bibfnamefont {T.~K.}\ \bibnamefont {Rajendra}},\ }\bibfield  {title} {\bibinfo {title} {Nuclear {Bodies}: {Random} {Aggregates} of {Sticky} {Proteins} or {Crucibles} of {Macromolecular} {Assembly}?},\ }\href {https://doi.org/10.1016/j.devcel.2009.10.017} {\bibfield  {journal} {\bibinfo  {journal} {Developmental Cell}\ }\textbf {\bibinfo {volume} {17}},\ \bibinfo {pages} {639} (\bibinfo {year} {2009})}\BibitemShut {NoStop}%
\bibitem [{\citenamefont {Banani}\ \emph {et~al.}(2017)\citenamefont {Banani}, \citenamefont {Lee}, \citenamefont {Hyman},\ and\ \citenamefont {Rosen}}]{banani_biomolecular_2017}%
  \BibitemOpen
  \bibfield  {author} {\bibinfo {author} {\bibfnamefont {S.~F.}\ \bibnamefont {Banani}}, \bibinfo {author} {\bibfnamefont {H.~O.}\ \bibnamefont {Lee}}, \bibinfo {author} {\bibfnamefont {A.~A.}\ \bibnamefont {Hyman}},\ and\ \bibinfo {author} {\bibfnamefont {M.~K.}\ \bibnamefont {Rosen}},\ }\bibfield  {title} {\bibinfo {title} {Biomolecular condensates: organizers of cellular biochemistry},\ }\href {https://doi.org/10.1038/nrm.2017.7} {\bibfield  {journal} {\bibinfo  {journal} {Nature Reviews Molecular Cell Biology}\ }\textbf {\bibinfo {volume} {18}},\ \bibinfo {pages} {285} (\bibinfo {year} {2017})},\ \bibinfo {note} {publisher: Nature Publishing Group}\BibitemShut {NoStop}%
\bibitem [{\citenamefont {Wang}\ \emph {et~al.}(2019)\citenamefont {Wang}, \citenamefont {Gao}, \citenamefont {Zheng}, \citenamefont {Liu}, \citenamefont {Dong}, \citenamefont {Li}, \citenamefont {Zhang}, \citenamefont {Wei}, \citenamefont {Qu}, \citenamefont {Li}, \citenamefont {Allis}, \citenamefont {Li}, \citenamefont {Li},\ and\ \citenamefont {Li}}]{wang_histone_2019}%
  \BibitemOpen
  \bibfield  {author} {\bibinfo {author} {\bibfnamefont {L.}~\bibnamefont {Wang}}, \bibinfo {author} {\bibfnamefont {Y.}~\bibnamefont {Gao}}, \bibinfo {author} {\bibfnamefont {X.}~\bibnamefont {Zheng}}, \bibinfo {author} {\bibfnamefont {C.}~\bibnamefont {Liu}}, \bibinfo {author} {\bibfnamefont {S.}~\bibnamefont {Dong}}, \bibinfo {author} {\bibfnamefont {R.}~\bibnamefont {Li}}, \bibinfo {author} {\bibfnamefont {G.}~\bibnamefont {Zhang}}, \bibinfo {author} {\bibfnamefont {Y.}~\bibnamefont {Wei}}, \bibinfo {author} {\bibfnamefont {H.}~\bibnamefont {Qu}}, \bibinfo {author} {\bibfnamefont {Y.}~\bibnamefont {Li}}, \bibinfo {author} {\bibfnamefont {C.~D.}\ \bibnamefont {Allis}}, \bibinfo {author} {\bibfnamefont {G.}~\bibnamefont {Li}}, \bibinfo {author} {\bibfnamefont {H.}~\bibnamefont {Li}},\ and\ \bibinfo {author} {\bibfnamefont {P.}~\bibnamefont {Li}},\ }\bibfield  {title} {\bibinfo {title} {Histone {Modifications} {Regulate} {Chromatin} {Compartmentalization} by {Contributing} to a {Phase} {Separation}
  {Mechanism}},\ }\href {https://doi.org/10.1016/j.molcel.2019.08.019} {\bibfield  {journal} {\bibinfo  {journal} {Molecular Cell}\ }\textbf {\bibinfo {volume} {76}},\ \bibinfo {pages} {646} (\bibinfo {year} {2019})}\BibitemShut {NoStop}%
\bibitem [{\citenamefont {Ladouceur}\ \emph {et~al.}(2020)\citenamefont {Ladouceur}, \citenamefont {Parmar}, \citenamefont {Biedzinski}, \citenamefont {Wall}, \citenamefont {Tope}, \citenamefont {Cohn}, \citenamefont {Kim}, \citenamefont {Soubry}, \citenamefont {Reyes-Lamothe},\ and\ \citenamefont {Weber}}]{ladouceur_clusters_2020}%
  \BibitemOpen
  \bibfield  {author} {\bibinfo {author} {\bibfnamefont {A.-M.}\ \bibnamefont {Ladouceur}}, \bibinfo {author} {\bibfnamefont {B.~S.}\ \bibnamefont {Parmar}}, \bibinfo {author} {\bibfnamefont {S.}~\bibnamefont {Biedzinski}}, \bibinfo {author} {\bibfnamefont {J.}~\bibnamefont {Wall}}, \bibinfo {author} {\bibfnamefont {S.~G.}\ \bibnamefont {Tope}}, \bibinfo {author} {\bibfnamefont {D.}~\bibnamefont {Cohn}}, \bibinfo {author} {\bibfnamefont {A.}~\bibnamefont {Kim}}, \bibinfo {author} {\bibfnamefont {N.}~\bibnamefont {Soubry}}, \bibinfo {author} {\bibfnamefont {R.}~\bibnamefont {Reyes-Lamothe}},\ and\ \bibinfo {author} {\bibfnamefont {S.~C.}\ \bibnamefont {Weber}},\ }\bibfield  {title} {\bibinfo {title} {Clusters of bacterial {RNA} polymerase are biomolecular condensates that assemble through liquid–liquid phase separation},\ }\href {https://doi.org/10.1073/pnas.2005019117} {\bibfield  {journal} {\bibinfo  {journal} {Proceedings of the National Academy of Sciences}\ }\textbf {\bibinfo {volume} {117}},\ \bibinfo
  {pages} {18540} (\bibinfo {year} {2020})},\ \bibinfo {note} {publisher: Proceedings of the National Academy of Sciences}\BibitemShut {NoStop}%
\bibitem [{\citenamefont {Dekker}\ \emph {et~al.}(2002)\citenamefont {Dekker}, \citenamefont {Rippe}, \citenamefont {Dekker},\ and\ \citenamefont {Kleckner}}]{dekker_capturing_2002}%
  \BibitemOpen
  \bibfield  {author} {\bibinfo {author} {\bibfnamefont {J.}~\bibnamefont {Dekker}}, \bibinfo {author} {\bibfnamefont {K.}~\bibnamefont {Rippe}}, \bibinfo {author} {\bibfnamefont {M.}~\bibnamefont {Dekker}},\ and\ \bibinfo {author} {\bibfnamefont {N.}~\bibnamefont {Kleckner}},\ }\bibfield  {title} {\bibinfo {title} {Capturing {Chromosome} {Conformation}},\ }\href {https://doi.org/10.1126/science.1067799} {\bibfield  {journal} {\bibinfo  {journal} {Science}\ }\textbf {\bibinfo {volume} {295}},\ \bibinfo {pages} {1306} (\bibinfo {year} {2002})},\ \bibinfo {note} {number: 5558 Publisher: American Association for the Advancement of Science}\BibitemShut {NoStop}%
\bibitem [{\citenamefont {Lieberman-Aiden}\ \emph {et~al.}(2009)\citenamefont {Lieberman-Aiden}, \citenamefont {van Berkum}, \citenamefont {Williams}, \citenamefont {Imakaev}, \citenamefont {Ragoczy}, \citenamefont {Telling}, \citenamefont {Amit}, \citenamefont {Lajoie}, \citenamefont {Sabo}, \citenamefont {Dorschner}, \citenamefont {Sandstrom}, \citenamefont {Bernstein}, \citenamefont {Bender}, \citenamefont {Groudine}, \citenamefont {Gnirke}, \citenamefont {Stamatoyannopoulos}, \citenamefont {Mirny}, \citenamefont {Lander},\ and\ \citenamefont {Dekker}}]{lieberman-aiden_comprehensive_2009}%
  \BibitemOpen
  \bibfield  {author} {\bibinfo {author} {\bibfnamefont {E.}~\bibnamefont {Lieberman-Aiden}}, \bibinfo {author} {\bibfnamefont {N.~L.}\ \bibnamefont {van Berkum}}, \bibinfo {author} {\bibfnamefont {L.}~\bibnamefont {Williams}}, \bibinfo {author} {\bibfnamefont {M.}~\bibnamefont {Imakaev}}, \bibinfo {author} {\bibfnamefont {T.}~\bibnamefont {Ragoczy}}, \bibinfo {author} {\bibfnamefont {A.}~\bibnamefont {Telling}}, \bibinfo {author} {\bibfnamefont {I.}~\bibnamefont {Amit}}, \bibinfo {author} {\bibfnamefont {B.~R.}\ \bibnamefont {Lajoie}}, \bibinfo {author} {\bibfnamefont {P.~J.}\ \bibnamefont {Sabo}}, \bibinfo {author} {\bibfnamefont {M.~O.}\ \bibnamefont {Dorschner}}, \bibinfo {author} {\bibfnamefont {R.}~\bibnamefont {Sandstrom}}, \bibinfo {author} {\bibfnamefont {B.}~\bibnamefont {Bernstein}}, \bibinfo {author} {\bibfnamefont {M.~A.}\ \bibnamefont {Bender}}, \bibinfo {author} {\bibfnamefont {M.}~\bibnamefont {Groudine}}, \bibinfo {author} {\bibfnamefont {A.}~\bibnamefont {Gnirke}}, \bibinfo {author}
  {\bibfnamefont {J.}~\bibnamefont {Stamatoyannopoulos}}, \bibinfo {author} {\bibfnamefont {L.~A.}\ \bibnamefont {Mirny}}, \bibinfo {author} {\bibfnamefont {E.~S.}\ \bibnamefont {Lander}},\ and\ \bibinfo {author} {\bibfnamefont {J.}~\bibnamefont {Dekker}},\ }\bibfield  {title} {\bibinfo {title} {Comprehensive mapping of long range interactions reveals folding principles of the human genome},\ }\href {https://doi.org/10.1126/science.1181369} {\bibfield  {journal} {\bibinfo  {journal} {Science (New York, N.Y.)}\ }\textbf {\bibinfo {volume} {326}},\ \bibinfo {pages} {289} (\bibinfo {year} {2009})},\ \bibinfo {note} {number: 5950}\BibitemShut {NoStop}%
\bibitem [{\citenamefont {Nichols}\ and\ \citenamefont {Corces}(2021)}]{nichols_principles_2021}%
  \BibitemOpen
  \bibfield  {author} {\bibinfo {author} {\bibfnamefont {M.~H.}\ \bibnamefont {Nichols}}\ and\ \bibinfo {author} {\bibfnamefont {V.~G.}\ \bibnamefont {Corces}},\ }\bibfield  {title} {\bibinfo {title} {Principles of {3D} compartmentalization of the human genome},\ }\href {https://doi.org/10.1016/j.celrep.2021.109330} {\bibfield  {journal} {\bibinfo  {journal} {Cell Reports}\ }\textbf {\bibinfo {volume} {35}},\ \bibinfo {pages} {109330} (\bibinfo {year} {2021})}\BibitemShut {NoStop}%
\bibitem [{\citenamefont {Rowley}\ and\ \citenamefont {Corces}(2018)}]{rowley_organizational_2018}%
  \BibitemOpen
  \bibfield  {author} {\bibinfo {author} {\bibfnamefont {M.~J.}\ \bibnamefont {Rowley}}\ and\ \bibinfo {author} {\bibfnamefont {V.~G.}\ \bibnamefont {Corces}},\ }\bibfield  {title} {\bibinfo {title} {Organizational principles of {3D} genome architecture},\ }\href {https://doi.org/10.1038/s41576-018-0060-8} {\bibfield  {journal} {\bibinfo  {journal} {Nature Reviews Genetics}\ }\textbf {\bibinfo {volume} {19}},\ \bibinfo {pages} {789} (\bibinfo {year} {2018})},\ \bibinfo {note} {number: 12 Publisher: Nature Publishing Group}\BibitemShut {NoStop}%
\bibitem [{\citenamefont {Rao}\ \emph {et~al.}(2014)\citenamefont {Rao}, \citenamefont {Huntley}, \citenamefont {Durand}, \citenamefont {Stamenova}, \citenamefont {Bochkov}, \citenamefont {Robinson}, \citenamefont {Sanborn}, \citenamefont {Machol}, \citenamefont {Omer}, \citenamefont {Lander},\ and\ \citenamefont {Aiden}}]{rao_3d_2014}%
  \BibitemOpen
  \bibfield  {author} {\bibinfo {author} {\bibfnamefont {S.~S.~P.}\ \bibnamefont {Rao}}, \bibinfo {author} {\bibfnamefont {M.~H.}\ \bibnamefont {Huntley}}, \bibinfo {author} {\bibfnamefont {N.~C.}\ \bibnamefont {Durand}}, \bibinfo {author} {\bibfnamefont {E.~K.}\ \bibnamefont {Stamenova}}, \bibinfo {author} {\bibfnamefont {I.~D.}\ \bibnamefont {Bochkov}}, \bibinfo {author} {\bibfnamefont {J.~T.}\ \bibnamefont {Robinson}}, \bibinfo {author} {\bibfnamefont {A.~L.}\ \bibnamefont {Sanborn}}, \bibinfo {author} {\bibfnamefont {I.}~\bibnamefont {Machol}}, \bibinfo {author} {\bibfnamefont {A.~D.}\ \bibnamefont {Omer}}, \bibinfo {author} {\bibfnamefont {E.~S.}\ \bibnamefont {Lander}},\ and\ \bibinfo {author} {\bibfnamefont {E.~L.}\ \bibnamefont {Aiden}},\ }\bibfield  {title} {\bibinfo {title} {A {3D} {Map} of the {Human} {Genome} at {Kilobase} {Resolution} {Reveals} {Principles} of {Chromatin} {Looping}},\ }\href {https://doi.org/10.1016/j.cell.2014.11.021} {\bibfield  {journal} {\bibinfo  {journal} {Cell}\ }\textbf
  {\bibinfo {volume} {159}},\ \bibinfo {pages} {1665} (\bibinfo {year} {2014})},\ \bibinfo {note} {publisher: Elsevier}\BibitemShut {NoStop}%
\bibitem [{\citenamefont {Hedström}\ \emph {et~al.}(2024{\natexlab{a}})\citenamefont {Hedström}, \citenamefont {Metzler},\ and\ \citenamefont {Lizana}}]{hedstrom_enhancer-insulator_2024}%
  \BibitemOpen
  \bibfield  {author} {\bibinfo {author} {\bibfnamefont {L.}~\bibnamefont {Hedström}}, \bibinfo {author} {\bibfnamefont {R.}~\bibnamefont {Metzler}},\ and\ \bibinfo {author} {\bibfnamefont {L.}~\bibnamefont {Lizana}},\ }\bibfield  {title} {\bibinfo {title} {Enhancer-{Insulator} {Pairing} {Reveals} {Heterogeneous} {Dynamics} in {Long}-{Distance} {3D} {Gene} {Regulation}},\ }\href {https://doi.org/10.1103/PRXLife.2.033008} {\bibfield  {journal} {\bibinfo  {journal} {PRX Life}\ }\textbf {\bibinfo {volume} {2}},\ \bibinfo {pages} {033008} (\bibinfo {year} {2024}{\natexlab{a}})},\ \bibinfo {note} {publisher: American Physical Society}\BibitemShut {NoStop}%
\bibitem [{\citenamefont {Ernst}\ and\ \citenamefont {Kellis}(2017)}]{ernstChromatinstateDiscoveryGenome2017}%
  \BibitemOpen
  \bibfield  {author} {\bibinfo {author} {\bibfnamefont {J.}~\bibnamefont {Ernst}}\ and\ \bibinfo {author} {\bibfnamefont {M.}~\bibnamefont {Kellis}},\ }\bibfield  {title} {\bibinfo {title} {Chromatin-state discovery and genome annotation with {{ChromHMM}}},\ }\href {https://doi.org/10.1038/nprot.2017.124} {\bibfield  {journal} {\bibinfo  {journal} {Nature Protocols}\ }\textbf {\bibinfo {volume} {12}},\ \bibinfo {pages} {2478} (\bibinfo {year} {2017})}\BibitemShut {NoStop}%
\bibitem [{\citenamefont {Boix}\ \emph {et~al.}(2021)\citenamefont {Boix}, \citenamefont {James}, \citenamefont {Park}, \citenamefont {Meuleman},\ and\ \citenamefont {Kellis}}]{boixRegulatoryGenomicCircuitry2021}%
  \BibitemOpen
  \bibfield  {author} {\bibinfo {author} {\bibfnamefont {C.~A.}\ \bibnamefont {Boix}}, \bibinfo {author} {\bibfnamefont {B.~T.}\ \bibnamefont {James}}, \bibinfo {author} {\bibfnamefont {Y.~P.}\ \bibnamefont {Park}}, \bibinfo {author} {\bibfnamefont {W.}~\bibnamefont {Meuleman}},\ and\ \bibinfo {author} {\bibfnamefont {M.}~\bibnamefont {Kellis}},\ }\bibfield  {title} {\bibinfo {title} {Regulatory genomic circuitry of human disease loci by integrative epigenomics},\ }\href {https://doi.org/10.1038/s41586-020-03145-z} {\bibfield  {journal} {\bibinfo  {journal} {Nature}\ }\textbf {\bibinfo {volume} {590}},\ \bibinfo {pages} {300} (\bibinfo {year} {2021})}\BibitemShut {NoStop}%
\bibitem [{\citenamefont {Wang}\ \emph {et~al.}(2021)\citenamefont {Wang}, \citenamefont {Zhang}, \citenamefont {Zhang}, \citenamefont {van Schaik}, \citenamefont {Zhang}, \citenamefont {Sasaki}, \citenamefont {Peric-Hupkes}, \citenamefont {Chen}, \citenamefont {Gilbert}, \citenamefont {van Steensel}, \citenamefont {Belmont},\ and\ \citenamefont {Ma}}]{wangSPINRevealsGenomewide2021}%
  \BibitemOpen
  \bibfield  {author} {\bibinfo {author} {\bibfnamefont {Y.}~\bibnamefont {Wang}}, \bibinfo {author} {\bibfnamefont {Y.}~\bibnamefont {Zhang}}, \bibinfo {author} {\bibfnamefont {R.}~\bibnamefont {Zhang}}, \bibinfo {author} {\bibfnamefont {T.}~\bibnamefont {van Schaik}}, \bibinfo {author} {\bibfnamefont {L.}~\bibnamefont {Zhang}}, \bibinfo {author} {\bibfnamefont {T.}~\bibnamefont {Sasaki}}, \bibinfo {author} {\bibfnamefont {D.}~\bibnamefont {Peric-Hupkes}}, \bibinfo {author} {\bibfnamefont {Y.}~\bibnamefont {Chen}}, \bibinfo {author} {\bibfnamefont {D.~M.}\ \bibnamefont {Gilbert}}, \bibinfo {author} {\bibfnamefont {B.}~\bibnamefont {van Steensel}}, \bibinfo {author} {\bibfnamefont {A.~S.}\ \bibnamefont {Belmont}},\ and\ \bibinfo {author} {\bibfnamefont {J.}~\bibnamefont {Ma}},\ }\bibfield  {title} {\bibinfo {title} {{{SPIN}} reveals genome-wide landscape of nuclear compartmentalization},\ }\href {https://doi.org/10.1186/s13059-020-02253-3} {\bibfield  {journal} {\bibinfo  {journal} {Genome Biology}\ }\textbf
  {\bibinfo {volume} {22}},\ \bibinfo {pages} {36} (\bibinfo {year} {2021})}\BibitemShut {NoStop}%
\bibitem [{\citenamefont {Xiong}\ and\ \citenamefont {Ma}(2019)}]{xiongRevealingHiCSubcompartments2019}%
  \BibitemOpen
  \bibfield  {author} {\bibinfo {author} {\bibfnamefont {K.}~\bibnamefont {Xiong}}\ and\ \bibinfo {author} {\bibfnamefont {J.}~\bibnamefont {Ma}},\ }\bibfield  {title} {\bibinfo {title} {Revealing {{Hi-C}} subcompartments by imputing inter-chromosomal chromatin interactions},\ }\href {https://doi.org/10.1038/s41467-019-12954-4} {\bibfield  {journal} {\bibinfo  {journal} {Nature Communications}\ }\textbf {\bibinfo {volume} {10}},\ \bibinfo {pages} {5069} (\bibinfo {year} {2019})}\BibitemShut {NoStop}%
\bibitem [{\citenamefont {Spracklin}\ \emph {et~al.}(2023)\citenamefont {Spracklin}, \citenamefont {Abdennur}, \citenamefont {Imakaev}, \citenamefont {Chowdhury}, \citenamefont {Pradhan}, \citenamefont {Mirny},\ and\ \citenamefont {Dekker}}]{spracklinDiverseSilentChromatin2023}%
  \BibitemOpen
  \bibfield  {author} {\bibinfo {author} {\bibfnamefont {G.}~\bibnamefont {Spracklin}}, \bibinfo {author} {\bibfnamefont {N.}~\bibnamefont {Abdennur}}, \bibinfo {author} {\bibfnamefont {M.}~\bibnamefont {Imakaev}}, \bibinfo {author} {\bibfnamefont {N.}~\bibnamefont {Chowdhury}}, \bibinfo {author} {\bibfnamefont {S.}~\bibnamefont {Pradhan}}, \bibinfo {author} {\bibfnamefont {L.~A.}\ \bibnamefont {Mirny}},\ and\ \bibinfo {author} {\bibfnamefont {J.}~\bibnamefont {Dekker}},\ }\bibfield  {title} {\bibinfo {title} {Diverse silent chromatin states modulate genome compartmentalization and loop extrusion barriers},\ }\href {https://doi.org/10.1038/s41594-022-00892-7} {\bibfield  {journal} {\bibinfo  {journal} {Nature Structural \& Molecular Biology}\ }\textbf {\bibinfo {volume} {30}},\ \bibinfo {pages} {38} (\bibinfo {year} {2023})}\BibitemShut {NoStop}%
\bibitem [{\citenamefont {Bianco}\ \emph {et~al.}(2018)\citenamefont {Bianco}, \citenamefont {Lupiáñez}, \citenamefont {Chiariello}, \citenamefont {Annunziatella}, \citenamefont {Kraft}, \citenamefont {Schöpflin}, \citenamefont {Wittler}, \citenamefont {Andrey}, \citenamefont {Vingron}, \citenamefont {Pombo}, \citenamefont {Mundlos},\ and\ \citenamefont {Nicodemi}}]{bianco_polymer_2018}%
  \BibitemOpen
  \bibfield  {author} {\bibinfo {author} {\bibfnamefont {S.}~\bibnamefont {Bianco}}, \bibinfo {author} {\bibfnamefont {D.~G.}\ \bibnamefont {Lupiáñez}}, \bibinfo {author} {\bibfnamefont {A.~M.}\ \bibnamefont {Chiariello}}, \bibinfo {author} {\bibfnamefont {C.}~\bibnamefont {Annunziatella}}, \bibinfo {author} {\bibfnamefont {K.}~\bibnamefont {Kraft}}, \bibinfo {author} {\bibfnamefont {R.}~\bibnamefont {Schöpflin}}, \bibinfo {author} {\bibfnamefont {L.}~\bibnamefont {Wittler}}, \bibinfo {author} {\bibfnamefont {G.}~\bibnamefont {Andrey}}, \bibinfo {author} {\bibfnamefont {M.}~\bibnamefont {Vingron}}, \bibinfo {author} {\bibfnamefont {A.}~\bibnamefont {Pombo}}, \bibinfo {author} {\bibfnamefont {S.}~\bibnamefont {Mundlos}},\ and\ \bibinfo {author} {\bibfnamefont {M.}~\bibnamefont {Nicodemi}},\ }\bibfield  {title} {\bibinfo {title} {Polymer physics predicts the effects of structural variants on chromatin architecture},\ }\href {https://doi.org/10.1038/s41588-018-0098-8} {\bibfield  {journal} {\bibinfo
  {journal} {Nature Genetics}\ }\textbf {\bibinfo {volume} {50}},\ \bibinfo {pages} {662} (\bibinfo {year} {2018})},\ \bibinfo {note} {number: 5 Publisher: Nature Publishing Group}\BibitemShut {NoStop}%
\bibitem [{\citenamefont {Di~Pierro}\ \emph {et~al.}(2017)\citenamefont {Di~Pierro}, \citenamefont {Cheng}, \citenamefont {Lieberman~Aiden}, \citenamefont {Wolynes},\ and\ \citenamefont {Onuchic}}]{di_pierro_novo_2017}%
  \BibitemOpen
  \bibfield  {author} {\bibinfo {author} {\bibfnamefont {M.}~\bibnamefont {Di~Pierro}}, \bibinfo {author} {\bibfnamefont {R.~R.}\ \bibnamefont {Cheng}}, \bibinfo {author} {\bibfnamefont {E.}~\bibnamefont {Lieberman~Aiden}}, \bibinfo {author} {\bibfnamefont {P.~G.}\ \bibnamefont {Wolynes}},\ and\ \bibinfo {author} {\bibfnamefont {J.~N.}\ \bibnamefont {Onuchic}},\ }\bibfield  {title} {\bibinfo {title} {De novo prediction of human chromosome structures: {Epigenetic} marking patterns encode genome architecture},\ }\href {https://doi.org/10.1073/pnas.1714980114} {\bibfield  {journal} {\bibinfo  {journal} {Proceedings of the National Academy of Sciences}\ }\textbf {\bibinfo {volume} {114}},\ \bibinfo {pages} {12126} (\bibinfo {year} {2017})},\ \bibinfo {note} {number: 46 Publisher: Proceedings of the National Academy of Sciences}\BibitemShut {NoStop}%
\bibitem [{\citenamefont {Jost}\ \emph {et~al.}(2014)\citenamefont {Jost}, \citenamefont {Carrivain}, \citenamefont {Cavalli},\ and\ \citenamefont {Vaillant}}]{jost_modeling_2014}%
  \BibitemOpen
  \bibfield  {author} {\bibinfo {author} {\bibfnamefont {D.}~\bibnamefont {Jost}}, \bibinfo {author} {\bibfnamefont {P.}~\bibnamefont {Carrivain}}, \bibinfo {author} {\bibfnamefont {G.}~\bibnamefont {Cavalli}},\ and\ \bibinfo {author} {\bibfnamefont {C.}~\bibnamefont {Vaillant}},\ }\bibfield  {title} {\bibinfo {title} {Modeling epigenome folding: formation and dynamics of topologically associated chromatin domains},\ }\href {https://doi.org/10.1093/nar/gku698} {\bibfield  {journal} {\bibinfo  {journal} {Nucleic Acids Research}\ }\textbf {\bibinfo {volume} {42}},\ \bibinfo {pages} {9553} (\bibinfo {year} {2014})}\BibitemShut {NoStop}%
\bibitem [{\citenamefont {Zhang}\ \emph {et~al.}(2024)\citenamefont {Zhang}, \citenamefont {Rosa},\ and\ \citenamefont {Sanguinetti}}]{zhang_bottom-up_2024}%
  \BibitemOpen
  \bibfield  {author} {\bibinfo {author} {\bibfnamefont {A.~C.~Y.}\ \bibnamefont {Zhang}}, \bibinfo {author} {\bibfnamefont {A.}~\bibnamefont {Rosa}},\ and\ \bibinfo {author} {\bibfnamefont {G.}~\bibnamefont {Sanguinetti}},\ }\bibfield  {title} {\bibinfo {title} {Bottom-up data integration in polymer models of chromatin organization},\ }\href {https://doi.org/10.1016/j.bpj.2023.12.006} {\bibfield  {journal} {\bibinfo  {journal} {Biophysical Journal}\ }\textbf {\bibinfo {volume} {123}},\ \bibinfo {pages} {184} (\bibinfo {year} {2024})}\BibitemShut {NoStop}%
\bibitem [{\citenamefont {Fudenberg}\ \emph {et~al.}(2020)\citenamefont {Fudenberg}, \citenamefont {Kelley},\ and\ \citenamefont {Pollard}}]{fudenberg_predicting_2020}%
  \BibitemOpen
  \bibfield  {author} {\bibinfo {author} {\bibfnamefont {G.}~\bibnamefont {Fudenberg}}, \bibinfo {author} {\bibfnamefont {D.~R.}\ \bibnamefont {Kelley}},\ and\ \bibinfo {author} {\bibfnamefont {K.~S.}\ \bibnamefont {Pollard}},\ }\bibfield  {title} {\bibinfo {title} {Predicting {3D} genome folding from {DNA} sequence with {Akita}},\ }\href {https://doi.org/10.1038/s41592-020-0958-x} {\bibfield  {journal} {\bibinfo  {journal} {Nature Methods}\ }\textbf {\bibinfo {volume} {17}},\ \bibinfo {pages} {1111} (\bibinfo {year} {2020})},\ \bibinfo {note} {number: 11 Publisher: Nature Publishing Group}\BibitemShut {NoStop}%
\bibitem [{\citenamefont {Schwessinger}\ \emph {et~al.}(2020)\citenamefont {Schwessinger}, \citenamefont {Gosden}, \citenamefont {Downes}, \citenamefont {Brown}, \citenamefont {Oudelaar}, \citenamefont {Telenius}, \citenamefont {Teh}, \citenamefont {Lunter},\ and\ \citenamefont {Hughes}}]{schwessinger_deepc_2020}%
  \BibitemOpen
  \bibfield  {author} {\bibinfo {author} {\bibfnamefont {R.}~\bibnamefont {Schwessinger}}, \bibinfo {author} {\bibfnamefont {M.}~\bibnamefont {Gosden}}, \bibinfo {author} {\bibfnamefont {D.}~\bibnamefont {Downes}}, \bibinfo {author} {\bibfnamefont {R.~C.}\ \bibnamefont {Brown}}, \bibinfo {author} {\bibfnamefont {A.~M.}\ \bibnamefont {Oudelaar}}, \bibinfo {author} {\bibfnamefont {J.}~\bibnamefont {Telenius}}, \bibinfo {author} {\bibfnamefont {Y.~W.}\ \bibnamefont {Teh}}, \bibinfo {author} {\bibfnamefont {G.}~\bibnamefont {Lunter}},\ and\ \bibinfo {author} {\bibfnamefont {J.~R.}\ \bibnamefont {Hughes}},\ }\bibfield  {title} {\bibinfo {title} {{DeepC}: predicting {3D} genome folding using megabase-scale transfer learning},\ }\href {https://doi.org/10.1038/s41592-020-0960-3} {\bibfield  {journal} {\bibinfo  {journal} {Nature Methods}\ }\textbf {\bibinfo {volume} {17}},\ \bibinfo {pages} {1118} (\bibinfo {year} {2020})},\ \bibinfo {note} {publisher: Nature Publishing Group}\BibitemShut {NoStop}%
\bibitem [{\citenamefont {Zhou}(2022)}]{zhou_sequence-based_2022}%
  \BibitemOpen
  \bibfield  {author} {\bibinfo {author} {\bibfnamefont {J.}~\bibnamefont {Zhou}},\ }\bibfield  {title} {\bibinfo {title} {Sequence-based modeling of three-dimensional genome architecture from kilobase to chromosome scale},\ }\href {https://doi.org/10.1038/s41588-022-01065-4} {\bibfield  {journal} {\bibinfo  {journal} {Nature Genetics}\ }\textbf {\bibinfo {volume} {54}},\ \bibinfo {pages} {725} (\bibinfo {year} {2022})},\ \bibinfo {note} {publisher: Nature Publishing Group}\BibitemShut {NoStop}%
\bibitem [{\citenamefont {Yang}\ \emph {et~al.}(2023)\citenamefont {Yang}, \citenamefont {Das}, \citenamefont {Gao}, \citenamefont {Karbalayghareh}, \citenamefont {Noble}, \citenamefont {Bilmes},\ and\ \citenamefont {Leslie}}]{yang_epiphany_2023}%
  \BibitemOpen
  \bibfield  {author} {\bibinfo {author} {\bibfnamefont {R.}~\bibnamefont {Yang}}, \bibinfo {author} {\bibfnamefont {A.}~\bibnamefont {Das}}, \bibinfo {author} {\bibfnamefont {V.~R.}\ \bibnamefont {Gao}}, \bibinfo {author} {\bibfnamefont {A.}~\bibnamefont {Karbalayghareh}}, \bibinfo {author} {\bibfnamefont {W.~S.}\ \bibnamefont {Noble}}, \bibinfo {author} {\bibfnamefont {J.~A.}\ \bibnamefont {Bilmes}},\ and\ \bibinfo {author} {\bibfnamefont {C.~S.}\ \bibnamefont {Leslie}},\ }\bibfield  {title} {\bibinfo {title} {Epiphany: predicting {Hi}-{C} contact maps from {1D} epigenomic signals},\ }\href {https://doi.org/10.1186/s13059-023-02934-9} {\bibfield  {journal} {\bibinfo  {journal} {Genome Biology}\ }\textbf {\bibinfo {volume} {24}},\ \bibinfo {pages} {134} (\bibinfo {year} {2023})}\BibitemShut {NoStop}%
\bibitem [{\citenamefont {Newman}(2016)}]{newman_networks_2016}%
  \BibitemOpen
  \bibfield  {author} {\bibinfo {author} {\bibfnamefont {M.~E.~J.}\ \bibnamefont {Newman}},\ }\href@noop {} {\emph {\bibinfo {title} {Networks: an introduction}}},\ \bibinfo {edition} {reprinted}\ ed.\ (\bibinfo  {publisher} {Oxford University Press},\ \bibinfo {address} {Oxford},\ \bibinfo {year} {2016})\BibitemShut {NoStop}%
\bibitem [{\citenamefont {Ashoor}\ \emph {et~al.}(2020)\citenamefont {Ashoor}, \citenamefont {Chen}, \citenamefont {Rosikiewicz}, \citenamefont {Wang}, \citenamefont {Cheng}, \citenamefont {Wang}, \citenamefont {Ruan},\ and\ \citenamefont {Li}}]{ashoorGraphEmbeddingUnsupervised2020}%
  \BibitemOpen
  \bibfield  {author} {\bibinfo {author} {\bibfnamefont {H.}~\bibnamefont {Ashoor}}, \bibinfo {author} {\bibfnamefont {X.}~\bibnamefont {Chen}}, \bibinfo {author} {\bibfnamefont {W.}~\bibnamefont {Rosikiewicz}}, \bibinfo {author} {\bibfnamefont {J.}~\bibnamefont {Wang}}, \bibinfo {author} {\bibfnamefont {A.}~\bibnamefont {Cheng}}, \bibinfo {author} {\bibfnamefont {P.}~\bibnamefont {Wang}}, \bibinfo {author} {\bibfnamefont {Y.}~\bibnamefont {Ruan}},\ and\ \bibinfo {author} {\bibfnamefont {S.}~\bibnamefont {Li}},\ }\bibfield  {title} {\bibinfo {title} {Graph embedding and unsupervised learning predict genomic sub-compartments from {{HiC}} chromatin interaction data},\ }\href {https://doi.org/10.1038/s41467-020-14974-x} {\bibfield  {journal} {\bibinfo  {journal} {Nature Communications}\ }\textbf {\bibinfo {volume} {11}},\ \bibinfo {pages} {1173} (\bibinfo {year} {2020})}\BibitemShut {NoStop}%
\bibitem [{\citenamefont {Cabreros}\ \emph {et~al.}(2015)\citenamefont {Cabreros}, \citenamefont {Abbe},\ and\ \citenamefont {Tsirigos}}]{cabreros_detecting_2015}%
  \BibitemOpen
  \bibfield  {author} {\bibinfo {author} {\bibfnamefont {I.}~\bibnamefont {Cabreros}}, \bibinfo {author} {\bibfnamefont {E.}~\bibnamefont {Abbe}},\ and\ \bibinfo {author} {\bibfnamefont {A.}~\bibnamefont {Tsirigos}},\ }\href {https://doi.org/10.48550/arXiv.1509.05121} {\bibinfo {title} {Detecting {Community} {Structures} in {Hi}-{C} {Genomic} {Data}}} (\bibinfo {year} {2015}),\ \bibinfo {note} {arXiv:1509.05121 [cs, q-bio, stat]}\BibitemShut {NoStop}%
\bibitem [{\citenamefont {Norton}\ \emph {et~al.}(2018)\citenamefont {Norton}, \citenamefont {Emerson}, \citenamefont {Huang}, \citenamefont {Kim}, \citenamefont {Titus}, \citenamefont {Gu}, \citenamefont {Bassett},\ and\ \citenamefont {Phillips-Cremins}}]{norton_detecting_2018}%
  \BibitemOpen
  \bibfield  {author} {\bibinfo {author} {\bibfnamefont {H.~K.}\ \bibnamefont {Norton}}, \bibinfo {author} {\bibfnamefont {D.~J.}\ \bibnamefont {Emerson}}, \bibinfo {author} {\bibfnamefont {H.}~\bibnamefont {Huang}}, \bibinfo {author} {\bibfnamefont {J.}~\bibnamefont {Kim}}, \bibinfo {author} {\bibfnamefont {K.~R.}\ \bibnamefont {Titus}}, \bibinfo {author} {\bibfnamefont {S.}~\bibnamefont {Gu}}, \bibinfo {author} {\bibfnamefont {D.~S.}\ \bibnamefont {Bassett}},\ and\ \bibinfo {author} {\bibfnamefont {J.~E.}\ \bibnamefont {Phillips-Cremins}},\ }\bibfield  {title} {\bibinfo {title} {Detecting hierarchical genome folding with network modularity},\ }\href {https://doi.org/10.1038/nmeth.4560} {\bibfield  {journal} {\bibinfo  {journal} {Nature Methods}\ }\textbf {\bibinfo {volume} {15}},\ \bibinfo {pages} {119} (\bibinfo {year} {2018})},\ \bibinfo {note} {number: 2 Publisher: Nature Publishing Group}\BibitemShut {NoStop}%
\bibitem [{\citenamefont {Hedström}\ \emph {et~al.}(2024{\natexlab{b}})\citenamefont {Hedström}, \citenamefont {Martínez},\ and\ \citenamefont {Lizana}}]{hedstrom_identifying_2024}%
  \BibitemOpen
  \bibfield  {author} {\bibinfo {author} {\bibfnamefont {L.}~\bibnamefont {Hedström}}, \bibinfo {author} {\bibfnamefont {A.~C.}\ \bibnamefont {Martínez}},\ and\ \bibinfo {author} {\bibfnamefont {L.}~\bibnamefont {Lizana}},\ }\href {https://doi.org/10.48550/arXiv.2405.05425} {\bibinfo {title} {Identifying stable communities in {Hi}-{C} data using a multifractal null model}} (\bibinfo {year} {2024}{\natexlab{b}}),\ \bibinfo {note} {arXiv:2405.05425 [physics, q-bio]}\BibitemShut {NoStop}%
\bibitem [{\citenamefont {Pancaldi}\ \emph {et~al.}(2016)\citenamefont {Pancaldi}, \citenamefont {Carrillo-de Santa-Pau}, \citenamefont {Javierre}, \citenamefont {Juan}, \citenamefont {Fraser}, \citenamefont {Spivakov}, \citenamefont {Valencia},\ and\ \citenamefont {Rico}}]{pancaldi_integrating_2016}%
  \BibitemOpen
  \bibfield  {author} {\bibinfo {author} {\bibfnamefont {V.}~\bibnamefont {Pancaldi}}, \bibinfo {author} {\bibfnamefont {E.}~\bibnamefont {Carrillo-de Santa-Pau}}, \bibinfo {author} {\bibfnamefont {B.~M.}\ \bibnamefont {Javierre}}, \bibinfo {author} {\bibfnamefont {D.}~\bibnamefont {Juan}}, \bibinfo {author} {\bibfnamefont {P.}~\bibnamefont {Fraser}}, \bibinfo {author} {\bibfnamefont {M.}~\bibnamefont {Spivakov}}, \bibinfo {author} {\bibfnamefont {A.}~\bibnamefont {Valencia}},\ and\ \bibinfo {author} {\bibfnamefont {D.}~\bibnamefont {Rico}},\ }\bibfield  {title} {\bibinfo {title} {Integrating epigenomic data and {3D} genomic structure with a new measure of chromatin assortativity},\ }\href {https://doi.org/10.1186/s13059-016-1003-3} {\bibfield  {journal} {\bibinfo  {journal} {Genome Biology}\ }\textbf {\bibinfo {volume} {17}},\ \bibinfo {pages} {152} (\bibinfo {year} {2016})},\ \bibinfo {note} {number: 1}\BibitemShut {NoStop}%
\bibitem [{\citenamefont {Pancaldi}(2021)}]{pancaldi_chromatin_2021}%
  \BibitemOpen
  \bibfield  {author} {\bibinfo {author} {\bibfnamefont {V.}~\bibnamefont {Pancaldi}},\ }\bibfield  {title} {\bibinfo {title} {Chromatin {Network} {Analyses}: {Towards} {Structure}-{Function} {Relationships} in {Epigenomics}},\ }\bibfield  {journal} {\bibinfo  {journal} {Frontiers in Bioinformatics}\ }\textbf {\bibinfo {volume} {1}},\ \href {https://doi.org/10.3389/fbinf.2021.742216} {10.3389/fbinf.2021.742216} (\bibinfo {year} {2021}),\ \bibinfo {note} {publisher: Frontiers}\BibitemShut {NoStop}%
\bibitem [{\citenamefont {Pancaldi}(2023)}]{pancaldi_network_2023}%
  \BibitemOpen
  \bibfield  {author} {\bibinfo {author} {\bibfnamefont {V.}~\bibnamefont {Pancaldi}},\ }\bibfield  {title} {\bibinfo {title} {Network models of chromatin structure},\ }\href {https://doi.org/10.1016/j.gde.2023.102051} {\bibfield  {journal} {\bibinfo  {journal} {Current Opinion in Genetics \& Development}\ }\textbf {\bibinfo {volume} {80}},\ \bibinfo {pages} {102051} (\bibinfo {year} {2023})}\BibitemShut {NoStop}%
\bibitem [{\citenamefont {Wang}\ and\ \citenamefont {Wong}(1987)}]{wang_stochastic_1987}%
  \BibitemOpen
  \bibfield  {author} {\bibinfo {author} {\bibfnamefont {Y.~J.}\ \bibnamefont {Wang}}\ and\ \bibinfo {author} {\bibfnamefont {G.~Y.}\ \bibnamefont {Wong}},\ }\bibfield  {title} {\bibinfo {title} {Stochastic {Blockmodels} for {Directed} {Graphs}},\ }\href {https://doi.org/10.1080/01621459.1987.10478385} {\bibfield  {journal} {\bibinfo  {journal} {Journal of the American Statistical Association}\ }\textbf {\bibinfo {volume} {82}},\ \bibinfo {pages} {8} (\bibinfo {year} {1987})}\BibitemShut {NoStop}%
\bibitem [{\citenamefont {Snijders}\ and\ \citenamefont {Nowicki}(1997)}]{snijders_estimation_1997}%
  \BibitemOpen
  \bibfield  {author} {\bibinfo {author} {\bibfnamefont {T.~A.}\ \bibnamefont {Snijders}}\ and\ \bibinfo {author} {\bibfnamefont {K.}~\bibnamefont {Nowicki}},\ }\bibfield  {title} {\bibinfo {title} {Estimation and {Prediction} for {Stochastic} {Blockmodels} for {Graphs} with {Latent} {Block} {Structure}},\ }\href {https://doi.org/10.1007/s003579900004} {\bibfield  {journal} {\bibinfo  {journal} {Journal of Classification}\ }\textbf {\bibinfo {volume} {14}},\ \bibinfo {pages} {75} (\bibinfo {year} {1997})},\ \bibinfo {note} {number: 1}\BibitemShut {NoStop}%
\bibitem [{\citenamefont {Lan}\ \emph {et~al.}(2012)\citenamefont {Lan}, \citenamefont {Witt}, \citenamefont {Katsumura}, \citenamefont {Ye}, \citenamefont {Wang}, \citenamefont {Bresnick}, \citenamefont {Farnham},\ and\ \citenamefont {Jin}}]{lanIntegrationHiCChIPseq2012}%
  \BibitemOpen
  \bibfield  {author} {\bibinfo {author} {\bibfnamefont {X.}~\bibnamefont {Lan}}, \bibinfo {author} {\bibfnamefont {H.}~\bibnamefont {Witt}}, \bibinfo {author} {\bibfnamefont {K.}~\bibnamefont {Katsumura}}, \bibinfo {author} {\bibfnamefont {Z.}~\bibnamefont {Ye}}, \bibinfo {author} {\bibfnamefont {Q.}~\bibnamefont {Wang}}, \bibinfo {author} {\bibfnamefont {E.~H.}\ \bibnamefont {Bresnick}}, \bibinfo {author} {\bibfnamefont {P.~J.}\ \bibnamefont {Farnham}},\ and\ \bibinfo {author} {\bibfnamefont {V.~X.}\ \bibnamefont {Jin}},\ }\bibfield  {title} {\bibinfo {title} {Integration of {{Hi-C}} and {{ChIP-seq}} data reveals distinct types of chromatin linkages},\ }\href {https://doi.org/10.1093/nar/gks501} {\bibfield  {journal} {\bibinfo  {journal} {Nucleic Acids Research}\ }\textbf {\bibinfo {volume} {40}},\ \bibinfo {pages} {7690} (\bibinfo {year} {2012})},\ \Eprint {https://arxiv.org/abs/22675074} {22675074} \BibitemShut {NoStop}%
\bibitem [{SMn()}]{SMnote}%
  \BibitemOpen
  \href@noop {} {}\bibinfo {note} {See Supplemental Material at http://XXX for details on the mathematical derivation of the variational algorithm and for supplementary figures and table}\BibitemShut {NoStop}%
\bibitem [{\citenamefont {Frieze}\ and\ \citenamefont {Karoński}(2015)}]{frieze_introduction_2015}%
  \BibitemOpen
  \bibfield  {author} {\bibinfo {author} {\bibfnamefont {A.}~\bibnamefont {Frieze}}\ and\ \bibinfo {author} {\bibfnamefont {M.}~\bibnamefont {Karoński}},\ }\href {https://doi.org/10.1017/CBO9781316339831} {\emph {\bibinfo {title} {Introduction to {Random} {Graphs}}}}\ (\bibinfo  {publisher} {Cambridge University Press},\ \bibinfo {address} {Cambridge},\ \bibinfo {year} {2015})\BibitemShut {NoStop}%
\bibitem [{\citenamefont {Bollobás}(2001)}]{bollobas_random_2001}%
  \BibitemOpen
  \bibfield  {author} {\bibinfo {author} {\bibfnamefont {B.}~\bibnamefont {Bollobás}},\ }\href {https://doi.org/10.1017/CBO9780511814068} {\emph {\bibinfo {title} {Random {Graphs}}}},\ \bibinfo {edition} {2nd}\ ed.,\ Cambridge {Studies} in {Advanced} {Mathematics}\ (\bibinfo  {publisher} {Cambridge University Press},\ \bibinfo {address} {Cambridge},\ \bibinfo {year} {2001})\BibitemShut {NoStop}%
\bibitem [{\citenamefont {Lee}\ and\ \citenamefont {Wilkinson}(2019)}]{lee_review_2019}%
  \BibitemOpen
  \bibfield  {author} {\bibinfo {author} {\bibfnamefont {C.}~\bibnamefont {Lee}}\ and\ \bibinfo {author} {\bibfnamefont {D.~J.}\ \bibnamefont {Wilkinson}},\ }\bibfield  {title} {\bibinfo {title} {A review of stochastic block models and extensions for graph clustering},\ }\href {https://doi.org/10.1007/s41109-019-0232-2} {\bibfield  {journal} {\bibinfo  {journal} {Applied Network Science}\ }\textbf {\bibinfo {volume} {4}},\ \bibinfo {pages} {1} (\bibinfo {year} {2019})},\ \bibinfo {note} {number: 1 Publisher: SpringerOpen}\BibitemShut {NoStop}%
\bibitem [{\citenamefont {Daudin}\ \emph {et~al.}(2008)\citenamefont {Daudin}, \citenamefont {Picard},\ and\ \citenamefont {Robin}}]{daudin_mixture_2008}%
  \BibitemOpen
  \bibfield  {author} {\bibinfo {author} {\bibfnamefont {J.-J.}\ \bibnamefont {Daudin}}, \bibinfo {author} {\bibfnamefont {F.}~\bibnamefont {Picard}},\ and\ \bibinfo {author} {\bibfnamefont {S.}~\bibnamefont {Robin}},\ }\bibfield  {title} {\bibinfo {title} {A mixture model for random graphs},\ }\href {https://doi.org/10.1007/s11222-007-9046-7} {\bibfield  {journal} {\bibinfo  {journal} {Statistics and Computing}\ }\textbf {\bibinfo {volume} {18}},\ \bibinfo {pages} {173} (\bibinfo {year} {2008})}\BibitemShut {NoStop}%
\bibitem [{\citenamefont {Abbe}(2018)}]{abbe_community_2018}%
  \BibitemOpen
  \bibfield  {author} {\bibinfo {author} {\bibfnamefont {E.}~\bibnamefont {Abbe}},\ }\bibfield  {title} {\bibinfo {title} {Community {Detection} and {Stochastic} {Block} {Models}: {Recent} {Developments}},\ }\href {http://jmlr.org/papers/v18/16-480.html} {\bibfield  {journal} {\bibinfo  {journal} {Journal of Machine Learning Research}\ }\textbf {\bibinfo {volume} {18}},\ \bibinfo {pages} {1} (\bibinfo {year} {2018})}\BibitemShut {NoStop}%
\bibitem [{\citenamefont {Grosberg}\ \emph {et~al.}(1993)\citenamefont {Grosberg}, \citenamefont {Rabin}, \citenamefont {Havlin},\ and\ \citenamefont {Neer}}]{Grosberg1993}%
  \BibitemOpen
  \bibfield  {author} {\bibinfo {author} {\bibfnamefont {A.}~\bibnamefont {Grosberg}}, \bibinfo {author} {\bibfnamefont {Y.}~\bibnamefont {Rabin}}, \bibinfo {author} {\bibfnamefont {S.}~\bibnamefont {Havlin}},\ and\ \bibinfo {author} {\bibfnamefont {A.}~\bibnamefont {Neer}},\ }\bibfield  {title} {\bibinfo {title} {Crumpled globule model of the three-dimensional structure of dna},\ }\href {http://stacks.iop.org/0295-5075/23/i=5/a=012} {\bibfield  {journal} {\bibinfo  {journal} {EPL (Europhysics Letters)}\ }\textbf {\bibinfo {volume} {23}},\ \bibinfo {pages} {373} (\bibinfo {year} {1993})}\BibitemShut {NoStop}%
\bibitem [{\citenamefont {M\"unkel}\ and\ \citenamefont {Langowski}(1998)}]{LangowskiPRE1998}%
  \BibitemOpen
  \bibfield  {author} {\bibinfo {author} {\bibfnamefont {C.}~\bibnamefont {M\"unkel}}\ and\ \bibinfo {author} {\bibfnamefont {J.}~\bibnamefont {Langowski}},\ }\bibfield  {title} {\bibinfo {title} {Chromosome structure predicted by a polymer model},\ }\href {https://link.aps.org/doi/10.1103/PhysRevE.57.5888} {\bibfield  {journal} {\bibinfo  {journal} {Phys. Rev. E}\ }\textbf {\bibinfo {volume} {57}},\ \bibinfo {pages} {5888} (\bibinfo {year} {1998})}\BibitemShut {NoStop}%
\bibitem [{\citenamefont {Rosa}\ and\ \citenamefont {Everaers}(2008)}]{RosaEveraersPlosCB2008}%
  \BibitemOpen
  \bibfield  {author} {\bibinfo {author} {\bibfnamefont {A.}~\bibnamefont {Rosa}}\ and\ \bibinfo {author} {\bibfnamefont {R.}~\bibnamefont {Everaers}},\ }\bibfield  {title} {\bibinfo {title} {Structure and dynamics of interphase chromosomes},\ }\href {https://doi.org/10.1371/journal.pcbi.1000153} {\bibfield  {journal} {\bibinfo  {journal} {PLOS Computational Biology}\ }\textbf {\bibinfo {volume} {4}},\ \bibinfo {pages} {1} (\bibinfo {year} {2008})}\BibitemShut {NoStop}%
\bibitem [{\citenamefont {Rosa}\ \emph {et~al.}(2010)\citenamefont {Rosa}, \citenamefont {Becker},\ and\ \citenamefont {Everaers}}]{rosa_looping_2010}%
  \BibitemOpen
  \bibfield  {author} {\bibinfo {author} {\bibfnamefont {A.}~\bibnamefont {Rosa}}, \bibinfo {author} {\bibfnamefont {N.~B.}\ \bibnamefont {Becker}},\ and\ \bibinfo {author} {\bibfnamefont {R.}~\bibnamefont {Everaers}},\ }\bibfield  {title} {\bibinfo {title} {Looping {Probabilities} in {Model} {Interphase} {Chromosomes}},\ }\href {https://www.ncbi.nlm.nih.gov/pmc/articles/PMC2877331/} {\bibfield  {journal} {\bibinfo  {journal} {Biophysical Journal}\ }\textbf {\bibinfo {volume} {98}},\ \bibinfo {pages} {2410} (\bibinfo {year} {2010})}\BibitemShut {NoStop}%
\bibitem [{\citenamefont {Grosberg}(2012)}]{grosberg_how_2012}%
  \BibitemOpen
  \bibfield  {author} {\bibinfo {author} {\bibfnamefont {A.~Y.}\ \bibnamefont {Grosberg}},\ }\bibfield  {title} {\bibinfo {title} {How two meters of {DNA} fit into a cell nucleus: {Polymer} models with topological constraints and experimental data},\ }\href {https://doi.org/10.1134/S1811238212070028} {\bibfield  {journal} {\bibinfo  {journal} {Polymer Science Series C}\ }\textbf {\bibinfo {volume} {54}},\ \bibinfo {pages} {1} (\bibinfo {year} {2012})}\BibitemShut {NoStop}%
\bibitem [{\citenamefont {Halverson}\ \emph {et~al.}(2014)\citenamefont {Halverson}, \citenamefont {Smrek}, \citenamefont {Kremer},\ and\ \citenamefont {Grosberg}}]{halverson_melt_2014}%
  \BibitemOpen
  \bibfield  {author} {\bibinfo {author} {\bibfnamefont {J.~D.}\ \bibnamefont {Halverson}}, \bibinfo {author} {\bibfnamefont {J.}~\bibnamefont {Smrek}}, \bibinfo {author} {\bibfnamefont {K.}~\bibnamefont {Kremer}},\ and\ \bibinfo {author} {\bibfnamefont {A.~Y.}\ \bibnamefont {Grosberg}},\ }\bibfield  {title} {\bibinfo {title} {From a melt of rings to chromosome territories: the role of topological constraints in genome folding},\ }\href@noop {} {\bibfield  {journal} {\bibinfo  {journal} {Reports on Progress in Physics}\ }\textbf {\bibinfo {volume} {77}},\ \bibinfo {pages} {022601} (\bibinfo {year} {2014})}\BibitemShut {NoStop}%
\bibitem [{\citenamefont {Mariadassou}\ \emph {et~al.}(2010)\citenamefont {Mariadassou}, \citenamefont {Robin},\ and\ \citenamefont {Vacher}}]{mariadassou_uncovering_2010}%
  \BibitemOpen
  \bibfield  {author} {\bibinfo {author} {\bibfnamefont {M.}~\bibnamefont {Mariadassou}}, \bibinfo {author} {\bibfnamefont {S.}~\bibnamefont {Robin}},\ and\ \bibinfo {author} {\bibfnamefont {C.}~\bibnamefont {Vacher}},\ }\bibfield  {title} {\bibinfo {title} {Uncovering latent structure in valued graphs: {A} variational approach},\ }\href {https://doi.org/10.1214/10-AOAS361} {\bibfield  {journal} {\bibinfo  {journal} {The Annals of Applied Statistics}\ }\textbf {\bibinfo {volume} {4}},\ \bibinfo {pages} {715} (\bibinfo {year} {2010})},\ \bibinfo {note} {publisher: Institute of Mathematical Statistics}\BibitemShut {NoStop}%
\bibitem [{\citenamefont {Airoldi}\ \emph {et~al.}(2008)\citenamefont {Airoldi}, \citenamefont {Blei}, \citenamefont {Fienberg},\ and\ \citenamefont {Xing}}]{airoldi_mixed_2008}%
  \BibitemOpen
  \bibfield  {author} {\bibinfo {author} {\bibfnamefont {E.~M.}\ \bibnamefont {Airoldi}}, \bibinfo {author} {\bibfnamefont {D.~M.}\ \bibnamefont {Blei}}, \bibinfo {author} {\bibfnamefont {S.~E.}\ \bibnamefont {Fienberg}},\ and\ \bibinfo {author} {\bibfnamefont {E.~P.}\ \bibnamefont {Xing}},\ }\bibfield  {title} {\bibinfo {title} {Mixed {Membership} {Stochastic} {Blockmodels}},\ }\href {https://www.ncbi.nlm.nih.gov/pmc/articles/PMC3119541/} {\bibfield  {journal} {\bibinfo  {journal} {Journal of machine learning research : JMLR}\ }\textbf {\bibinfo {volume} {9}},\ \bibinfo {pages} {1981} (\bibinfo {year} {2008})}\BibitemShut {NoStop}%
\bibitem [{\citenamefont {Johnson}\ \emph {et~al.}(2007)\citenamefont {Johnson}, \citenamefont {Mortazavi}, \citenamefont {Myers},\ and\ \citenamefont {Wold}}]{johnson_genome-wide_2007}%
  \BibitemOpen
  \bibfield  {author} {\bibinfo {author} {\bibfnamefont {D.~S.}\ \bibnamefont {Johnson}}, \bibinfo {author} {\bibfnamefont {A.}~\bibnamefont {Mortazavi}}, \bibinfo {author} {\bibfnamefont {R.~M.}\ \bibnamefont {Myers}},\ and\ \bibinfo {author} {\bibfnamefont {B.}~\bibnamefont {Wold}},\ }\bibfield  {title} {\bibinfo {title} {Genome-{Wide} {Mapping} of in {Vivo} {Protein}-{DNA} {Interactions}},\ }\href {https://doi.org/10.1126/science.1141319} {\bibfield  {journal} {\bibinfo  {journal} {Science}\ }\textbf {\bibinfo {volume} {316}},\ \bibinfo {pages} {1497} (\bibinfo {year} {2007})},\ \bibinfo {note} {number: 5830 Publisher: American Association for the Advancement of Science}\BibitemShut {NoStop}%
\bibitem [{\citenamefont {Allenby}\ \emph {et~al.}(2005)\citenamefont {Allenby}, \citenamefont {Rossi},\ and\ \citenamefont {McCulloch}}]{allenby_hierarchical_2005}%
  \BibitemOpen
  \bibfield  {author} {\bibinfo {author} {\bibfnamefont {G.~M.}\ \bibnamefont {Allenby}}, \bibinfo {author} {\bibfnamefont {P.~E.}\ \bibnamefont {Rossi}},\ and\ \bibinfo {author} {\bibfnamefont {R.~E.}\ \bibnamefont {McCulloch}},\ }\href {https://doi.org/10.2139/ssrn.655541} {\bibinfo {title} {Hierarchical {Bayes} {Models}: {A} {Practitioners} {Guide}}} (\bibinfo {year} {2005})\BibitemShut {NoStop}%
\bibitem [{\citenamefont {Gelman}\ \emph {et~al.}(2014)\citenamefont {Gelman}, \citenamefont {Carlin}, \citenamefont {Stern}, \citenamefont {Dunson}, \citenamefont {Vehtari},\ and\ \citenamefont {Rubin}}]{gelman_bayesian_2014}%
  \BibitemOpen
  \bibfield  {author} {\bibinfo {author} {\bibfnamefont {A.}~\bibnamefont {Gelman}}, \bibinfo {author} {\bibfnamefont {J.~B.}\ \bibnamefont {Carlin}}, \bibinfo {author} {\bibfnamefont {H.~S.}\ \bibnamefont {Stern}}, \bibinfo {author} {\bibfnamefont {D.~B.}\ \bibnamefont {Dunson}}, \bibinfo {author} {\bibfnamefont {A.}~\bibnamefont {Vehtari}},\ and\ \bibinfo {author} {\bibfnamefont {D.~B.}\ \bibnamefont {Rubin}},\ }\href@noop {} {\emph {\bibinfo {title} {Bayesian data analysis}}},\ \bibinfo {edition} {third edition}\ ed.,\ Texts in statistical science series\ (\bibinfo  {publisher} {CRC Press, Taylor and Francis Group},\ \bibinfo {address} {Boca Raton London New York},\ \bibinfo {year} {2014})\BibitemShut {NoStop}%
\bibitem [{\citenamefont {Aitchison}(1982)}]{aitchison_statistical_1982}%
  \BibitemOpen
  \bibfield  {author} {\bibinfo {author} {\bibfnamefont {J.}~\bibnamefont {Aitchison}},\ }\bibfield  {title} {\bibinfo {title} {The {Statistical} {Analysis} of {Compositional} {Data}},\ }\href {https://doi.org/10.1111/j.2517-6161.1982.tb01195.x} {\bibfield  {journal} {\bibinfo  {journal} {Journal of the Royal Statistical Society: Series B (Methodological)}\ }\textbf {\bibinfo {volume} {44}},\ \bibinfo {pages} {139} (\bibinfo {year} {1982})}\BibitemShut {NoStop}%
\bibitem [{\citenamefont {Blei}\ and\ \citenamefont {Lafferty}(2007)}]{blei_correlated_2007}%
  \BibitemOpen
  \bibfield  {author} {\bibinfo {author} {\bibfnamefont {D.~M.}\ \bibnamefont {Blei}}\ and\ \bibinfo {author} {\bibfnamefont {J.~D.}\ \bibnamefont {Lafferty}},\ }\bibfield  {title} {\bibinfo {title} {A correlated topic model of {Science}},\ }\bibfield  {journal} {\bibinfo  {journal} {The Annals of Applied Statistics}\ }\textbf {\bibinfo {volume} {1}},\ \href {https://doi.org/10.1214/07-AOAS114} {10.1214/07-AOAS114} (\bibinfo {year} {2007}),\ \bibinfo {note} {arXiv:0708.3601 [stat]}\BibitemShut {NoStop}%
\bibitem [{\citenamefont {Esposito}\ \emph {et~al.}(2022)\citenamefont {Esposito}, \citenamefont {Bianco}, \citenamefont {Chiariello}, \citenamefont {Abraham}, \citenamefont {Fiorillo}, \citenamefont {Conte}, \citenamefont {Campanile},\ and\ \citenamefont {Nicodemi}}]{esposito_polymer_2022}%
  \BibitemOpen
  \bibfield  {author} {\bibinfo {author} {\bibfnamefont {A.}~\bibnamefont {Esposito}}, \bibinfo {author} {\bibfnamefont {S.}~\bibnamefont {Bianco}}, \bibinfo {author} {\bibfnamefont {A.~M.}\ \bibnamefont {Chiariello}}, \bibinfo {author} {\bibfnamefont {A.}~\bibnamefont {Abraham}}, \bibinfo {author} {\bibfnamefont {L.}~\bibnamefont {Fiorillo}}, \bibinfo {author} {\bibfnamefont {M.}~\bibnamefont {Conte}}, \bibinfo {author} {\bibfnamefont {R.}~\bibnamefont {Campanile}},\ and\ \bibinfo {author} {\bibfnamefont {M.}~\bibnamefont {Nicodemi}},\ }\bibfield  {title} {\bibinfo {title} {Polymer physics reveals a combinatorial code linking {3D} chromatin architecture to {1D} chromatin states},\ }\href {https://doi.org/10.1016/j.celrep.2022.110601} {\bibfield  {journal} {\bibinfo  {journal} {Cell Reports}\ }\textbf {\bibinfo {volume} {38}},\ \bibinfo {pages} {110601} (\bibinfo {year} {2022})}\BibitemShut {NoStop}%
\bibitem [{\citenamefont {Lloyd}(1982)}]{lloydLeastSquaresQuantization1982}%
  \BibitemOpen
  \bibfield  {author} {\bibinfo {author} {\bibfnamefont {S.}~\bibnamefont {Lloyd}},\ }\bibfield  {title} {\bibinfo {title} {Least squares quantization in {{PCM}}},\ }\href {https://doi.org/10.1109/TIT.1982.1056489} {\bibfield  {journal} {\bibinfo  {journal} {IEEE Transactions on Information Theory}\ }\textbf {\bibinfo {volume} {28}},\ \bibinfo {pages} {129} (\bibinfo {year} {1982})}\BibitemShut {NoStop}%
\bibitem [{\citenamefont {Schreiber}\ \emph {et~al.}(2020)\citenamefont {Schreiber}, \citenamefont {Singh}, \citenamefont {Bilmes},\ and\ \citenamefont {Noble}}]{schreiber_pitfall_2020}%
  \BibitemOpen
  \bibfield  {author} {\bibinfo {author} {\bibfnamefont {J.}~\bibnamefont {Schreiber}}, \bibinfo {author} {\bibfnamefont {R.}~\bibnamefont {Singh}}, \bibinfo {author} {\bibfnamefont {J.}~\bibnamefont {Bilmes}},\ and\ \bibinfo {author} {\bibfnamefont {W.~S.}\ \bibnamefont {Noble}},\ }\bibfield  {title} {\bibinfo {title} {A pitfall for machine learning methods aiming to predict across cell types},\ }\href {https://doi.org/10.1186/s13059-020-02177-y} {\bibfield  {journal} {\bibinfo  {journal} {Genome Biology}\ }\textbf {\bibinfo {volume} {21}},\ \bibinfo {pages} {282} (\bibinfo {year} {2020})}\BibitemShut {NoStop}%
\bibitem [{\citenamefont {Rao}\ \emph {et~al.}(2017)\citenamefont {Rao}, \citenamefont {Huang}, \citenamefont {Glenn St~Hilaire}, \citenamefont {Engreitz}, \citenamefont {Perez}, \citenamefont {Kieffer-Kwon}, \citenamefont {Sanborn}, \citenamefont {Johnstone}, \citenamefont {Bascom}, \citenamefont {Bochkov}, \citenamefont {Huang}, \citenamefont {Shamim}, \citenamefont {Shin}, \citenamefont {Turner}, \citenamefont {Ye}, \citenamefont {Omer}, \citenamefont {Robinson}, \citenamefont {Schlick}, \citenamefont {Bernstein}, \citenamefont {Casellas}, \citenamefont {Lander},\ and\ \citenamefont {Aiden}}]{rao_cohesin_2017}%
  \BibitemOpen
  \bibfield  {author} {\bibinfo {author} {\bibfnamefont {S.~S.~P.}\ \bibnamefont {Rao}}, \bibinfo {author} {\bibfnamefont {S.-C.}\ \bibnamefont {Huang}}, \bibinfo {author} {\bibfnamefont {B.}~\bibnamefont {Glenn St~Hilaire}}, \bibinfo {author} {\bibfnamefont {J.~M.}\ \bibnamefont {Engreitz}}, \bibinfo {author} {\bibfnamefont {E.~M.}\ \bibnamefont {Perez}}, \bibinfo {author} {\bibfnamefont {K.-R.}\ \bibnamefont {Kieffer-Kwon}}, \bibinfo {author} {\bibfnamefont {A.~L.}\ \bibnamefont {Sanborn}}, \bibinfo {author} {\bibfnamefont {S.~E.}\ \bibnamefont {Johnstone}}, \bibinfo {author} {\bibfnamefont {G.~D.}\ \bibnamefont {Bascom}}, \bibinfo {author} {\bibfnamefont {I.~D.}\ \bibnamefont {Bochkov}}, \bibinfo {author} {\bibfnamefont {X.}~\bibnamefont {Huang}}, \bibinfo {author} {\bibfnamefont {M.~S.}\ \bibnamefont {Shamim}}, \bibinfo {author} {\bibfnamefont {J.}~\bibnamefont {Shin}}, \bibinfo {author} {\bibfnamefont {D.}~\bibnamefont {Turner}}, \bibinfo {author} {\bibfnamefont {Z.}~\bibnamefont {Ye}}, \bibinfo {author}
  {\bibfnamefont {A.~D.}\ \bibnamefont {Omer}}, \bibinfo {author} {\bibfnamefont {J.~T.}\ \bibnamefont {Robinson}}, \bibinfo {author} {\bibfnamefont {T.}~\bibnamefont {Schlick}}, \bibinfo {author} {\bibfnamefont {B.~E.}\ \bibnamefont {Bernstein}}, \bibinfo {author} {\bibfnamefont {R.}~\bibnamefont {Casellas}}, \bibinfo {author} {\bibfnamefont {E.~S.}\ \bibnamefont {Lander}},\ and\ \bibinfo {author} {\bibfnamefont {E.~L.}\ \bibnamefont {Aiden}},\ }\bibfield  {title} {\bibinfo {title} {Cohesin {Loss} {Eliminates} {All} {Loop} {Domains}},\ }\href {https://doi.org/10.1016/j.cell.2017.09.026} {\bibfield  {journal} {\bibinfo  {journal} {Cell}\ }\textbf {\bibinfo {volume} {171}},\ \bibinfo {pages} {305} (\bibinfo {year} {2017})}\BibitemShut {NoStop}%
\bibitem [{noa(2024{\natexlab{a}})}]{noauthor_encode_2024}%
  \BibitemOpen
  \href {https://www.encodeproject.org/} {\bibinfo {title} {{ENCODE}}} (\bibinfo {year} {2024}{\natexlab{a}})\BibitemShut {NoStop}%
\bibitem [{\citenamefont {Davidson}\ and\ \citenamefont {Peters}(2021)}]{Davidson2021}%
  \BibitemOpen
  \bibfield  {author} {\bibinfo {author} {\bibfnamefont {I.~F.}\ \bibnamefont {Davidson}}\ and\ \bibinfo {author} {\bibfnamefont {J.-M.}\ \bibnamefont {Peters}},\ }\bibfield  {title} {\bibinfo {title} {Genome folding through loop extrusion by smc complexes},\ }\href {https://doi.org/10.1038/s41580-021-00349-7} {\bibfield  {journal} {\bibinfo  {journal} {Nature Reviews Molecular Cell Biology}\ }\textbf {\bibinfo {volume} {22}},\ \bibinfo {pages} {445} (\bibinfo {year} {2021})}\BibitemShut {NoStop}%
\bibitem [{\citenamefont {Leidescher}\ \emph {et~al.}()\citenamefont {Leidescher}, \citenamefont {Ribisel}, \citenamefont {Ullrich}, \citenamefont {Feodorova}, \citenamefont {Hildebrand}, \citenamefont {Galitsyna}, \citenamefont {Bultmann}, \citenamefont {Link}, \citenamefont {Thanisch}, \citenamefont {Mulholland}, \citenamefont {Dekker}, \citenamefont {Leonhardt}, \citenamefont {Mirny},\ and\ \citenamefont {Solovei}}]{leidescherSpatialOrganizationTranscribed2022}%
  \BibitemOpen
  \bibfield  {author} {\bibinfo {author} {\bibfnamefont {S.}~\bibnamefont {Leidescher}}, \bibinfo {author} {\bibfnamefont {J.}~\bibnamefont {Ribisel}}, \bibinfo {author} {\bibfnamefont {S.}~\bibnamefont {Ullrich}}, \bibinfo {author} {\bibfnamefont {Y.}~\bibnamefont {Feodorova}}, \bibinfo {author} {\bibfnamefont {E.}~\bibnamefont {Hildebrand}}, \bibinfo {author} {\bibfnamefont {A.}~\bibnamefont {Galitsyna}}, \bibinfo {author} {\bibfnamefont {S.}~\bibnamefont {Bultmann}}, \bibinfo {author} {\bibfnamefont {S.}~\bibnamefont {Link}}, \bibinfo {author} {\bibfnamefont {K.}~\bibnamefont {Thanisch}}, \bibinfo {author} {\bibfnamefont {C.}~\bibnamefont {Mulholland}}, \bibinfo {author} {\bibfnamefont {J.}~\bibnamefont {Dekker}}, \bibinfo {author} {\bibfnamefont {H.}~\bibnamefont {Leonhardt}}, \bibinfo {author} {\bibfnamefont {L.}~\bibnamefont {Mirny}},\ and\ \bibinfo {author} {\bibfnamefont {I.}~\bibnamefont {Solovei}},\ }\bibfield  {title} {\bibinfo {title} {Spatial organization of transcribed eukaryotic genes},\ }\href
  {https://doi.org/10.1038/s41556-022-00847-6} {\bibfield  {journal} {\bibinfo  {journal} {Nature Cell Biology}\ }\textbf {\bibinfo {volume} {24}},\ \bibinfo {pages} {327}}\BibitemShut {NoStop}%
\bibitem [{\citenamefont {Jerkovic´}\ and\ \citenamefont {Cavalli}(2021)}]{jerkovic_understanding_2021}%
  \BibitemOpen
  \bibfield  {author} {\bibinfo {author} {\bibfnamefont {I.}~\bibnamefont {Jerkovic´}}\ and\ \bibinfo {author} {\bibfnamefont {G.}~\bibnamefont {Cavalli}},\ }\bibfield  {title} {\bibinfo {title} {Understanding {3D} genome organization by multidisciplinary methods},\ }\href {https://doi.org/10.1038/s41580-021-00362-w} {\bibfield  {journal} {\bibinfo  {journal} {Nature Reviews Molecular Cell Biology}\ }\textbf {\bibinfo {volume} {22}},\ \bibinfo {pages} {511} (\bibinfo {year} {2021})},\ \bibinfo {note} {number: 8 Publisher: Nature Publishing Group}\BibitemShut {NoStop}%
\bibitem [{\citenamefont {Schoenfelder}\ \emph {et~al.}(2018)\citenamefont {Schoenfelder}, \citenamefont {Javierre}, \citenamefont {Furlan-Magaril}, \citenamefont {Wingett},\ and\ \citenamefont {Fraser}}]{schoenfelder_promoter_2018}%
  \BibitemOpen
  \bibfield  {author} {\bibinfo {author} {\bibfnamefont {S.}~\bibnamefont {Schoenfelder}}, \bibinfo {author} {\bibfnamefont {B.-M.}\ \bibnamefont {Javierre}}, \bibinfo {author} {\bibfnamefont {M.}~\bibnamefont {Furlan-Magaril}}, \bibinfo {author} {\bibfnamefont {S.~W.}\ \bibnamefont {Wingett}},\ and\ \bibinfo {author} {\bibfnamefont {P.}~\bibnamefont {Fraser}},\ }\bibfield  {title} {\bibinfo {title} {Promoter {Capture} {Hi}-{C}: {High}-resolution, {Genome}-wide {Profiling} of {Promoter} {Interactions}},\ }\href {https://doi.org/10.3791/57320} {\bibfield  {journal} {\bibinfo  {journal} {Journal of Visualized Experiments : JoVE}\ ,\ \bibinfo {pages} {57320}} (\bibinfo {year} {2018})}\BibitemShut {NoStop}%
\bibitem [{\citenamefont {Li}\ \emph {et~al.}(2010)\citenamefont {Li}, \citenamefont {Fullwood}, \citenamefont {Xu}, \citenamefont {Mulawadi}, \citenamefont {Velkov}, \citenamefont {Vega}, \citenamefont {Ariyaratne}, \citenamefont {Mohamed}, \citenamefont {Ooi}, \citenamefont {Tennakoon}, \citenamefont {Wei}, \citenamefont {Ruan},\ and\ \citenamefont {Sung}}]{li_chia-pet_2010}%
  \BibitemOpen
  \bibfield  {author} {\bibinfo {author} {\bibfnamefont {G.}~\bibnamefont {Li}}, \bibinfo {author} {\bibfnamefont {M.~J.}\ \bibnamefont {Fullwood}}, \bibinfo {author} {\bibfnamefont {H.}~\bibnamefont {Xu}}, \bibinfo {author} {\bibfnamefont {F.~H.}\ \bibnamefont {Mulawadi}}, \bibinfo {author} {\bibfnamefont {S.}~\bibnamefont {Velkov}}, \bibinfo {author} {\bibfnamefont {V.}~\bibnamefont {Vega}}, \bibinfo {author} {\bibfnamefont {P.~N.}\ \bibnamefont {Ariyaratne}}, \bibinfo {author} {\bibfnamefont {Y.~B.}\ \bibnamefont {Mohamed}}, \bibinfo {author} {\bibfnamefont {H.-S.}\ \bibnamefont {Ooi}}, \bibinfo {author} {\bibfnamefont {C.}~\bibnamefont {Tennakoon}}, \bibinfo {author} {\bibfnamefont {C.-L.}\ \bibnamefont {Wei}}, \bibinfo {author} {\bibfnamefont {Y.}~\bibnamefont {Ruan}},\ and\ \bibinfo {author} {\bibfnamefont {W.-K.}\ \bibnamefont {Sung}},\ }\bibfield  {title} {\bibinfo {title} {{ChIA}-{PET} tool for comprehensive chromatin interaction analysis with paired-end tag sequencing},\ }\href
  {https://doi.org/10.1186/gb-2010-11-2-r22} {\bibfield  {journal} {\bibinfo  {journal} {Genome Biology}\ }\textbf {\bibinfo {volume} {11}},\ \bibinfo {pages} {R22} (\bibinfo {year} {2010})}\BibitemShut {NoStop}%
\bibitem [{\citenamefont {Li}\ \emph {et~al.}(2014)\citenamefont {Li}, \citenamefont {Cai}, \citenamefont {Chang}, \citenamefont {Hong}, \citenamefont {Zhou}, \citenamefont {Kulakova}, \citenamefont {Kolchanov},\ and\ \citenamefont {Ruan}}]{li_chromatin_2014}%
  \BibitemOpen
  \bibfield  {author} {\bibinfo {author} {\bibfnamefont {G.}~\bibnamefont {Li}}, \bibinfo {author} {\bibfnamefont {L.}~\bibnamefont {Cai}}, \bibinfo {author} {\bibfnamefont {H.}~\bibnamefont {Chang}}, \bibinfo {author} {\bibfnamefont {P.}~\bibnamefont {Hong}}, \bibinfo {author} {\bibfnamefont {Q.}~\bibnamefont {Zhou}}, \bibinfo {author} {\bibfnamefont {E.~V.}\ \bibnamefont {Kulakova}}, \bibinfo {author} {\bibfnamefont {N.~A.}\ \bibnamefont {Kolchanov}},\ and\ \bibinfo {author} {\bibfnamefont {Y.}~\bibnamefont {Ruan}},\ }\bibfield  {title} {\bibinfo {title} {Chromatin {Interaction} {Analysis} with {Paired}-{End} {Tag} ({ChIA}-{PET}) sequencing technology and application},\ }\href {https://doi.org/10.1186/1471-2164-15-S12-S11} {\bibfield  {journal} {\bibinfo  {journal} {BMC Genomics}\ }\textbf {\bibinfo {volume} {15}},\ \bibinfo {pages} {S11} (\bibinfo {year} {2014})}\BibitemShut {NoStop}%
\bibitem [{\citenamefont {Wang}\ and\ \citenamefont {Blei}(2013)}]{wang_variational_2013}%
  \BibitemOpen
  \bibfield  {author} {\bibinfo {author} {\bibfnamefont {C.}~\bibnamefont {Wang}}\ and\ \bibinfo {author} {\bibfnamefont {D.~M.}\ \bibnamefont {Blei}},\ }\href {http://arxiv.org/abs/1209.4360} {\bibinfo {title} {Variational {Inference} in {Nonconjugate} {Models}}} (\bibinfo {year} {2013}),\ \bibinfo {note} {issue: arXiv:1209.4360 arXiv:1209.4360 [stat]}\BibitemShut {NoStop}%
\bibitem [{\citenamefont {Nocedal}\ and\ \citenamefont {Wright}(2006)}]{nocedal_numerical_2006}%
  \BibitemOpen
  \bibfield  {author} {\bibinfo {author} {\bibfnamefont {J.}~\bibnamefont {Nocedal}}\ and\ \bibinfo {author} {\bibfnamefont {S.~J.}\ \bibnamefont {Wright}},\ }\href {https://doi.org/10.1007/978-0-387-40065-5} {\emph {\bibinfo {title} {Numerical {Optimization}}}},\ Springer {Series} in {Operations} {Research} and {Financial} {Engineering}\ (\bibinfo  {publisher} {Springer New York},\ \bibinfo {year} {2006})\BibitemShut {NoStop}%
\bibitem [{noa(2024{\natexlab{b}})}]{noauthor_4dn_2024}%
  \BibitemOpen
  \href {https://data.4dnucleome.org/} {\bibinfo {title} {{4DN} {Data} {Portal}}} (\bibinfo {year} {2024}{\natexlab{b}})\BibitemShut {NoStop}%
\bibitem [{\citenamefont {Durand}\ \emph {et~al.}(2016)\citenamefont {Durand}, \citenamefont {Robinson}, \citenamefont {Shamim}, \citenamefont {Machol}, \citenamefont {Mesirov}, \citenamefont {Lander},\ and\ \citenamefont {Aiden}}]{durand_juicebox_2016}%
  \BibitemOpen
  \bibfield  {author} {\bibinfo {author} {\bibfnamefont {N.~C.}\ \bibnamefont {Durand}}, \bibinfo {author} {\bibfnamefont {J.~T.}\ \bibnamefont {Robinson}}, \bibinfo {author} {\bibfnamefont {M.~S.}\ \bibnamefont {Shamim}}, \bibinfo {author} {\bibfnamefont {I.}~\bibnamefont {Machol}}, \bibinfo {author} {\bibfnamefont {J.~P.}\ \bibnamefont {Mesirov}}, \bibinfo {author} {\bibfnamefont {E.~S.}\ \bibnamefont {Lander}},\ and\ \bibinfo {author} {\bibfnamefont {E.~L.}\ \bibnamefont {Aiden}},\ }\bibfield  {title} {\bibinfo {title} {Juicebox {Provides} a {Visualization} {System} for {Hi}-{C} {Contact} {Maps} with {Unlimited} {Zoom}},\ }\href {https://doi.org/10.1016/j.cels.2015.07.012} {\bibfield  {journal} {\bibinfo  {journal} {Cell Systems}\ }\textbf {\bibinfo {volume} {3}},\ \bibinfo {pages} {99} (\bibinfo {year} {2016})}\BibitemShut {NoStop}%
\bibitem [{noa(2024{\natexlab{c}})}]{noauthor_deeptoolspybigwig_2024}%
  \BibitemOpen
  \href {https://github.com/deeptools/pyBigWig} {\bibinfo {title} {deeptools/{pyBigWig}}} (\bibinfo {year} {2024}{\natexlab{c}}),\ \bibinfo {note} {original-date: 2015-07-31T12:07:36Z}\BibitemShut {NoStop}%
\bibitem [{\citenamefont {Vinh}\ \emph {et~al.}(2010)\citenamefont {Vinh}, \citenamefont {Epps},\ and\ \citenamefont {Bailey}}]{vinhInformationTheoreticMeasures2010}%
  \BibitemOpen
  \bibfield  {author} {\bibinfo {author} {\bibfnamefont {N.~X.}\ \bibnamefont {Vinh}}, \bibinfo {author} {\bibfnamefont {J.}~\bibnamefont {Epps}},\ and\ \bibinfo {author} {\bibfnamefont {J.}~\bibnamefont {Bailey}},\ }\bibfield  {title} {\bibinfo {title} {Information {{Theoretic Measures}} for {{Clusterings Comparison}}: {{Variants}}, {{Properties}}, {{Normalization}} and {{Correction}} for {{Chance}}},\ }\href {http://jmlr.org/papers/v11/vinh10a.html} {\bibfield  {journal} {\bibinfo  {journal} {Journal of Machine Learning Research}\ }\textbf {\bibinfo {volume} {11}},\ \bibinfo {pages} {2837} (\bibinfo {year} {2010})}\BibitemShut {NoStop}%
\bibitem [{\citenamefont {Cover}\ and\ \citenamefont {Thomas}(2006)}]{coverElementsInformationTheory2006}%
  \BibitemOpen
  \bibfield  {author} {\bibinfo {author} {\bibfnamefont {T.~M.}\ \bibnamefont {Cover}}\ and\ \bibinfo {author} {\bibfnamefont {J.~A.}\ \bibnamefont {Thomas}},\ }\href@noop {} {\emph {\bibinfo {title} {Elements of Information Theory}}},\ \bibinfo {edition} {2nd}\ ed.\ (\bibinfo  {publisher} {Wiley-Interscience},\ \bibinfo {year} {2006})\BibitemShut {NoStop}%
\bibitem [{\citenamefont {Kuhn}(1955)}]{kuhnHungarianMethodAssignment1955}%
  \BibitemOpen
  \bibfield  {author} {\bibinfo {author} {\bibfnamefont {H.~W.}\ \bibnamefont {Kuhn}},\ }\bibfield  {title} {\bibinfo {title} {The {{Hungarian}} method for the assignment problem},\ }\href {https://doi.org/10.1002/nav.3800020109} {\bibfield  {journal} {\bibinfo  {journal} {Naval Research Logistics Quarterly}\ }\textbf {\bibinfo {volume} {2}},\ \bibinfo {pages} {83} (\bibinfo {year} {1955})}\BibitemShut {NoStop}%
\end{thebibliography}%

\widetext
\clearpage
\begin{center}
\textbf{\Large Supplemental Material \\ \vspace*{1.5mm} bioSBM: a random graph model to integrate epigenomic data in chromatin structure prediction} \\
\vspace*{5mm}
Alex Chen Yi Zhang, Angelo Rosa, Guido Sanguinetti
\vspace*{10mm}
\end{center}

\setcounter{equation}{0}
\setcounter{figure}{0}
\setcounter{table}{0}
\setcounter{page}{1}
\setcounter{section}{0}
\setcounter{page}{1}
\makeatletter
\renewcommand{\theequation}{S\arabic{equation}}
\renewcommand{\thefigure}{S\arabic{figure}}
\renewcommand{\thetable}{S\arabic{table}}
\renewcommand{\thesection}{S\arabic{section}}
\renewcommand{\thepage}{S\arabic{page}}

\onecolumngrid
\tableofcontents

\clearpage

\section{Derivation of the Variational Bayes Expectation Maximization (VBEM) algorithm for bioSBM}\label{sec:VEM_derivation}
In this Section, we provide a more detailed derivation of the equations at the core of the Variational Bayes Expectation Maximization (VBEM) algorithm for the model introduced in the paper.
The quantities in our problem are:
\begin{itemize}
    \item Observed data $Y \in \mathbb{R}^{N \times N}$: matrix representing Hi-C maps with a log-observed-over-expected normalization.
    \item Known node covariates $X \in \mathbb{R}^{P \times N}$: each row $x_i$ is the P-dimensional feature vector associated to node $i$.
    \item Latent variables: $\theta_{i=1 \ldots N}$ and $z_{ij}$ and $z_{ji}$ for $i=1 \ldots N,\,\,\,j=1 \ldots i-1$, where $\theta_i$ is the latent membership vector for node $i$ and $z_{ji}$ is the membership assignment of node $i$ in the interaction with node $j$. 
    \item Hyperparameters: $(\Sigma,\Gamma,B,\sigma^2)\equiv\Psi$.
\end{itemize}

A lower bound for the evidence $P(Y|\Sigma,\Gamma,B,X)$ is given by the ELBO (literally \emph{Evidence Lower BOund}):
\begin{align}
    \mathcal{L}(q,\Psi)
    = & \mathbb{E}_q\left[\log P(Y,\eta_{1:N}, Z|\Psi, X) - \log q(\eta_{1:N}, Z)\right] = \nonumber \\
    = & \mathbb{E}_{q}\left[\log P(\eta_{1:N}|\Sigma, \Gamma,X) + \log P(Z|\eta_{1:N}) + \log P(Y|Z, B) - \log q(\eta_{1:N}, Z) \right] \nonumber \\
    = & \sum_{i=1}^N \mathbb{E}_q\left[ \log P(\eta_i|\mu_i=\Gamma x_i, \Sigma]) \right] + \sum_{i=1}^N \sum_{j=1}^{i-1} \mathbb{E}_q\left[ \log P(z_{ij}|\theta_i)P(z_{ji}|\theta_j) \right] \\
    & + \sum_{i=1}^N \sum_{j=1}^{i-1} \mathbb{E}_q\left[ \log P(Y_{ij}|z_{ij},z_{ji},B)\right] + H(q) \leq P(Y|\Psi, X) \nonumber 
\end{align} 
The inequality becomes an equality if the distribution $q(\eta_{1:N}, Z)$ is exactly the posterior distribution.
The variational Bayes \emph{Expectation-Maximization} (VBEM) method consists of iteratively tightening this bound to approximate the true posterior.
In the E-step of the algorithm, one needs to maximize the ELBO with respect to the variational distribution $q(\eta_{1:N}, Z)$ while in the M-step one carries out the maximization with respect to the hyperparameters.

\subsection{Variational E-step}
To carry out the optimization task we take a mean-field assumption on the shape of the variational distribution $q(\eta, Z)$, where we assume the factorized form $q(\eta, Z) = q(\eta) q(Z)= \prod_{i} q(\eta_i) \prod_{i \neq j}q(z_{ij})$.  
By taking the functional derivatives of the ELBO with respect to the variational distribution and by fixing them to zero one finds the update equations for the E-step, which in the general form read:
\begin{eqnarray}
q(\eta_i) 
& \propto & \exp \left\{ \log P(\eta_i|\mu_i, \Sigma ) + \mathbb{E}_{q(Z)}[\log P(Z|\eta_i )] \right\} \\ \label{eq:eta_update}
q(z_{ij})
& \propto & \exp \left\{ \mathbb{E}_{q(z_{ji})} [ \log P(Y_{ij} |z_{ij} , z_{ji}, B) ] + \mathbb{E}_{q(\eta_i )}[\log P( z_{ij}|\eta_i )] \right\} \label{eq:zeta_update}
\end{eqnarray}
Because of the non-conjugacy of the logistic normal prior with the multinomial likelihood, we will resort to the approximation technique introduced by \cite{wang_variational_2013}.
More explicitly:
\begin{equation}
    q(\eta_i) \propto \exp \{ \log P(\eta_i | \mu_i, \Sigma) + \sum_{j (\neq i)} \mathbb{E}_{ q(z_{ij})} [\log P(z_{ij} | \eta_i)] \} := \exp \{ f (\eta_i)\}
\end{equation}
Let $\hat{\eta}_i$ be the maximum a-posteriori value of the parameter. The method introduced by \cite{wang_variational_2013} consists in approximating the function $ f(\eta_i)$ with its expansion up to second order around $\hat{\eta}_i$. This gives the following Gaussian approximation for the variational approximation $q(\eta_i)$:
\begin{equation}
    q(\eta_i) \simeq \mathcal{N}(\eta_i|\, \hat{\eta}_i, -\nabla^2 f(\hat{\eta}_i)^{-1})
\end{equation}
Therefore we need the function at the exponent to be doubly differentiable and to find the maximum a posterior (variational) estimate $\hat{\eta}_i$ we can use standard gradient ascent techniques.
If we write out explicitly the function we have that
\begin{equation}
    f(\eta_i) = -\frac{1}{2}(\eta_i - \mu_i)^T (\Sigma)^{-1} (\eta_i - \mu_i) + \sum_{j (\neq i)}^N \sum_{k=1}^K \left( \eta_{i,k} - \log \sum_{k'=1}^K \exp(\eta_{i,k'})\right) q(z_{ij} = k)
\end{equation}
Defining $\theta(\eta)$ as the \emph{softmax} of $\eta$ i.e.  $\theta(\eta)_{k} = \exp(\eta_{k})/ \sum_{k'}\exp(\eta_{k'})$ we can obtain the gradient and the Hessian matrix of $f(\eta_i)$.
\begin{equation}
    \nabla f(\eta_i) = \sum_{j (\neq i)}^N \left( \mathbb{E}_{q(z_{ij})} [z_{ij}] - \theta(\eta_i) \sum_{k=1}^K  q(z_{ij} = k) \right) -\Sigma^{-1}(\eta_i - \mu_i)
\end{equation}
\begin{equation}
    \nabla^2 f(\eta_i) = \sum_{j (\neq i)}^N \left( ( \text{diag} (\theta(\eta_i)) - \theta(\eta_i) \theta(\eta_i)^T ) \sum_{k=1}^K  q(z_{ij} = k) \right) - \Sigma^{-1} 
\end{equation}
For the community assignments distributions we have $q(z_{ij}) = \text{Mult}(\phi_{ij})$ with $\phi_{ij}$ as variational parameters, then we can explicit the expectations in the above expressions as:
\begin{equation}
    \nabla f(\eta_i) =  - (N-1)\theta(\eta_i) + \sum_{j (\neq i)}^N \phi_{ij} -\Sigma^{-1}(\eta_i - \mu_i)
\end{equation}
\begin{equation}
    \nabla^2 f(\eta_i) =  -(N - 1) ( \text{diag} (\theta(\eta_i)) - \theta(\eta_i) \theta(\eta_i)^T ) - \Sigma^{-1} 
\end{equation}
Where $\text{diag}(\theta)$ is the diagonal matrix where the non-zero elements are the components of $\theta$.
It can be shown that $-\nabla^2 f(\eta_i)$ is always a positive definite matrix, therefore the method always gives a valid covariance matrix for the variational approximation. 
We use the Broyden–Fletcher–Goldfarb–Shanno (BFGS) algorithm with second-order backtracking line-search~\cite{nocedal_numerical_2006}. The optimization procedure returns $(\hat{\eta}_i, -\nabla^2 f(\hat{\eta}_i)^{-1})$, thereby updating the variational distribution on $\eta$, $q(\eta_i) \simeq \mathcal{N}(\eta_i|\, \hat{\eta}_i, -\nabla^2 f(\hat{\eta}_i)^{-1})$.
This is the E-step for the membership distributions per node. This step, as for most variational mean field methods for mixed membership block models, has a cost which is $\sim \mathcal{O}(N^2)$.

For the community assignments posteriors we explicitly write Eq. \eqref{eq:zeta_update} and we get that for every $(i,j)$ and community $k$
\begin{align}
    \phi_{ij,k} 
    \propto & \exp \left\{ \sum_{g=1}^K \phi_{ji,g} \log P(Y_{ij} | B_{kg}, \sigma^2_{kg}) + \mathbb{E}_{q(\eta_i)} [ \log P(z_{ij} | \eta_i)] \right\} \\
    = & \exp \{ \mathbb{E}_{q(\eta_i)} [ \log P(z_{ij} | \eta_i)] \} \prod_{g=1}^K P(Y_{ij} | B_{kg}, \sigma^2_{kg})^{\phi_{ji,g}} \nonumber 
\end{align}
The term in the exponent reads:
\begin{align}
    \mathbb{E}_{q(\eta_i)} [ \log P(z_{ij} | \eta_i)]
    = & \,\, \mathbb{E}_{q(\eta_i)} [ \eta_{i,k} - \log \sum_{g=1}^K \exp(\eta_{i,g}) ] \\
    = & \,\, \lambda_{i,k} - \int d\eta_i \,\mathcal{N}(\lambda_i , \nu_i) \log \sum_{g=1}^K \exp(\eta_{i,g}) \nonumber 
\end{align}
By imposing normalization $\sum_k \phi_{ij,k} =1$ for all pairs $(i,j)$ the updates for the community assignments posteriors become
\begin{equation}
    \phi_{ij,k} \propto \exp(\lambda_{i,k}) \prod_{g=1}^K P(Y_{ij} | B_{kg} , \sigma^2_{kg})^{\phi_{ji,g}}
    \label{eq::Estep_phi}
\end{equation}
Now, if we substitute in the Gaussian edge likelihood, we get the explicit update equations:
\begin{equation}
    \log \phi_{ij,k} \,=\, \lambda_{i,k} - \sum_{g=1}^K \left[ \frac{(Y_{ij} - B_{kg})^2}{2 \sigma^2_{kg}} + \frac{\log \sigma^2_{kg}}{2}\right] \phi_{ji,g} + const.
\end{equation}
Everything described until now represents the E-step of the VBEM algorithm and the updates are performed with fixed $(\Sigma,\Gamma,B,\sigma^2)$.
The values for these parameters are updated in the M-step where instead we fix the variational parameters $\{\lambda_i, \nu_i\} \text{ for } i=1 \ldots N$ and  $\phi_{ij} \text{ for } i\neq j$ that are optimized in the E-step.

\subsection{Variational M-step}
For the global covariance matrix, we need to take the matrix derivative of the terms in the ELBO that contain $\Sigma$. By exploiting some properties of matrix derivatives we find that:
\begin{align}
    \frac{\partial \mathcal{L}(q,\Psi)}{\partial \Sigma^{-1}} 
    = & \frac{1}{2} \frac{\partial }{\partial \Sigma^{-1}} \sum_{i=1}^N \mathbb{E}_{q(\eta_i)} [(\eta_i - \mu_i)^T (\Sigma)^{-1} (\eta_i - \mu_i)  + \log |\Sigma^{-1}| ] = \nonumber \\
    = & \frac{1}{2} \frac{\partial }{\partial \Sigma^{-1}} \sum_{i=1}^N \log |\Sigma^{-1}| - (\lambda_i - \mu_i)^T \Sigma^{-1} (\lambda_i - \mu_i) + \text{Tr}(\Sigma^{-1} \nu_i) = \\
    = & \frac{N}{2}\Sigma^T - \frac{1}{2} \sum_{i=1}^N \nu_i^T + (\lambda_i - \mu_i) (\lambda_i - \mu_i)^T =0 \nonumber
    \label{eq::Mstep_Sigma_derivative}
\end{align}
Then the covariance matrix that maximizes the ELBO is given by 
\begin{equation}
\label{eq:Sigma_update}
    \hat{\Sigma} = \frac{1}{N} \sum_{i=1}^N \nu_i^T + (\lambda_i - \Gamma(x_i)) (\lambda_i - \Gamma(x_i))^T
\end{equation}
Something additional that we can do is to add a regularization to the norm of $\Sigma$ for numerical stability.
In particular, something simple we can do is to regularize using the \emph{nuclear norm}.
For a generic matrix, the nuclear norm is given by the sum of the singular values; for a positive definite matrix such as $\Sigma$ it is easy to show that the norm corresponds to the trace of the matrix $\text{Tr}(\Sigma)$.
We then introduce an additional term $-R \cdot \text{Tr}(\Sigma)$ in the ELBO where $R=rN$ (selected with cross-validation) is the regularization strength and we write the explicit $N$ dependence in such a way that all the terms of the ELBO containing $\Sigma$ are \emph{extensive}.
The results analyzed in the paper are obtained using $r=0.1$ With the additional term, Eq.~\eqref{eq:Sigma_update} reads:
\begin{align}
    & \frac{N}{2}\Sigma - \frac{1}{2} \sum_{i=1}^N \nu_i^T + (\lambda_i - \mu_i) (\lambda_i - \mu_i)^T -R \frac{\partial \text{Tr}(\Sigma)}{\partial \Sigma^{-1}} = \nonumber \\
    = \, & \frac{N}{2}\Sigma - \frac{1}{2} \sum_{i=1}^N \nu_i^T + (\lambda_i - \mu_i) (\lambda_i - \mu_i)^T + R \Sigma^{2} = 0
\end{align}
This equation is solved by the following matrix:
\begin{equation}
    \hat{\Sigma} = \frac{1}{4r} \left[ -\mathbb{I} + \sqrt{ \mathbb{I} + \frac{8r}{N} \sum_{i=1}^N \nu_i^T + (\lambda_i - \Gamma(x_i)) (\lambda_i - \Gamma(x_i))^T }\right]
\end{equation}
Which is the M-step update concerning the covariance matrix $\Sigma$.

The maximization of the ELBO with respect to $B_{kg}$ and $\sigma_{kg}$ entails:
\begin{equation}
    \frac{\partial \mathcal{L}(q,\Psi)}{\partial B_{kg}} = \sum_{i \neq j}^N \phi_{ij,k} \phi_{ji,g} \frac{Y_{ij} - B_{kg}}{\sigma^2_{kg}}
\end{equation}
\begin{equation}
    \frac{\partial \mathcal{L}(q,\Psi)}{\partial \sigma_{kg}} = \sum_{i \neq j}^N \phi_{ij,k} \phi_{ji,g} \left[ \frac{(Y_{ij} - B_{kg})^2}{2 \sigma_{kg}^2} - \frac{1}{2 \sigma_{kg}} \right]  
\end{equation}
By setting the above derivatives to zero we obtain: 
\begin{equation}
    \hat{B}_{kg} = \frac{\sum_{i \neq j}^N \phi_{ij,k} \phi_{ji,g} Y_{ij}}{ \sum_{i \neq j}^N \phi_{ij,k} \phi_{ji,g}}
\end{equation}
\begin{equation}
    \hat{\sigma}_{kg} = \frac{\sum_{i \neq j}^N \phi_{ij,k} \phi_{ji,g} (Y_{ij} - B_{kg})^2 }{ \sum_{i \neq j}^N \phi_{ij,k} \phi_{ji,g}}
\end{equation}
These are the last update equations needed to specify the full VBEM algorithm for bioSBM.

Following these calculations, the explicit expression for the ELBO that we compute to monitor the progression of the optimization reads:
\begin{align}
    \mathcal{L} 
    = & \frac{N}{2} \log |\Sigma^{-1}| - \frac{1}{2}\sum_{i=1}^N [ (\lambda_i - \mu_i)^T \Sigma^{-1} (\lambda_i - \mu_i) + \text{Tr}(\Sigma^{-1} \nu_i) ] + \sum_{i \neq j}^N \sum_{k=1}^K \phi_{ij,k} \lambda_{i,k} \nonumber \\
    & - (N-1)\sum_{i=1}^N \left( \log \sum_{k'=1}^K \exp (\lambda_{i,k'}) + \frac{1}{2} \text{Tr}\{ ( \text{diag} (\theta(\lambda_i)) - \theta(\lambda_i) \theta(\lambda_i)^T ) \nu_i \} \right) \\
    & + \sum_{i=1}^N \sum_{j=1}^{i-1} \sum_{k,g=1}^K \phi_{ij,k} \phi_{ji,g} \log P (Y_{ij}| B_{kg}, \sigma_{kg}) + \frac{1}{2} \sum_{i=1}^N \log |\nu_i| -\sum_{i \neq j }^N \sum_{k=1}^K \phi_{ij,k} \log \phi_{ij,k} \nonumber
\end{align}

\section{Method details}\label{sec:MethodsDetails}
\subsection{Dataset preprocessing}\label{sec:data_processing}
Chromatin contact files were downloaded from the 4DN consortium data portal~\cite{noauthor_4dn_2024}, and log-observed-over-expected Hi-C maps were extracted from these files using the \textit{hic-straw}~\cite{durand_juicebox_2016} Python toolkit.
We have processed each arm of each chromosome separately, with the exception of the very short $p$ arms of the acrocentric chromosomes 13,14, 15, 21 and 22.
To remove anomalously low reads from the Hi-C maps, we remove all rows and columns where the average value of the normalized Hi-C matrix has a modified Z-score smaller than $-3.0$.

ChIP-seq tracks for the input covariates were downloaded from the ENCODE Project database~\cite{noauthor_encode_2024}, and from these epigenetic data was extracted using the \textit{pyBigWiG}~\cite{noauthor_deeptoolspybigwig_2024} Python toolkit.
These tracks were then denoised using a standard lowpass order one Butterworth filter with critical frequency = 0.5 and normalized with a standard scaler.

\subsection{Simulation of predicted maps}\label{sec:maps_generation}
Starting from the global model parameter inferred from training data, we can generate predictions for an unseen chromosome with covariate matrix $X^{test}=(x_1^{test},\ldots , x_N^{test})$.
In particular, we compute the average membership proportions for each node as:
\begin{equation}
    \theta_i^{test} = \Gamma x_i^{test}
\end{equation}
Then for the prediction of edge weight $Y_{ij}^{test}$, we average over all possible pairs of community assignments of nodes $i$ and $j$, weighting with the membership proportions:
\begin{equation}
    Y_{ij}^{test} = \sum_{k,g}^K \theta_{ik} B_{kg} \theta_{jg}
\end{equation}
The prediction accuracy is then computed as the Pearson correlation between the model-generated maps and the experimental logarithmic observed over expected Hi-C maps.
For modeling the depletion of RAD21 we simply set the row of $X^{test}$ corresponding to RAD21 to zero.

Note that while the training of the model has been carried out looking only at intra-arm contacts data, the predicitons are made on whole chromosomes, including the contact between $p$ and $q$ arm.

\subsection{Log-fold enrichment}\label{sec:logfold_enrichment}
Given a partition of a set of $N$ elements $U = \{ u_1, \ldots, u_N\}$, where $u_i$ is the label assigned to the $i$-th element, and given a feature vector $F = \{ f_1, \ldots, f_N\}$, the log-fold enrichment of a certain cluster $k$ in the feature is defined as:
\begin{equation*}
    \ln \Big[  \frac{\sum_{i=1}^N f_i \cdot \delta(u_i,k)}{\sum_{i=1}^N f_i} \Big]
\end{equation*}
with $\delta(u_i,k)=1$ if $u_i = k$ and zero otherwise.

\subsection{Adjusted Mutual Information (AMI)}\label{sec:AMI}
Consider two partitions or clusterings of a set of $N$ elements, $U = \{ u_1, \ldots, u_N\}$ and $V = \{ v_1, \ldots, v_N\}$.
The Adjusted Mutual Information (AMI) between $U$ and $V$ is a measure of the similarity between the two clusterings and it is corrected for the effect of agreements solely by chance~\cite{vinhInformationTheoreticMeasures2010}.
It is defined as follows:
\begin{equation}
    AMI(U,V) = \frac{MI(U,V) - \mathbb{E}[MI(U,V)]}{\text{max}\{H(U), H(V)\}  - \mathbb{E}[MI(U,V)]}
\end{equation}
where $H(\cdot)$ and $MI(\cdot,\cdot)$ are the Shannon entropy and the mutual information respectively~\cite{coverElementsInformationTheory2006}.
The AMI takes its maximum value $1$ when the two partitions are identical, and it takes its minimum value $AMI=0$ when the two partitions have MI equal to the expected value solely due to chance.

\subsection{Clustering of membership vectors}\label{sec:clustering}
The inference procedure by bioSBM yields for each node a membership proportion vector of dimension $K$, rather than a single discrete label.
To make a direct comparison with known biological annotations such as AB compartments, subcompartments~\cite{rao_3d_2014}, or the interaction profile groups (IPGs) by Spracklin {\it et al.}~\cite{spracklinDiverseSilentChromatin2023}, we perform $k$-means clustering~\cite{lloydLeastSquaresQuantization1982} of the membership proportion vectors from our model.
To compare with AB compartment, we set the number of clusters to $2$, while to compare with subcompartments and IPGs, we set the number of clusters to 6.
Then, to evaluate the clustering results, we need to deal with the fact that the numerical labels assigned to the clusters are arbitrary, due to permutation ambiguity.
To match the numerical labels found with two clustering procedures, we solve a linear assignment problem, where the goal is to maximize the overlap between clusters.
We solve the matching problem using the so-called Hungarian algorithm~\cite{kuhnHungarianMethodAssignment1955}.

\newpage

\section*{Supplemental Table and Figures}\label{app:figs}

\newpage

%
\begin{table}[ht]
    \centering
    \begin{tabular}{|c|c|c c|}
    \hline
    \hline
    Cell Line & Assay & Experiment accession code & Data file code \\
    \hline
            & H3K4me3 (ChIP-seq)  & ENCSR057BWO & ENCFF287HAO \\
            & H3K27ac (ChIP-seq)  & ENCSR000AKC & ENCFF469WVA \\
            & H3K27me3 (ChIP-seq) & ENCSR000AKD & ENCFF919DOR \\
            & H3K4me1 (ChIP-seq)  & ENCSR000AKF & ENCFF564KBE \\
            & H3K36me3 (ChIP-seq) & ENCSR000AKE & ENCFF312MUY \\
            & H3K9me3 (ChIP-seq)  & ENCSR000AOX & ENCFF683HCZ \\
    GM12878 & H3K9ac (ChIP-seq)   & ENCSR000AKH & ENCFF599TRR \\
            & H3K4me2 (ChIP-seq)  & ENCSR000AKG & ENCFF627OKN \\
            & H4K20me1 (ChIP-seq) & ENCSR000AKI & ENCFF479XIQ \\
            & H2AFZ (ChIP-seq)    & ENCSR000AOV & ENCFF935EGN \\
            & H3K79me2 (ChIP-seq) & ENCSR000AOW & ENCFF931USZ \\
            & CTCF (ChIP-seq)     & ENCSR000DZN & ENCFF485CGE \\
            & POLR2A (ChIP-seq)   & ENCSR000EAD & ENCFF328MMS \\
            & RAD21 (ChIP-seq)    & ENCSR000EAC & ENCFF571ZJJ \\
    \hline
            & H3K4me3 (ChIP-seq)  & ENCSR333OPW & ENCFF649ZLF \\
            & H3K27ac (ChIP-seq)  & ENCSR661KMA & ENCFF787LMI \\
            & H3K27me3 (ChIP-seq) & ENCSR810BDB & ENCFF717ZKL \\ 
            & H3K4me1 (ChIP-seq)  & ENCSR161MXP & ENCFF239FXT \\
            & H3K36me3 (ChIP-seq) & ENCSR091QXP & ENCFF024LGD\\
    HCT116  & H3K9me3 (ChIP-seq)  & ENCSR179BUC & ENCFF254TIW\\
{\it and}   & H3K9ac (ChIP-seq)   & ENCSR093SHE & ENCFF743PJP\\
    HCT116 RAD21-  & H3K4me2 (ChIP-seq)  & ENCSR794ULT & ENCFF693HVA\\
            & H4K20me1 (ChIP-seq)  & ENCSR474DOV & ENCFF349NQI\\
            & H2AFZ (ChIP-seq)    & ENCSR227XNT & ENCFF863EHT\\
            & H3K79me2 (ChIP-seq)  & ENCSR494CCN & ENCFF631DLM\\
            & CTCF (ChIP-seq)     & ENCSR450DXU & ENCFF787LAV\\
            & POLR2A (ChIP-seq)   & ENCSR000EUU & ENCFF802CGI\\
            & RAD21 (ChIP-seq)    & ENCSR000BSB & ENCFF776IXR\\
    \hline
            & H3K4me3 (ChIP-seq)  & ENCSR668LDD & ENCFF253TOF \\
            & H3K27ac (ChIP-seq)  & ENCSR000AKP & ENCFF381NDD \\
            & H3K27me3 (ChIP-seq) & ENCSR000AKQ & ENCFF242ENK \\
            & H3K4me1 (ChIP-seq)  & ENCSR000AKS & ENCFF607SUJ \\
            & H3K36me3 (ChIP-seq) & ENCSR000AKR & ENCFF317VHO \\
            & H3K9me3 (ChIP-seq)  & ENCSR000APE & ENCFF601JGK \\
K562 & H3K9ac (ChIP-seq)   & ENCSR000AKV & ENCFF286WRJ \\
            & H3K4me2 (ChIP-seq)  & ENCSR000AKT & ENCFF959YJV \\
            & H4K20me1 (ChIP-seq) & ENCSR000AKX & ENCFF605FAF \\
            & H2AFZ (ChIP-seq)    & ENCSR000APC & ENCFF621DJP \\
            & H3K79me2 (ChIP-seq) & ENCSR000APD & ENCFF544AVW \\
            & CTCF (ChIP-seq)     & ENCSR000AKO & ENCFF979PWH \\
            & POLR2A (ChIP-seq)   & ENCSR031TFS & ENCFF124WLE \\
            & RAD21 (ChIP-seq)    & ENCSR000FAD & ENCFF320RTQ \\
    \hline
    \hline
    \hline
    GM12878 & \textit{in situ} Hi-C  & 4DNES3JX38V5 & 4DNFI1UEG1HD \\
    \hline
    HCT116 RAD21- & \textit{in situ} Hi-C  & 4DNESJV9TH8Q & 4DNFILIM6FDL \\
    \hline
    HCT116 & \textit{in situ} Hi-C & 4DNES3QAGOZZ & 4DNFIP71EWXC\\
    \hline 
    K562 & \textit{in situ} Hi-C & 4DNESI7DEJTM & 4DNFITUOMFUQ\\
    \hline
    \end{tabular}
    \caption{
    Accession code for experiments and data files used throughout the paper.
    }
    \label{tab:InputData}
\end{table}

\begin{figure}[htbp]
    \centering
    \includegraphics[width=0.95\textwidth]{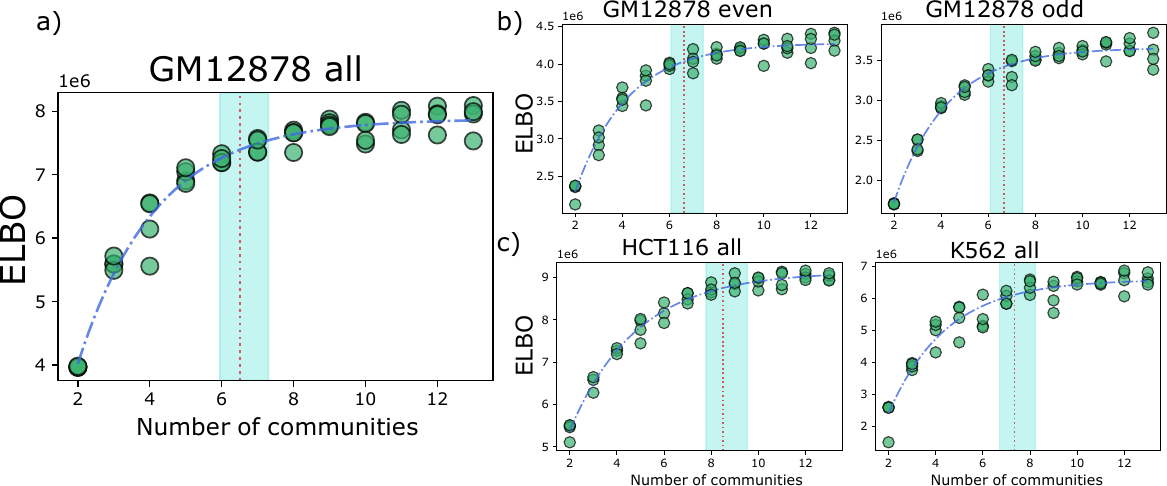}
    \caption[Model selection for block model]{
    The Evidence Lower BOund (ELBO) was used as a criterion for Bayesian model selection. For each dataset and each value for the number of communities, the variational algorithm was run multiple times with random initialization. Each run was stopped when the relative increase of the ELBO went below $10^{-5}$. The final ELBO values as a function of K was fitted with a saturating exponential of the form $a(1-e^{-K/\tilde{K}}) + b$ and the turquoise band represents a saturation in the range $93.5-96.5\%$. a) shows that for the dataset we used to perform predictions (all chromosomes of the GM12878 cell line), we identify the number of communities to be between 6 and 7. b) shows that the model selection is robust to subsampling the dataset, as the model selection of the odd numbered chromosomes and even numbered chromosomes yields consistent results c) shows that the number of communities to describe the interaction structures are different in different cell lines.
    }
    \label{fig:number_of_classes}
\end{figure}

\begin{figure}
    \centering
    \includegraphics[width=0.95\linewidth]{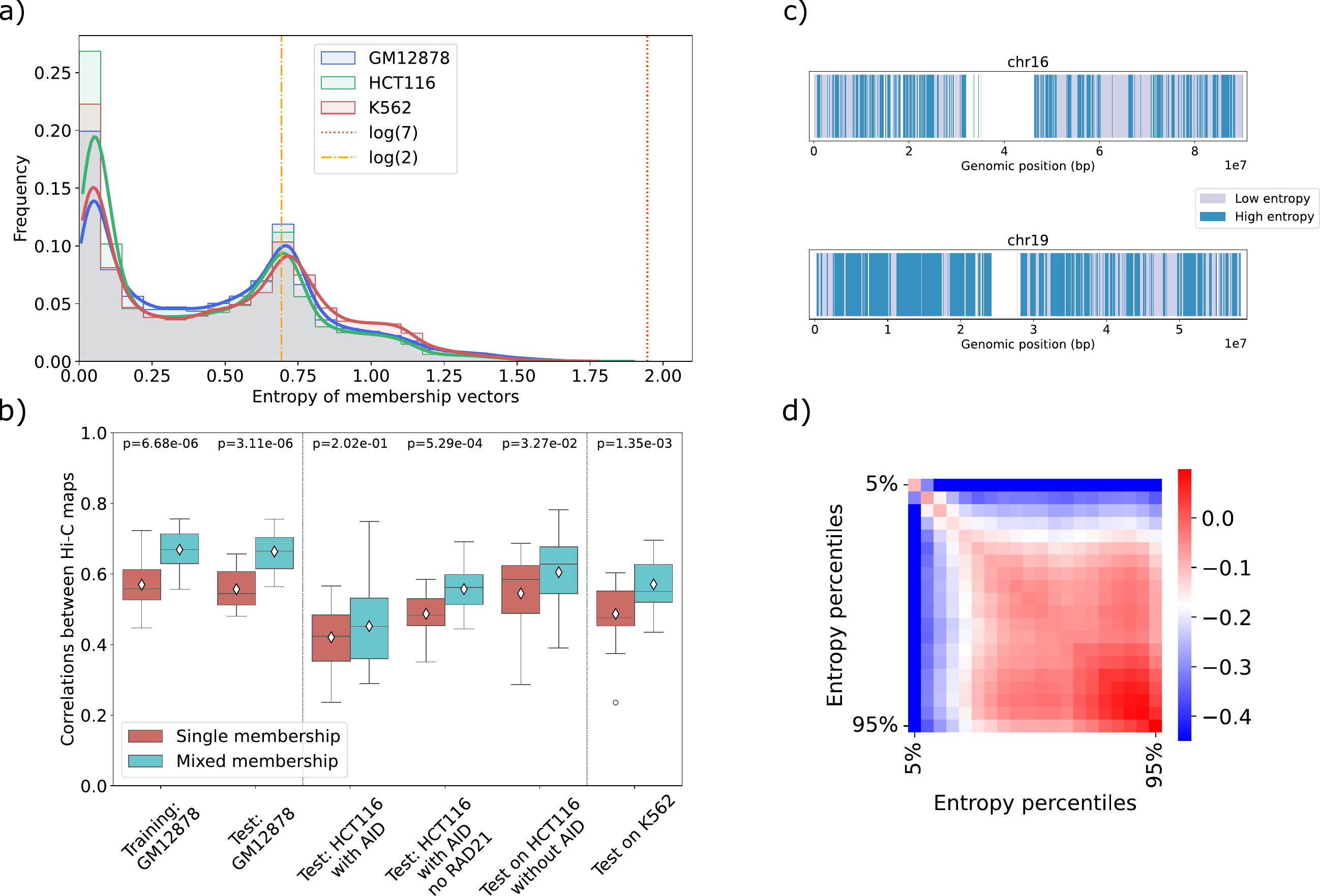}
    \caption{
    (a)
    Shannon entropies of the MAP estimates of the membership vectors for all analyzed genomic regions. The peak at zero shows that a big fraction of nodes displays a single membership behavior, however, there is still a considerable degree of {\it mixed-membershipness} if one looks at the entirety of the genomic regions under study.
    The pattern appears to be conserved across different cell lines.
    (b) Blue box plots show the prediction correlations reported in Fig.~5 in main text.
    Red box plots represent the same predictions, but with the membership vectors reduced to the most probable (single) community assignment.
    (c) Examples of patterns of genomic regions with high or low entropy in membership proportions, classified with the 50th percentile of the distribution from panel (a).
    (d) Saddle plot computed on genome-wide Hi-C contacts in GM12878, showing contact differences between regions stratified by entropy levels.
    } 
    \label{fig:mixed-membershipness}
\end{figure}

\begin{figure}
    \centering
    \includegraphics[width=0.8\linewidth]{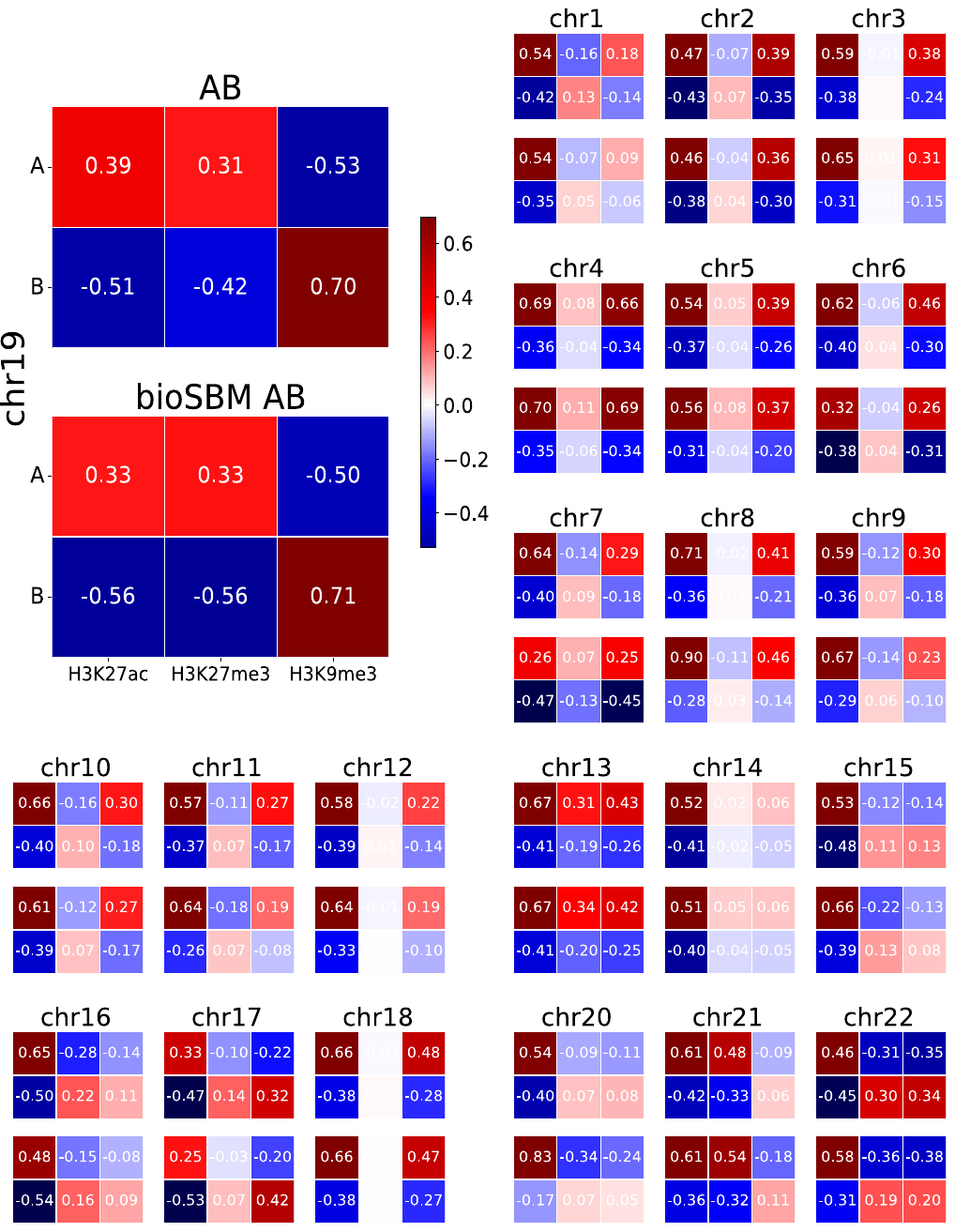}
    \caption{Log-fold enrichment in key biochemical features for all somatic chromosomes. For each chromosome, the top heatmap represents enrichment for A/B compartments, while the bottom one is for the binary clustering obtained from bioSBM membership proportion vectors.}
    \label{fig:AB_enrich_all_chroms}
\end{figure}

\begin{figure}
    \centering
    \includegraphics[width=0.78\linewidth]{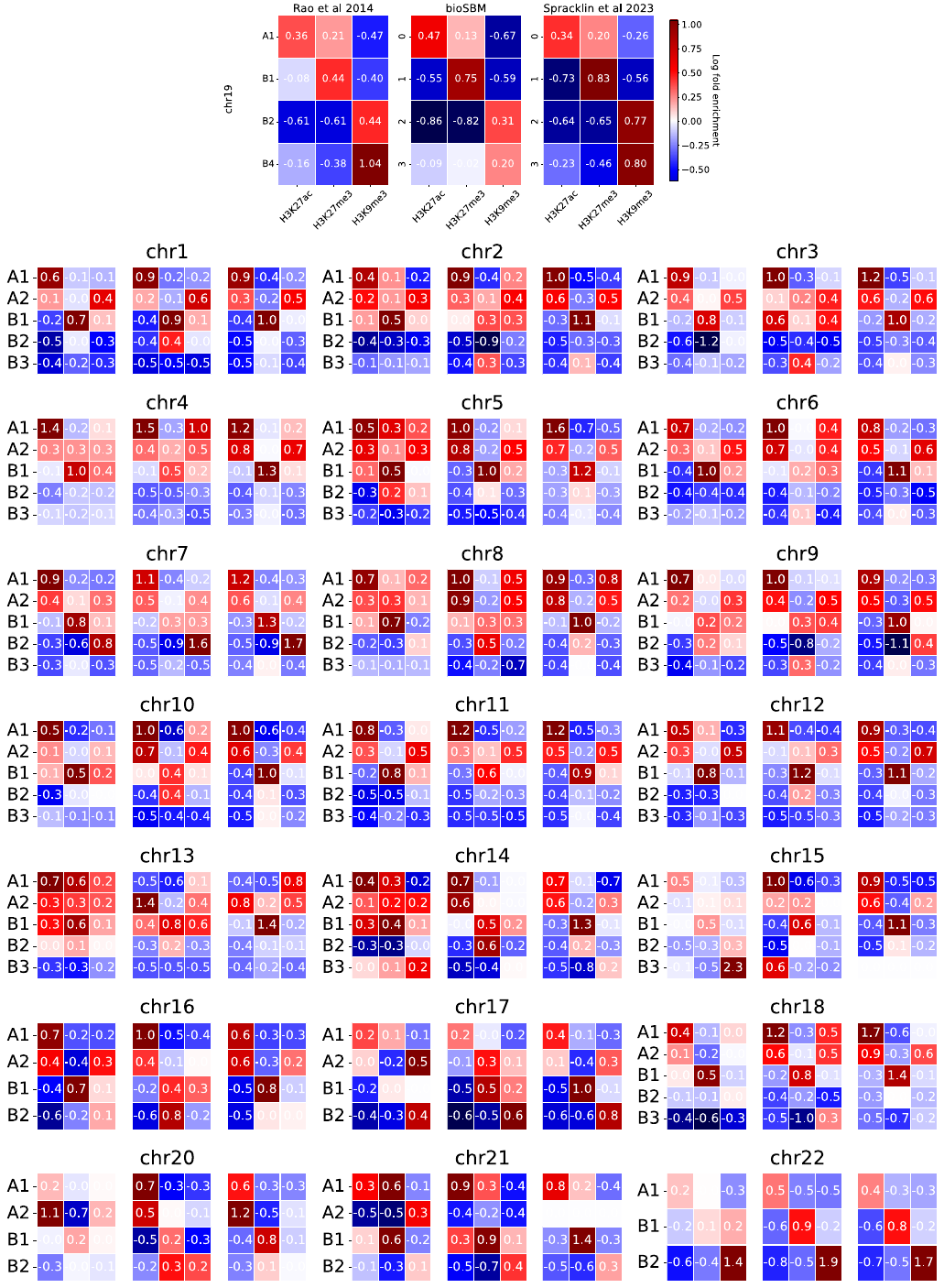}
    \caption{Log-fold enrichment in key biochemical features for all somatic chromosomes. For each chromosome, the left heatmap represents enrichment for subcompartments~\cite{rao_3d_2014}, the right one is for the clusters obtained with the spectral clustering algorithm from~\cite{spracklinDiverseSilentChromatin2023} using 6 groups, and the central one is for the clustering obtained from bioSBM membership proportion vectors setting the number of clusters equal to 6.}
    \label{fig:6_enrich_all_chroms}
\end{figure}

\begin{figure}
    \centering
    \includegraphics[width=0.94\linewidth]{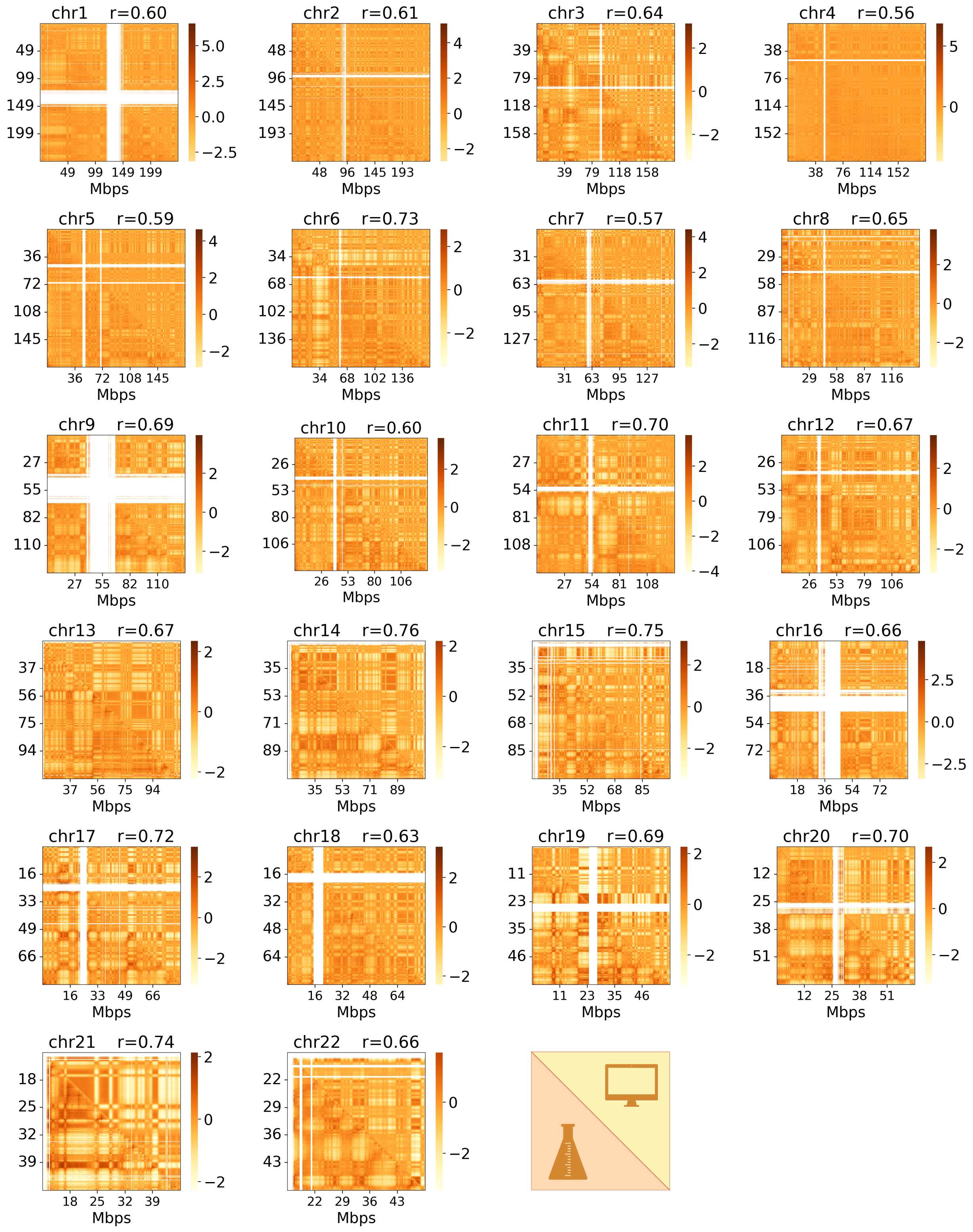}
    \caption{Hi-C maps predicted for GM12878. Bottom triangles depict experimental log O/E Hi-C maps for the GM12878 cell line. The top triangles represent model predictions. The prediction for each even-numbered chromosome is carried out by the model trained on odd-numbered chromosomes alone and vice versa.}
    \label{fig:hic_maps_all_chroms}
\end{figure}

\begin{figure}
    \centering
    \includegraphics[width=0.65\linewidth]{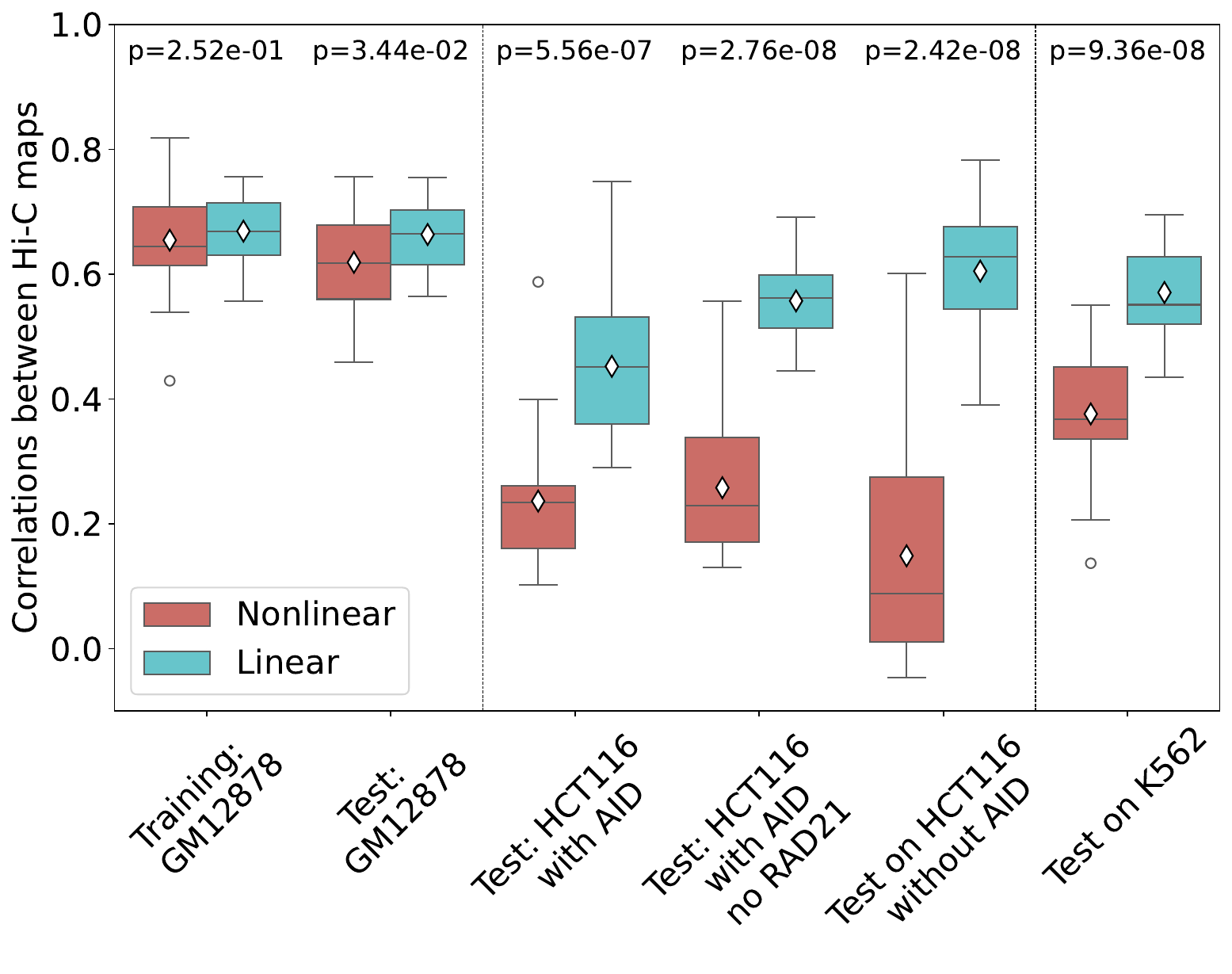}
    \caption{
    Correlations between model-generated maps and experimental log O/E Hi-C maps for two models with a different map from biochemical covariates to membership vectors.
    The turquoise plots correspond to the case where $\Gamma$ is the simple $14 \times 7$ matrix representing a linear mapping, discussed in main text.
    The red boxes correspond to the case where $\Gamma()$ is a feed-forward deep neural network with 4 hidden layers of size $(64,64,64,32)$ with the standard \textit{ReLU} (rectified linear unit)
    non-linear activations.
    While on the training set and on test chromosomes from the same cell line of the training data, the non-linear model maintains a high accuracy, for predictions on other cell lines, the correlations decrease drastically.
    }
    \label{fig:correlations_linear_vs_non_linear}
\end{figure}

\begin{figure}
    \centering
    \includegraphics[width=0.8\linewidth]{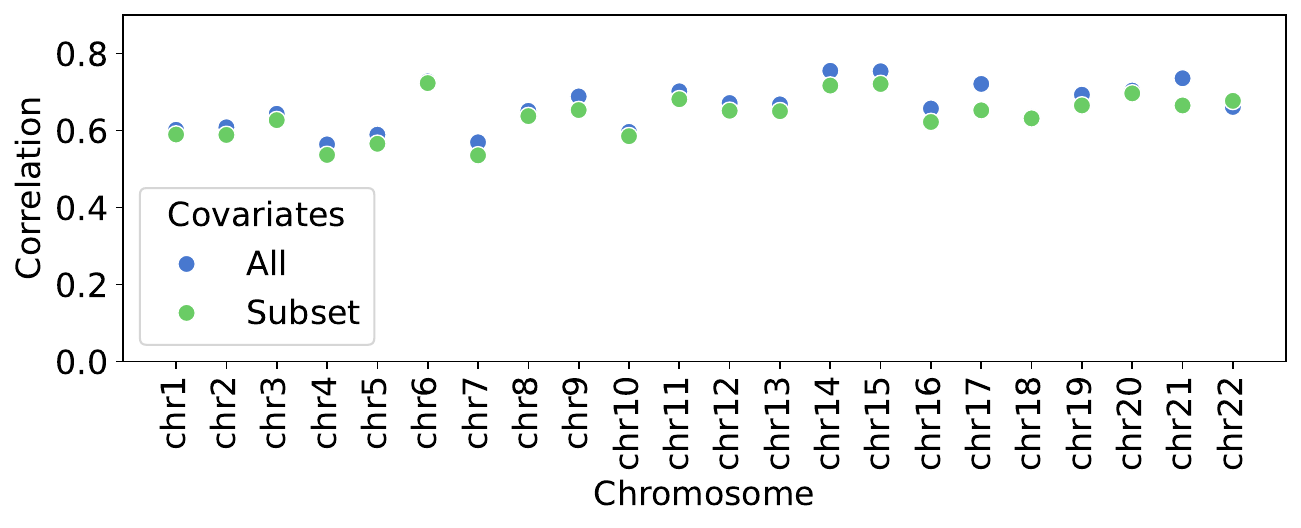}
    \caption{Prediction accuracies for GM12878 using a subset of covariates. The blue dots refer to Pearson correlation coefficients between experimental log O/E maps and model generated contacts using the full set of 14 covariates. The green dots refer to predictions made with {\it H3k27ac, H3k27me3, H3k9me3, H3k79me2, H3k9ac, RAD21, POLR2A}. The predictions on the even-numbered chromosomes are done by re-training the regression weights $\Gamma$~(Sec.~\ref{sec:VEM_derivation}), with the membership proportion vectors inferred for odd-numbered chromosomes. The reverse is done for the odd-numbered chromosomes.}
    \label{fig:prediction_accuracies_less_covariates}
\end{figure}

\begin{figure}
    \centering
    \includegraphics[width=0.9\linewidth]{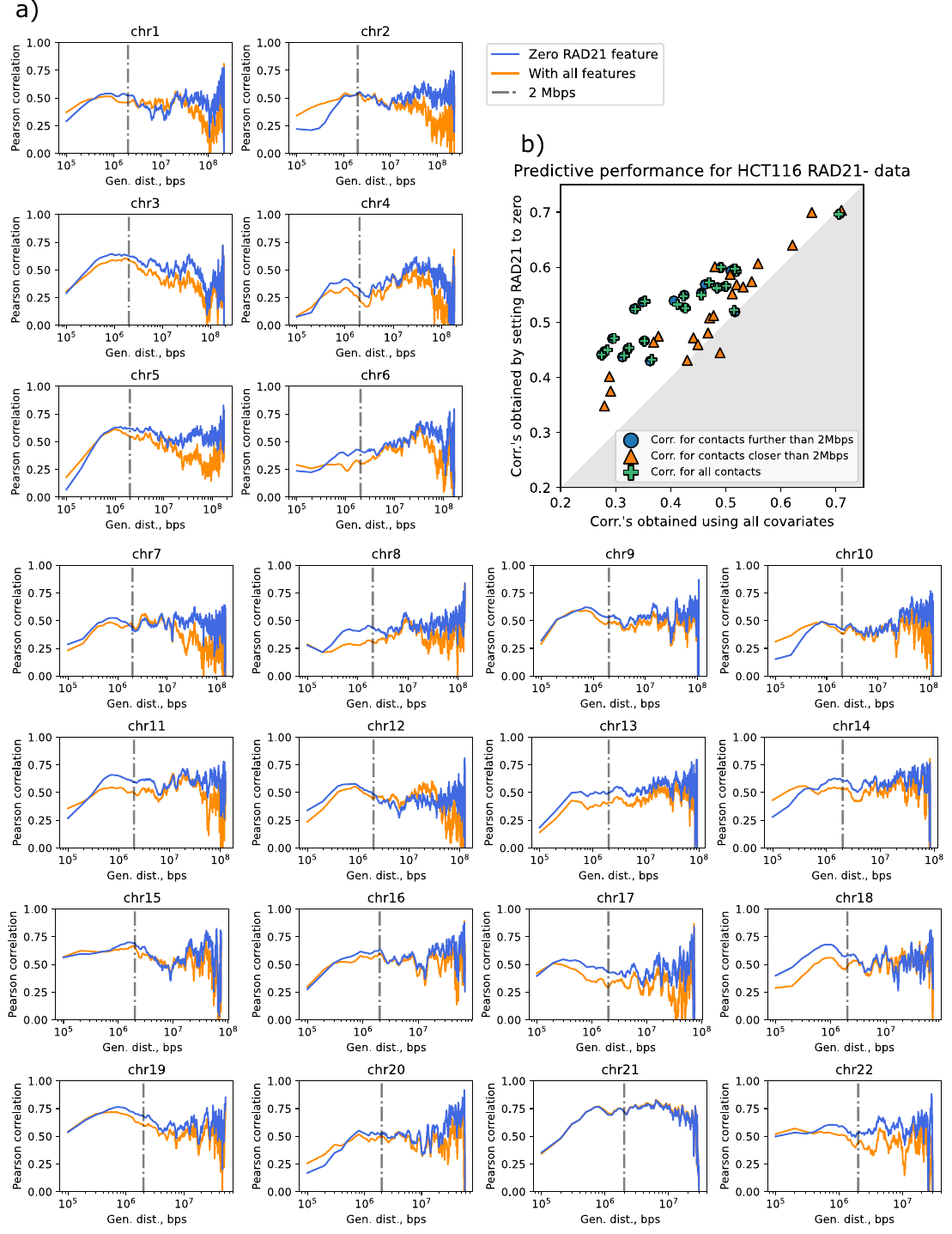}
    \caption{(a) Correlations between experimental maps for HCT116 cells after RAD21-degradataion (\cite{rao_cohesin_2017}), and prediction made with or without the input RAD21 track, for each chromosome, stratified by genomic distance. (b) Correlations computed using: (i) all contacts, (ii) short-range contacts at distances shorter than 2Mbps, (iii) long-range contacts at higher distances. }
    \label{fig:corrs_vs_gen_dist_hct}
\end{figure}

\begin{figure}
    \centering
    \includegraphics[width=0.85\linewidth]{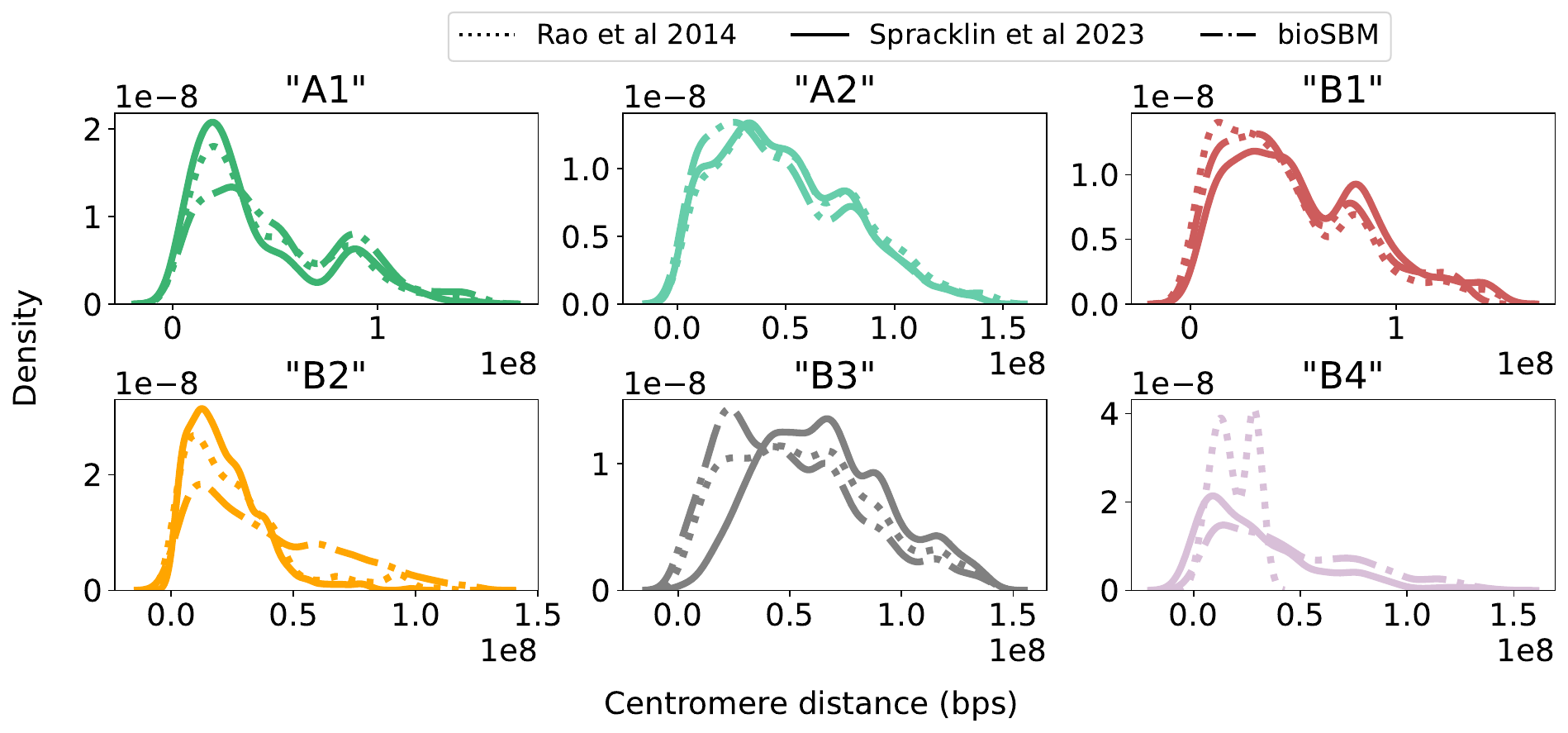}
    \caption{Histograms of centromere distance for different clusters. Dotted lines represent distributions for the subcompartments defined in~\cite{rao_3d_2014}. Solid lines correspond to the six clusters obtained using the spectral clustering algorithm from~\cite{spracklinDiverseSilentChromatin2023}, matched post hoc to the most similar subcompartment (see Sec.~\ref{sec:clustering}). Dash-dotted lines indicate clusters derived by applying k-Means to the membership vectors inferred by bioSBM. The absence of clearly distinct ``centromere-proximal'' or ``centromere-distal'' clusters suggests that these groupings do not reflect trivial spatial biases, which could otherwise indicate overestimation of the number of biologically meaningful clusters.}
    \label{fig:centrome-distances}
\end{figure}

\clearpage


\end{document}